\documentclass[a4paper,12pt]{article}
\pdfoutput=1
\usepackage{setspace}
\usepackage[USenglish]{babel}
\usepackage[utf8]{inputenc}
\usepackage{csquotes}
\usepackage{amsmath}

\usepackage{subfiles}
\usepackage{nameref}
\usepackage{array,multirow,graphicx}
\usepackage{authblk}
\usepackage{stmaryrd}

\usepackage{subfloat}
\usepackage{subcaption}
\usepackage[labelfont=bf,justification=justified,singlelinecheck=true]{caption}
\usepackage{mathtools}
\usepackage{float}
\usepackage{graphicx} 
\graphicspath{{Figures/}}

\usepackage[square,numbers,sort&compress]{natbib}
\usepackage[hyperindex,breaklinks]{hyperref}

\usepackage{xcolor}

\newcommand{\norm}[1]{\left\lVert#1\right\rVert}

\title{An efficient and quantitative phase-field model for elastically heterogeneous two-phase solids based on a partial rank-one homogenization scheme}
\author{\small{Sourav Chatterjee$^{a,b,\star}$, Daniel Schwen$^c$, Nele Moelans$^a$}}
\date{%
   \small{$^a$Department of Materials Engineering, KU Leuven, Kasteelpark Arenberg 44, Leuven BE-3001, Belgium}\\%
    \small{$^b$\textcolor{black}{Department of Materials Science and Engineering, University of Florida, Gainesville, 32611, FL,  USA}}\\%
   \small{$^c$Computational Mechanics and Materials Department, Idaho National Laboratory, Idaho Falls, ID 83415, United States}\\
    \small{$^*$Corresponding author. 
    E-mail addresses:}\\
    \small{soran.chatterjee@gmail.com (S. Chatterjee), daniel.schwen@inl.gov (D. Schwen), nele.moelans@kuleuven.be (N. Moelans)}
}

\begin{document}
\maketitle
\section*{\center Abstract}
This paper presents an efficient and quantitative phase-field model for elastically heterogeneous alloys that ensures the two mechanical compatibilities---static and kinematic, in conjunction with chemical equilibrium within the interfacial region. Our model contrasts with existing phase-field models that either violate static compatibility or interfacial chemical equilibrium or are computationally costly. For computational efficiency, the partial rank-one homogenization (PRH) scheme is employed to enforce both static and kinematic compatibilities at the interface. Moreover, interfacial chemical equilibrium is ensured by replacing the composition field with the diffusion potential field as the independent variable of the model. Its performance is demonstrated by simulating four single-particle and one multi-particle cases for two binary two-phase alloys: Ni-Al $\gamma^{\prime}/\gamma$ and UO$_2$/void. Its accuracy is then investigated against analytical solutions. For the single-particle $\gamma^{\prime}/\gamma$ alloy, we find that the accuracy of the phase-field results remains unaffected for both planar and non-planar geometries when the PRH scheme is employed. Fortuitously, in the UO$_2$/void simulations, despite a strong elastic heterogeneity---the ratio of Young's modulus of the void phase to that of the UO$_2$ phase is $10^{-4}$---we find that the PRH scheme shows significantly better convergence compared to the Voigt-Taylor scheme (VTS) for both planar and non-planar geometries. Nevertheless, for the same interface width range as in the $\gamma^{\prime}/\gamma$ case, the interface migration in these simulations shows dependence on interface width. Contrary to the $\gamma^{\prime}/\gamma$ simulations, we also find that the simulated elastic fields show deviations from the analytical solution in the non-planar UO$_2$/void case using the PRH scheme.\\
\textit{Keywords:} Alloy; Interface; Micro-mechanics; Non-homogeneous media; Thermodynamics of solids

\section{Introduction}
Several two-phase alloys are elastically heterogeneous, anisotropic and multicomponent. Their mechanical properties are usually controlled by the microstructure formed during solid-state phase transformation. It has been found both experimentally \cite{Ardell1966, Miyazaki1989, Marquis2001, Lund2002, Sudbrack2008} and numerically \cite{Johnson1984,  Johnson1990, Voorhees1992, Socrate1992, Abinandanan19931, CHSu1996, Jou1997, Akaiwa2001, Thornton2004a, Thornton2004b, Li1998, Vaithyanathan2002, Zhu2004, Guru2007} that the microstructure in an elastically constrained alloy system is significantly different from an unstressed alloy system. Moreover, LSW (Lifshitz-Slyozov-Wagner) type coarsening theories \cite{Ardell1972, Morral1994} for unstressed non-dilute alloys indicate that thermochemical properties and particle-matrix interfacial free energy are the two primary factors controlling the rate of transformation. While, in an elastically constrained alloy, factors—such as elastic anisotropy, heterogeneity and misfit strains—also controls the transformation kinetics (see \cite{Doi1996, Fratzl1999, Ardell2014} for reviews). However, a quantitative understanding of the combined effect of thermochemical properties, interfacial energy and elastic factors on multicomponent two-phase microstructures is not well-established.

The classical formulation of a two-phase alloy undergoing coherent first-order solid-state transformation is a free-boundary problem \cite{Fried1999, Gurtin1993}, where the precipitate-matrix interface is treated as a zero-thickness surface. It requires solving a coupled set of diffusion and mechanical equilibrium equations within the bulk domains subjected to interfacial jump conditions as boundary conditions (see  \cite{Voorhees2004}, \cite{Gurtin2010} for details). But for arbitrary precipitate morphologies, it can only be solved using numerical techniques \cite{Voorhees1992}. Few such studies \cite{Voorhees1992}, \cite{Abinandanan19932}, \cite{Jou1997},  \cite{CHSu1996}, \cite{Thornton2004a}, \cite{Duddu2011}, \cite{Zhao2015} have successfully simulated the elastically constrained two-phase evolution using this classical formulation. This is mainly because of the numerical difficulty of tracking the interface position. To overcome this limitation, these studies have made the following simplifying assumptions: fixed particle shape  \cite{Abinandanan19931}; quasi-static binary diffusion  \cite{CHSu1996}, \cite{Jou1997}, \cite{Thornton2004a}; and elastic homogeneity \cite{Thornton2004a}, \cite{Duddu2011}, \cite{Zhao2015}. 

In a phase-field model (PFM), however, explicit interface tracking is not needed, as the interface is defined implicitly by a scalar field variable. Hence, Chen, Wang and Khachaturyan's approach \cite{Wang1993, LQChen2002,Khachaturyan1983} combined with the Fourier spectral method \cite{Chen1998}, \cite{SYHu2001} has so far been the mainstay for simulating periodically stressed microstructures (see \cite{Li1998, Vaithyanathan2002, Zhu2004, Guru2007} as few applications). Unfortunately, a drawback of this PFM is that the diffuse interface width cannot always be treated as an independent parameter for numerical convenience \cite{Plapp2011}. This is often needed since the desired microstructural length scale, say in micrometers, is at least $10^{3}$ times that of a \textit{real} solid-solid interface width \cite{Karma2001}, \cite{Steinbach2009}. Consequently, it is difficult to simultaneously resolve both these length scales in a computationally efficient manner unless the bulk and interfacial properties are independent \cite{Plapp2007}. Therefore, two equivalent alloy PFMs for unstressed solids — one a Helmholtz-based functional \cite{Tiaden1998}, \cite{Kim1999} and another a grand-potential-based functional \cite{Plapp2011}—have been developed. Depending on numerical convenience, the interface width in these models can be treated as an independent parameter without affecting the simulation accuracy.  This assertion is, however, only valid when the driving force due to bulk contributions is small \cite{NProvatas2011}. Such models are called  ``thin-interface" PFMs \cite{Karma2001}.

Steinbach and Apel \cite{Steinbach2006} first attempted to formulate such a thin-interface multi-phase field model for elastically stressed alloys starting from a Helmholtz-based functional. Analogous to their chemical PFM \cite{Eiken2006}, they assumed equality of elastic stresses in the diffuse interface, thus ensuring local mechanical equilibrium \cite{Steinbach2006}. However, Durga \textit{et al.} \cite{DurgaA2013}  showed that in this model, the simulation results depend on the interface width and attributed this to the ``excess" interfacial bulk driving force that depends on discontinuous elastic fields. They also found that schemes that only ensure kinematic compatibility (continuity of displacement fields), such as Voigt-Taylor and Khachaturyan schemes, are interface-width dependent.  Subsequently, they suggested a Helmholtz-based model \cite{Durga2015}  that ensures both kinematic compatibility and mechanical equilibrium (continuity of traction vectors) in the interfacial region. Motivated by the same principle, Schneider \textit{et al.}  then suggested another equivalent phase-field model for coherent two-phase \cite{Schneider2015} and multi-phase \cite{Schneider2017} solids without the chemical contribution. Following this, Svendsen \textit{et al.} \cite{Svendsen2018} generalized Helmholtz-based formulations that ensure static and kinematic compatibilities to multiphase multicomponent solids undergoing finite deformations. Contrary to both Durga's \cite{DurgaA2013} and Svendsen's \cite{Svendsen2018} formulations, Tschukin \textit{et al.} \cite{Tschukin2017a}  extended Schneider's linear elastic model \cite{Schneider2015}  for two-phase solids by incorporating chemical effects starting from a grand-potential functional. This model was later applied to study bainitic transformation in steels \cite{KAmos2018, Schoof2020}. However,  the simulations presented in these works \cite{Tschukin2017a, KAmos2018, Schoof2020} have two limitations: i) stresses due to compositional heterogeneity within the bulk phases are not included, i.e., the eigenstrains are independent of the composition (or diffusion potential) field; and ii) the system is elastically isotropic. It bears emphasis that the first attempt to extend the grand-potential-based phase-field models to elastically stressed alloys was by Mushonegra \textit{et al.} \cite{Mushongera2015}. Their model, however, violates interfacial static compatibility. Nevertheless, Durga \cite{Durga2015} and Tschukin \cite{Tschukin2017a} showed that their simulated results are independent of interface width provided both mechanical compatibilities and chemical equilibrium are satisfied in the interfacial region, since then the interfacial bulk driving force is a function of only continuous fields.

To obtain the continuous elastic field components, however, these models  \cite{Durga2015},  \cite{Schneider2015}, \cite{Tschukin2017a} require several coordinate transformations of elastic fields from the sample reference frame to a local reference frame which is defined at each grid point, and are therefore computationally intensive \cite{Tschukin2017}. Particularly, when the inclusion is heterogeneous, anisotropic, and non-planar, such transformations would be costly since the global and local reference frames are not aligned. It is plausible that because of this, such approaches have so far not been implemented in any general finite element framework. As a solution, Tschukin \cite{Tschukin2017} suggested to circumvent this by using a projection tensor to split the elastic fields into their tangential and normal components relative to the interface. Nevertheless, his approach still requires the decomposition of elastic fields into continuous and discontinuous components. An alternative approach to these ``thin-interface" models is the partial rank-one homogenization scheme proposed by Mosler  \textit{et al.} \cite{Mosler2014}. Its advantage is that it satisfies the mechanical jump conditions but does not require the decomposition of elastic fields. Recently, Bartels \textit{et al.} \cite{Bartels2021} used this scheme to formulate a phase field coupled with diffusion and mechanical equilibrium equations.

However, Bartels' model \cite{Bartels2021} does not guarantee interfacial chemical equilibrium since it tacitly assumes equal composition in the interfacial region, similar to the model proposed by Wheeler \textit{et al.} \cite{Wheeler1992}. As a consequence of this assumption, Kim \textit{et al.} \cite{Kim1998} showed that there is an upper limit on the interfacial thickness that varies with bulk chemical properties. This paper, therefore, focusses on formulating a PFM that ensures mechanical compatibilities and interfacial chemical equilibrium. It should be noted that since the mechanical driving force remains the same, this could be achieved either starting from a Helmholtz-based functional  \cite{Tiaden1998} \cite{Kim1999} or a grand-potential-based functional \cite{Plapp2011}. In this work, we extend the latter approach because of the potential computational advantage of not explicitly solving for the equality of diffusion potentials (see \cite{Plapp2011} for details). Moreover, to model stresses engendered by compositional heterogeneity, we show that both the eigenstrains and the stiffness tensors can be formulated as functions of solute diffusion potentials and, thus, are indirectly composition-dependent. We further investigate to what extent the interface width in this model can be adjusted for numerical convenience without affecting the simulation accuracy. 

This work also compares the PRH scheme with the Voigt-Taylor scheme (VTS). Since it is difficult to implement and compare all existing schemes \cite{Durga2015}, our choice was motivated by the following two reasons: first, Ammar \cite{Ammar2010}, based on Cahn-Larche's \cite{Cahn1984} analytical solution, found that the Reuss-Sachs scheme does not predict the equilibrium two-phase compositions in coherently stressed solids correctly. Therefore, we did not compare PRH with this scheme. Second, Khachturayan (KHS) and VTS schemes \cite{Ammar2010}, \cite{DurgaA2013}  are equivalent because the total (or compatible) strains $\boldsymbol{\epsilon}$ are assumed to be locally equal in both phases, and thus they satisfy kinematic compatibility at the interface. But the mechanical stresses and interfacial driving force at the interface are slightly different due to differences in schemes of interpolation. Consequently, the field equations are also different. Specifically, for a two phase-system, it can be shown that the ``interfacial driving forces'' due to mechanical contributions are \cite{Durga2015}, \cite{Ammar2010}:
\begin{align}
\frac{\partial f^{khs}_{el}}{\partial \phi} &= -h^{\prime}(\phi)\left\lbrace  \frac{1}{2}\llbracket \boldsymbol{\mathcal{C}}\rrbracket:\left(\boldsymbol{\epsilon} - \langle \boldsymbol{\epsilon}^{\star}\rangle \right)- \left[\langle\boldsymbol{\mathcal{C}}\rangle:\left(\boldsymbol{\epsilon} - \langle\boldsymbol{\epsilon}^{\star}\rangle\right)
\right]: \llbracket\boldsymbol{\epsilon}^{\star}\rrbracket\right\rbrace, \label{EqnRC1}\\
\frac{\partial f^{vts}_{el}}{\partial \phi} &= -h^{\prime}(\phi)\left\lbrace  \frac{1}{2}\llbracket \boldsymbol{\mathcal{C}}\rrbracket:\boldsymbol{\epsilon}  - \frac{1}{2}\llbracket\boldsymbol{\mathcal{C}}:\boldsymbol{\epsilon}^{\star}\rrbracket\right\rbrace\label{EqnRC2},
\end{align}
where $f_{el}$ is the elastic strain energy density; $\phi$ is the phase-field variable such that the subdomain where $\phi=1(0)$ is the $\beta\text{-inclusion} (\alpha\text{-matrix})$ phase; $h^{\prime}(\phi)$ is the first derivative of the interpolation function $h(\phi)$ with respect to $\phi$; $\boldsymbol{\mathcal{C}}$ is the elastic modulus; and $\boldsymbol{\epsilon}^{\star}$ is the eigenstrain. For sake of brevity, we have used the short-hand notations: $\llbracket \boldsymbol{\Phi}\rrbracket = \left(\boldsymbol{\Phi}^{\alpha} - \boldsymbol{\Phi}^{\beta}\right)$  and $\langle\boldsymbol{\Phi} \rangle = \boldsymbol{\Phi}^{\beta}h(\phi) + [1-h(\phi)] \boldsymbol{\Phi}^{\alpha}$, to indicate the jump and interpolation of the quantity $\boldsymbol{\Phi}$ across the interface. It is apparent from Eqs. (\ref{EqnRC1}) \& (\ref{EqnRC2}) that the interfacial driving forces become equal for the special case when there are no eigenstrains in the system. It can also be shown that the mechanical stresses are equal, and hence the schemes are identical for this special case. Recently, Aagesen \textit{et al.} \cite{Aagensen2017} and Simon \textit{et al.}  \cite{Simon2020} compared these two schemes. They recommended the KHS scheme over the VTS scheme due to a comparatively smaller absolute ``excess" energy contribution in the former compared to the latter. However, they also found changes in interfacial energy with variation in interface width in both of these schemes. Due to the aforementioned reasons, we think that our contribution in this paper of comparing with only VTS is incomplete but reasonable.

The paper is organized as follows. In Section 2, we formulate an efficient and quantitative PFM based on the partial rank-one homogenization scheme. We also detail the parameters and material properties required to simulate two model binary alloy systems—$\gamma/\gamma^{\prime}$ and UO$_2$/void. Next, the performance of the model is demonstrated by simulating five test cases, which include both planar and non-planar geometries. Finally, the paper is concluded in Section 4.

\section{Formulation}
\subsection{Notations}
In this paper, we have denoted vectors, tensors and an array of scalar variables with boldface letters. We have employed Einstein summation convention, i.e., when an index is repeated in a term, it implies a summation from $1$ to $3$. A free index implies a range from $1$ to $3$, unless stated otherwise.

Let $Oxyz$ be a Cartesian reference frame, then  a vector field $\boldsymbol{v}$ at a point $\boldsymbol{x}$ and time $t$ is compactly written as $\boldsymbol{v}(\boldsymbol{x},t)= v_{i}\boldsymbol{e}_{i}$, where $v_{i}(\boldsymbol{x},t) = \boldsymbol{v}\cdot\boldsymbol{e}_{i}$ are the components of $\boldsymbol{v}$ relative to the chosen global orthonormal  basis $\left\{ \boldsymbol{e}_{i}\right\}$. We have written the dot and tensor products between two vectors $\boldsymbol{a}$ and $\boldsymbol{b}$ as: $\boldsymbol{a}\cdot \boldsymbol{b} = a_{i}b_{i}$ and $\boldsymbol{a}\otimes \boldsymbol{b} = (a_{i}b_{j})\boldsymbol{e}_{i} \otimes \boldsymbol{e}_{j}$, respectively. A tensor field is compactly written as $\boldsymbol{\sigma}(\boldsymbol{x},t) = \sigma_{ij}\boldsymbol{e}_{i}\otimes \boldsymbol{e}_{j}$, where $\sigma_{ij} = \boldsymbol{e}_{i}\cdot\boldsymbol{\sigma}\boldsymbol{e}_{j}$ are its components. The inner product of two tensors, say $\boldsymbol{\sigma}$ and $\boldsymbol{\epsilon}$, is written as $\boldsymbol{\sigma}:\boldsymbol{\epsilon} = \sigma_{ij}\epsilon_{ij}$. The transpose of a tensor $\boldsymbol{\sigma}$ is written as $\left(\boldsymbol{\sigma}\right)^{T}$. The gradient of a scalar field $\phi$ is written as $\nabla \phi = \phi_{,i}\boldsymbol{e}_{i}$. Similarly, the gradient of vector field $\boldsymbol{u}$ is compactly written as $\nabla \boldsymbol{u} = u_{j,i}\boldsymbol{e}_{i} \otimes \boldsymbol{e}_{j}$, where the comma before the index denotes the derivative with respect to $\boldsymbol{x}$.

\subsection{Field variables, jump and interpolation function}
We have restricted ourselves to an isothermal elastically heterogeneous $n$-component system consisting of two solid bulk phases, namely $\alpha$ and $\beta$. Let $\boldsymbol{u}(\boldsymbol{x}, t) $ denote the displacement field that is equal to the difference in the position of a particle $P$ at any instant $t$ in the deformed state to its position in the reference state, i.e., $\boldsymbol{u}(\boldsymbol{x},t) = \boldsymbol{x}(t) - \boldsymbol{x}^{\prime}$. The reference state is here assumed to be the stress-free $\alpha$ state. A scalar field variable, phase-field $\phi(\mathbf{x},t)$, distinguishes the $\alpha$ and $\beta$ phases. Specifically, when $\phi = 1$, it indicates the $\beta$ phase, and when $\phi=0$, it denotes the $\alpha$ region. The overall composition field of a solute $k$ is denoted as $c_{k}(\boldsymbol{x},t)$. Moreover, its conjugate field, the diffusion potential field, which is equal to the difference between the chemical potentials of solute $k$ and the solvent $n$, is written as $\tilde{\mu}_{k}(\boldsymbol{x},t)$. For multicomponent systems, to denote a list of $(n-1)$ solute diffusion potentials, the symbol $\boldsymbol{\tilde{\mu}} = \left\{\tilde{\mu}_{k=1,2\hdots n-1}\right\}$ is used. In addition,  the jump, $\llbracket \Psi \rrbracket$, of a field quantity $\Psi$ is defined as the difference between the values taken by the field in the $\beta$ phase from the $\alpha$ phase. These $\textit{phase}$ dependent fields are denoted with a superscript indicating the phases, for example $\Psi^{\theta=\alpha,\beta}$

Finally, an interpolation function, denoted by $h(\phi)$, is used to smoothly interpolate properties between the two bulk phases. We have taken $h(\phi)$ to be $ \phi^{3}\left(6\phi^{2} -15\phi + 10\right)$ \cite{Kim1999}, such that its value varies from $h(1) =1$ in the $\beta$ phase to $h(0)= 0$ in the $\alpha$ phase. 

\subsection{Kinematic and constitutive equations}
For small deformations, the total (or constrained) strain  $\boldsymbol{\epsilon}(\boldsymbol{u})$  is equal to $(1/2)\left[\text{grad}\,\boldsymbol{u} + \left(\text{grad}\, \boldsymbol{u}\right)^{T}\right]$ \cite{Eshelby1957}, \cite{Mura1991}. Following Mosler's work \cite{Mosler2014}, \cite{Kiefer2017}, \cite{Bartels2017}, we assume that the total strain $\boldsymbol{\epsilon}(\boldsymbol{u})$ varies smoothly between $\alpha$ and $\beta$ phases as 
\begin{align}
\epsilon_{ij}(\boldsymbol{u}) = [1-h(\phi)]\epsilon_{ij}^{\alpha} + h(\phi)\epsilon_{ij}^{\beta},
\label{Eqn3}
\end{align}
where $\boldsymbol{\epsilon}^{\theta=\alpha,\beta}$ are the total \textit{phase} strains. It should be noted that, in contrast to Mosler's original approach \cite{Mosler2014}, we have used the interpolation function $h(\phi)$ instead of the phase-field variable $\phi$ to smoothly interpolate the \textit{phase} strains across the interface. Our choice is motivated by following previous quantitative phase-field models for alloy solidification; specifically the works of Kim \textit{et al.} \cite{Kim1999} and Plapp \cite{Plapp2011},  where this function is used to interpolate both the bulk free energies and \textit{phase} compositions. Their argument for using this function is that it guarantees thermodynamic consistency of the model since $\left.h^{\prime}(\phi)\right|_{\phi=0,1} = \left.h^{\prime\prime}(\phi)\right|_{\phi=0,1} =0$ \cite{Wang1993}. Moreover, by using $\llbracket \boldsymbol{\epsilon}\rrbracket =  \left(\boldsymbol{\epsilon}^{\alpha} - \boldsymbol{\epsilon}^{\beta}\right)$, Eq. (\ref{Eqn3}) and solving for $\boldsymbol{\epsilon}^{\alpha}$ and $\boldsymbol{\epsilon}^{\beta}$, we see that \cite{Mosler2014,Kiefer2017}:
\begin{align}
\epsilon_{ij}^{\alpha}(\boldsymbol{\epsilon}, \llbracket \boldsymbol{\epsilon} \rrbracket,\phi) = e_{ij}^{\alpha} +\epsilon_{ij}^{\star\alpha} &=  \epsilon_{ij}(\boldsymbol{u}) + h(\phi)\llbracket \epsilon_{ij} \rrbracket,  \label{Eqn4} \\
\epsilon_{ij}^{\beta}(\boldsymbol{\epsilon}, \llbracket \boldsymbol{\epsilon} \rrbracket,\phi)  = e_{ij}^{\beta} + \epsilon_{ij}^{\star\beta}&=  \epsilon_{ij}(\boldsymbol{u}) - [1- h(\phi)]\llbracket \epsilon_{ij} \rrbracket ,
\label{Eqn5}
\end{align}
where $\llbracket \boldsymbol{\epsilon}\rrbracket$ is the jump in \textit{phase} strains, $\boldsymbol{e}^{\theta=\alpha,\beta}$ are elastic strains, and $\boldsymbol{\epsilon}^{\star\theta=\alpha,\beta}$ are eigenstrains. As noted in the Introduction, since $h(0)=0$ and $h(1) = 1$, according to Eqs. (\ref{Eqn4})-(\ref{Eqn5}), both $\boldsymbol{\epsilon}^{\alpha}$ and $\boldsymbol{\epsilon}^{\beta}$ become equal to the total strain $\boldsymbol{\epsilon}(\boldsymbol{u})$ within the bulk $\alpha$ and $\beta$ regions, respectively. However, their definition varies within the interfacial region depending on the value of $\llbracket\boldsymbol{\epsilon}\rrbracket$, which in turn depends on the homogenization assumption. Concretely, for the partial rank-one homogenization scheme \cite{Mosler2014}, the jump in total \textit{phase} strains $\llbracket \boldsymbol{\epsilon} \rrbracket$ must satisfy the Hadamard jump conditions \cite{Mosler2014}, \cite{Kiefer2017}, \cite{Bartels2017}. This yields
\begin{align}
\llbracket \epsilon_{ij} \rrbracket = (1/2)\left[a_{i}(\phi, \boldsymbol{\epsilon})n_{j}(\nabla \phi) + n_{i}(\nabla \phi)a_{j}(\phi, \boldsymbol{\epsilon}) \right],
\label{Eqn6}
\end{align}
where $\boldsymbol{n}\left(\boldsymbol{x}, t \right) = -\nabla \phi/|\nabla \phi|$ is the unit normal vector field directed from $\beta$ to $\alpha$ and $\boldsymbol{a}$ is a measure of strain jump. Contrary to this scheme, when all strain jump components are assumed to be identically zero, i.e., $\llbracket\boldsymbol{\epsilon}\rrbracket = 0$, the scheme reduces to the Voigt-Taylor \cite{Mosler2014} or the Khachaturyan schemes \cite{Durga2015}. Moreover, in the partial rank-one scheme, to find the unknown $\boldsymbol{a}$ in Eq. (\ref{Eqn6}) one must ensure local mechanical equilibrium at each point within the interface. This is equivalent to minimization of total energy with respect to $\llbracket\boldsymbol{\epsilon}\rrbracket$, as shown by Mosler \textit{et al.} \cite{Mosler2014}. Concretely, local mechanical equilibrium implies that the traction vectors $\boldsymbol{t}^{\theta}(\boldsymbol{x}, \boldsymbol{n})$ must be continuous in the interface
\begin{align}
t_{k}^{\alpha} - t_{k}^{\beta}= 0 \implies \left\{\sigma_{ik}^{\alpha}(\boldsymbol{e}^{\alpha}) - \sigma_{ik}^{\beta}(\boldsymbol{e}^{\beta})\right\}n_{i}\left(\nabla \phi\right) = 0,
\label{Eqn7}
\end{align}
where the elastic phase stresses $\boldsymbol{\sigma}^{\theta = \alpha, \beta}$ are given by Hooke's law
\begin{align}
\sigma_{ik}^{\theta} = \mathcal{C}_{ikjl}^{\theta}e^{\theta}_{jl}(\boldsymbol{\epsilon}, \llbracket \boldsymbol{\epsilon} \rrbracket, \phi, \boldsymbol{\tilde{\mu}}) = \mathcal{C}_{ikjl}\left(\epsilon_{jl}^{\theta} - \epsilon_{jl}^{\star\theta}\right).
\label{Eqn8}
\end{align}
By solving Eq. (\ref{Eqn7}) for the unknown $\boldsymbol{a}$, it can be shown that an expression for $\boldsymbol{a}$ can be analytically obtained (see Appendix A). Precisely,
\begin{align}
a_{j}(\phi, \boldsymbol{\epsilon}, \boldsymbol{n}) = - \mathcal{K}_{jk}^{-1}\left\{\llbracket \mathcal{C}_{kijl}\rrbracket\epsilon_{jl} - \left(\mathcal{C}_{kijl}^{\alpha}\epsilon_{jl}^{\star\alpha}-\mathcal{C}_{kijl}^{\beta}\epsilon_{jl}^{\star\beta}\right)\right\}n_{i}.
\label{Eqn9}
\end{align}
For brevity's sake, the exact expression for $\boldsymbol{\mathcal{K}}$ is given in Appendix A. From the right-hand side of Eq. (\ref{Eqn9}), we find that the magnitude of $\boldsymbol{a}$ depends on two phase-specific elastic properties. In the first term, it is the difference in stiffness tensors of the matrix and the precipitate phases, and thus the elastic heterogeneity. While, due to the term in the brackets, the magnitude of $\boldsymbol{a}$ also depends on the strength of \textit{phase} eigenstrains. It must also be emphasized that Mura \cite{Mura1991} has given similar expressions for $\boldsymbol{a}$ in a sharp-interface setting (see Eqs. $(6.8)$ \& $(6.12.5)$ in \cite{Mura1991}).  

Since the value $\boldsymbol{\mathcal{K}}$ is directly related to gradient of $\phi$, $\nabla \phi$, (see Eq. (\ref{EqnA2}) for exact formula), its value is \textit{nearly} zero in the bulk phases. Consequently, its inverse is very large in the bulk. However, these bulk values are redundant in the calculation; since it can be noticed from Eq. (\ref{Eqn9}) that $\boldsymbol{a}$ is directly dependent on $\nabla \phi$. Moreover, our experience shows that this inverse calculation is computationally costly and it affects the convergence of the simulation.  Therefore, we have set a cut-off value to distinguish the bulk from the interfacial region. Concretely, if the  $\norm{\nabla \phi}^{2} <$ $1\mathrm{e}{-18}$, then the inverse of $\boldsymbol{\mathcal{K}}$, the magnitude of $\boldsymbol{a}$ and the strain jump $\llbracket \boldsymbol{\epsilon} \rrbracket$ are all set to zero.

Next, to formulate the evolution equations, we make a constitutive assumption about the total energy within the two-phase system. This energy is based on a grand-potential functional $\Omega$ in which the independent variables are the continuous solute diffusion potentials $\boldsymbol{\tilde{\mu}}$ instead of the discontinuous solute compositions $\boldsymbol{c}$. The main incentive to start from a grand-potential functional is that the bulk chemical and interfacial contributions are independent in this model, as demonstrated by Plapp \cite{Plapp2011} for binary alloys and by Choudhury and Nestler \cite{Choudhary2012} for multicomponent alloys. To extend their idea to stressed multicomponent alloys, we assume that this functional is a function of the displacement field $\boldsymbol{u}$ as well. Specifically, 
\begin{align}
\Omega [\phi,\boldsymbol{\tilde{\mu}},\boldsymbol{u}] = \int_{V}\left[\omega_{bulk}(\phi,\boldsymbol{\tilde{\mu}},\boldsymbol{\epsilon}) + \omega_{int}(\phi,\nabla \phi)\right]dv,
\label{Eqn10}
\end{align}
where the bulk $\omega_{bulk}(\phi,\boldsymbol{\tilde{\mu}},\boldsymbol{\epsilon})$ and interfacial contributions $\omega_{int}(\phi, \nabla \phi)$ take the following forms
\begin{align}
\omega_{bulk}(\phi,\boldsymbol{\tilde{\mu}},\boldsymbol{\epsilon}, \boldsymbol{a}) &= h(\phi)\omega_{bulk}^{\beta}(\boldsymbol{\tilde{\mu}}, \boldsymbol{\epsilon}^{\beta}) + [1-h(\phi)]\omega_{bulk}^{\alpha}(\boldsymbol{\tilde{\mu}}, \boldsymbol{\epsilon}^{\alpha}),\label{Eqn11}\\
\omega_{int}(\phi, \nabla \phi) &= \kappa/2\norm{\nabla \phi}^{2} + mg(\phi) = \kappa/2\norm{\nabla \phi}^{2} +  m\phi^2(1-\phi)^2.
\label{Eqn12}
\end{align}
By definition, the bulk grand-potential density $\omega_{bulk}^{\theta}$ is  the Legendre transform of the Helmholtz free energy, that is
\begin{align}
\omega_{bulk}^{\theta}\left(\boldsymbol{\tilde{\mu}}, \boldsymbol{\epsilon}^{\theta}\right) = f_{bulk}^{\theta}(\boldsymbol{c}^{\theta}, \boldsymbol{\epsilon}^{\theta}) - \sum_{k=1}^{n-1}\tilde{\mu}_{k}c_{k}^{\theta}(\boldsymbol{\tilde{\mu}}),
\label{Eqn13}
\end{align}
where $f_{bulk}^{\theta}$ is the (Helmholtz) free energy density of phase $\theta$. It bears emphasis that Eq. (\ref{Eqn13}) is only valid when $f_{bulk}^{\theta}$ is a convex function of $\boldsymbol{c}$ \cite{Plapp2011}, and hence is not applicable for spinodal transformations. Moreover,  using $df_{bulk}^{\theta} = \boldsymbol{\sigma}^{\theta}:d\boldsymbol{\epsilon}^{\theta} + \sum_{k=1}^{n-1}\tilde{\mu}_{k}dc_{k}^{\theta}$, the differential of Eq. (\ref{Eqn13}) can be written as $d\omega_{bulk}^{\theta} = \boldsymbol{\sigma}^{\theta}:d\boldsymbol{\epsilon}^{\theta}  - \sum_{k=1}^{n-1}c_{k}^{\theta}d\tilde{\mu}_{k}$, where
\begin{align}
\sigma_{ij}^{\theta}(\boldsymbol{\epsilon}^{\theta}, \boldsymbol{\tilde{\mu}}) = \left.\frac{\partial \omega_{bulk}^{\theta}}{\partial \epsilon_{ij}^{\theta}}\right|_{\boldsymbol{\tilde{\mu}}}, \quad c_{t}^{\theta}(\boldsymbol{\epsilon}^{\theta}, \boldsymbol{\tilde{\mu}}) = \left.-\frac{\partial \omega_{bulk}^{\theta}}{\partial \tilde{\mu}_{t}}\right|_{\boldsymbol{\epsilon}^{\theta}}.
\label{Eqn14}
\end{align}
It should be noted that Eqs. (\ref{Eqn14}) can be used to calculate the \textit{phase} stress $\sigma_{ij}^{\theta}$ and the \textit{phase} composition $c_{t}^{\theta}$, provided an expression of $\omega_{bulk}^{\theta}$ is known. Such an expression can be obtained by Taylor expansion. Precisely,  
\begin{align}
\omega_{bulk}^{\theta}(\boldsymbol{\epsilon}^{\theta}, \boldsymbol{\tilde{\mu}}) = \omega_{chem}^{\theta}(\tilde{\boldsymbol{\mu}}) + (1/2)\mathcal{C}_{ijkl}^{\theta}(\boldsymbol{\tilde{\mu}})\left[\epsilon_{kl}^{\theta} - \epsilon_{kl}^{\star\theta}(\boldsymbol{\tilde{\mu}})\right]\left[\epsilon_{ij}^{\theta} - \epsilon_{ij}^{\star\theta}(\boldsymbol{\tilde{\mu}})\right],
\label{Eqn15}
\end{align}
where $\omega_{chem}^{\theta}(\tilde{\boldsymbol{\mu}})$ is the chemical grand-potential expressed as a function of diffusion potentials and the second term is the elastic strain energy contribution to the total bulk grand-potential of a phase $\theta$.  Here, $\omega_{chem}^{\theta}(\boldsymbol{\tilde{\mu}})$ is defined as the ratio of the molar grand-potential to molar volume, i.e., $\Omega_{m}^{\theta}(\boldsymbol{\tilde{\mu}})/V_{m}$. It should be noted that the molar grand-potential can be calculated from molar Gibbs energy $G_{m}^{\theta}$ using $\Omega_{m}^{\theta} = G_{m}^{\theta} - \sum_{k=1}^{n-1}\tilde{\mu}_{k}X_{k}^{\theta}$, provided the relation between \textit{phase} mole fraction and diffusion potential is invertible \cite{Plapp2011}, \cite{Choudhary2012}, \cite{Chatterjee2021}. Moreover, motivated by Fried and Gurtin \cite{Fried1999}, we assume that the elastic modulus $\boldsymbol{\mathcal{C}}(\boldsymbol{\tilde\mu})$ and the eigenstrain $\boldsymbol{\epsilon}^{\star}(\boldsymbol{\tilde\mu})$ to be dependent on the continuous diffusion-potentials. This assumption is made because the independent variables that represent solute components in the model are the diffusion potentials instead of compositions. This is only needed when elastic stresses due to compositional heterogeneity within the \textit{bulk phases} are included in the formulation.

Simon \textit{et al.} 
 \cite{Simon2020} argues that a possible ``limitation" of the grand-potential approach is that it is difficult to include stresses due to compositional heterogeneity, since the relationship between composition and diffusion potential may not be easily invertible when the elastic moduli and the eigenstrains depend directly on the composition variable. To overcome this, we assume a direct dependence of these constants on diffusion potential instead of composition to include such stresses in the formulation. 
 
Next, using the second relation in Eq. (\ref{Eqn14}) and using the expression given in Eq. (\ref{Eqn15}), the \textit{phase} molar density $c_{r}$ [mol/m$^{3}$] of the $r^{th}$ solute in phase $\theta$ can be explicitly written as
\begin{align}
c_{r}^{\theta}\left(\boldsymbol{\epsilon}^{\theta}, \boldsymbol{\tilde{\mu}}\right) &= \frac{X_{r}^{\theta}(\boldsymbol{\tilde{\mu}})}{V_{m}^{\theta}} - \frac{1}{2}\frac{\partial \mathcal{C}_{ijkl}^{\theta}}{\partial \tilde{\mu}_{r}}\left[\epsilon_{kl}^{\theta} -\epsilon_{kl}^{\star\theta}\right]\left[\epsilon_{ij}^{\theta} - \epsilon_{ij}^{\star\theta}\right] + \frac{\partial \epsilon_{ij}^{\star\theta}}{\partial \tilde{\mu}_{r}}\sigma_{ij}^{\theta}(\boldsymbol{\epsilon}^{\theta}) \label{Eqn16},
\end{align}
where $X_{r}^{\theta}(\boldsymbol{\tilde{\mu}}) = -\partial \Omega_{m}^{\theta}/\partial \tilde{\mu}_{r}$. To calculate the \textit{phase} mole fractions $X_{r}^{\theta}$ for  non-dilute and non-ideal alloys, the interested reader can refer to Refs. \cite{Choudhary2015}, \cite{Chatterjee2021}. It is apparent from Eq. (\ref{Eqn16}) that when the partial molar volumes are assumed to be equal, i.e., the eigenstrain and the elastic modulus are treated as constants, the solute \textit{phase} molar density, $c_{r}^{\theta}$, becomes independent of the elastic fields in the bulk phase $\theta$. To verify the thermodynamic consistency of Eq. (\ref{Eqn16}), it can be shown that the following Maxwell relation holds 
\begin{align}
\left.\frac{\partial \sigma_{ij}^{\theta}}{\partial \tilde{\mu}_{r}}\right|_{\boldsymbol{\epsilon}^{\theta}} = - \left.\frac{\partial c_{r}^{\theta}}{\partial \epsilon_{ij}^{\theta}}\right|_{\boldsymbol{\mu}}.
\label{Eqn17}
\end{align}
However, it is still unclear how the diffusion-potential dependence of eigenstrain can be calculated either analytically or experimentally. To show this, we rewrite Eq. (\ref{Eqn16}) by assuming a constant stiffness tensor, 
\begin{align}
c_{r}^{\theta}(\boldsymbol{\epsilon}^{\theta},\boldsymbol{\tilde{\mu}}) = \frac{X_{r}^{\theta}(\boldsymbol{\tilde{\mu}})}{V_{m}^{\theta}} + \left(\sum_{m=1}^{n-1}\frac{\partial \epsilon_{ij}^{\star\theta}}{\partial c_{m}^{\theta}}\chi_{mr}^{\theta}(\boldsymbol{\tilde{\mu}})\right)\sigma_{ij}^{\theta}(\boldsymbol{\epsilon}^{\theta}),
\label{Eqn18}
\end{align}
where we have used chain rule to express
\begin{align}
\frac{\partial \epsilon_{ij}^{\star\theta}}{\partial \tilde{\mu}_{r}} = \sum_{m=1}^{n-1}\frac{\partial \epsilon_{ij}^{\star}}{\partial c_{m}^{\theta}}\frac{\partial c_{m}^{\theta}}{\partial \tilde{\mu}_{r}} = \sum_{m=1}^{n-1}\frac{\partial \epsilon_{ij}^{\star\theta}}{\partial c_{m}^{\theta}}\chi_{mr}^{\theta}(\boldsymbol{\tilde{\mu}}).
\label{Eqn19}
\end{align}
Eq. (\ref{Eqn19}) shows that diffusion potential dependence of the eigenstrain can be determined, if the eigenstrain is known as a function of solute composition. In general, composition-dependent expression for eigenstrain can be obtained experimentally \cite{Ardell2014}. It should be noted that we have followed the notation introduced by Plapp \cite{Plapp2011} to write the inverse of thermodynamic factor matrix $\partial c_{m}^{\theta}/\partial \tilde{\mu}_{r}$  as $\chi_{mr}^{\theta}(\boldsymbol{\tilde\mu})$ in Eq. (\ref{Eqn19}). Moreover, it can be shown that for isotropic $\epsilon_{ij}^{\star}$ and binary alloys, Eq. (\ref{Eqn18}) reduces to Eq. (5.10) in Ref. \cite{Larche1985}. 
\subsection{Governing equations}
By taking the first variation of the total grand-potential functional, i.e., Eq. (\ref{Eqn10}), yields the following system of equations for $k=1\hdots (n-1)$
\begin{align}
h(\phi)c_{k}^{\beta}(\tilde{\boldsymbol{\mu}}) + \left[1-h(\phi)\right]c_{k}^{\alpha}(\tilde{\boldsymbol{\mu}}) - c_{k} &= 0,
\label{Eqn20}\\
\text{div}\left[h(\phi)\sigma_{ij}^{\beta}\left(\boldsymbol{e}^{\beta}\right) + [1-h(\phi)]\sigma_{ij}^{\alpha}(\boldsymbol{e}^{\alpha})\right] &= 0,
\label{Eqn21}\\
\dot{\phi} + L_{\phi}\left[mg^{\prime}(\phi) - \kappa\Delta\phi_{\theta} + \frac{\partial \omega_{bulk}}{\partial \phi} -\text{div}\left(\frac{\partial \omega_{bulk}}{\partial \nabla \phi}\right)\right]  &= 0,
\label{Eqn23}
\end{align}
where (see Appendix C)
\begin{align}
\frac{\partial \omega_{bulk}}{\partial \phi} &= 
h^{\prime}(\phi)\left[\left\{\omega_{b}^{\beta}\left(\boldsymbol{\tilde{\mu}},\boldsymbol{e}^{\beta}\right) - \omega_{b}^{\alpha}(\boldsymbol{\tilde{\mu}},\boldsymbol{e}^{\alpha})\right\}
+ \left\langle \sigma_{ij}\right\rangle\llbracket\epsilon_{ij} \rrbracket\right],
\label{Eqn24a}\\
\frac{\partial \omega_{bulk}}{\partial \nabla \phi} &=  h(\phi)[1-h(\phi)]\left\{\sigma_{jk}^{\alpha}\left(\boldsymbol{e}^{\alpha}\right) - \sigma_{jk}^{\beta}\left(\boldsymbol{e}^{\beta}\right)\right\}\frac{\partial n_{k}}{\partial \phi_{,i}}a_{j}.\label{Eqn25a}
\end{align}
Here $\left\langle \sigma_{ij}\right\rangle$ denotes $h(\phi)\sigma_{ij}^{\beta} + [1-h]\sigma_{ij}^{\alpha}$; $L_{\phi}$ is phase-field mobility; and $\dot{\phi}$ is rate of change of $\phi$. Notice that in Eq. (\ref{Eqn23}) we have used the Allen-Cahn equation \cite{Allen1979} to directly write the evolution equation for the phase-field variable $\phi$. Moreover, the evolution equation for the overall molar density of a solute $k$ is given by \cite{Agren1992}
\begin{align}
\dot{c}_{k} - \nabla\left(\sum_{j=1}^{n-1}L_{kj}\nabla \tilde{\mu}_{j}\right) &= 0,
\label{Eqn22}
\end{align} 
where $L_{kj}(\boldsymbol{\tilde{\mu}}, \phi) = L_{kj}^{\alpha}(\boldsymbol{\tilde{\mu}})[1-h(\phi)] + L_{kj}^{\beta}(\boldsymbol{\tilde{\mu}})h(\phi)$ are the components of the overall Onsager mobilities that varies smoothly across the interface. These components relate the flux of component $k$ to the gradient of diffusion potential of solute $j$. Also, notice that the components of  \textit{phase} Onsager mobilities $L_{kj}^{\theta=\alpha,\beta}$ are defined as functions of the continuous solute diffusion potentials, since they are the independent variables in this model. However, in this work, we have assumed constant Onsager mobilities in both phases (see Table \ref{Table1.1}) 

\subsection{Numerical method}
Although the derivation is variationally consistent, we neglected the last term in Eq. (\ref{Eqn23}) since it was leading to non-convergent solution in one of the simulated cases (see Appendix F).  Consequently, our implementation is non-variational. Nevertheless, it was found that the accuracy of the phase-field results remained unchanged by this choice (Appendix F).  Thus, Eqs. (\ref{Eqn20})-(\ref{Eqn23}) and (\ref{Eqn22}) were numerically solved to determine the unknown field variables $\boldsymbol{\tilde{\mu}}$, $\boldsymbol{u}$, $\phi$ and $\boldsymbol{c}$ subjected to initial and boundary conditions. We have used the MOOSE framework \cite{Gaston2009, Schwen2017} to solve these system of equations. To this end, we first non-dimensionalized these equations by introducing the following dimensionless variables: $\overline{\boldsymbol{x}} = \boldsymbol{x}/l_{c}$; $\overline{t} = t/t_{c}$; $\overline{\tilde{\mu}} = \tilde{\mu}/RT$; and $\overline{\boldsymbol{u}} =\boldsymbol{u}/l_{c} $, where $l_{c}$ is characteristic length, $t_{c}$ is characteristic time, $R$ is gas constant, and $T$ is simulation temperature. After change of variables and ignoring the last term in Eq. (\ref{Eqn23}), yields
\begin{align}
h(\phi)X_{k}^{\beta}(\overline{\tilde{\boldsymbol{\mu}}}) + \left[1-h(\phi)\right]X_{k}^{\alpha}(\overline{\tilde{\boldsymbol{\mu}}}) - X_{k} &= 0,
\label{Eqn24}\\
\overline{\text{div}}\left[h(\phi)\overline{\sigma_{ij}^{\beta}}\left(\boldsymbol{e}^{\beta}\right) + [1-h(\phi)]\overline{\sigma_{ij}^{\alpha}}(\boldsymbol{e}^{\alpha})\right] &= 0,
\label{Eqn25}\\
\overline{\dot{X}_{k}} - \overline{\nabla}\left(\sum_{j=1}^{n-1}\overline{L_{kj}}\overline{\nabla} \overline{\tilde{\mu}_{j}}\right) &= 0,
\label{Eqn26}\\
\overline{\dot{\phi}} + \overline{L_{\phi}}\left[g^{\prime}(\phi) -\overline{\kappa}\phi_{\theta} + \lambda_{1}\frac{\partial \omega_{chem}}{\partial \phi}   +\lambda_{2}\frac{\partial \omega_{elastic}}{\partial \phi} \right]  &= 0,
\label{Eqn27}
\end{align}
where $\overline{\left(\cdot\right)}$ indicates dimensionless quantities, and
\begin{align}
X_{k}^{\theta=\alpha,\beta} &= c_{k}^{\theta}V_{m},\\
\overline{L_{kj}} &= \left(L_{kj}t_{c}RT\right)/l_{c}^{2},\\
\overline{\kappa} &=  (\kappa/m\,l_{c}^{2}),\\
\lambda_{1} &= (RT/mV_{m}),\\
\lambda_{2} &= (\mu_{el}/m),\\
\partial \omega_{chem}/\partial \phi &= \frac{h^{\prime}(\phi)}{RT}\left[\Omega_{m}^{\beta},(\overline{\tilde{\boldsymbol{\mu}}}) - \Omega_{m}^{\alpha}(\overline{\tilde{\boldsymbol{\mu}}})\right],\label{EqnRV33}\\
\partial \omega_{elastic}/\partial \phi &= \frac{h^{\prime}(\phi)}{\mu_{el}}\left\{(1/2)\left[\sigma_{ij}^{\beta}e_{ij}^{\beta} - \sigma_{ij}^{\alpha}e_{ij}^{\alpha}\right] +\left(\sigma_{ij}^{\beta}h + \sigma_{ij}^{\alpha}(1-h)\right)\llbracket\epsilon_{ij} \rrbracket \right\}\label{EqnRV34}.
\end{align}
Further,  for the Ni-Al $\gamma/\gamma^{\prime}$ and UO$_2$/void  simulations, $\mu_{el}$ was taken to be equal to the shear modulus of the $\gamma^{\prime}$ phase  and UO$_2$ phase, respectively. Following this, the weak form (see Supplementary Material S$2$)  was formulated and the residuals resulting from Eqs. (\ref{Eqn24})-(\ref{Eqn27}) were implemented as ``Kernels'' in this software package. In MOOSE, since the discretized equations are solved iteratively, the Jacobian matrix is needed. In Appendices B and C, we have also derived the Jacobian terms specific to Eqs. (\ref{Eqn21}) \& (\ref{Eqn23}), respectively. We believe these Jacobian terms are most relevant for this study from the viewpoint of numerical convergence. We also note that in Eq. (\ref{Eqn23}) only the second last term is unique to our implementation, and therefore only its derivatives with respect to strain and phase-field are given in Appendix C. It is worth mentioning that manually coding the Jacobian matrix can be avoided using the automatic differentiation feature in MOOSE \cite{Schwen2017}. 

It is worth emphasizing that contrary to existing grand-potential based works \cite{Plapp2011}, \cite{Choudhary2012}, \cite{Mushongera2015}, \cite{Schoof2020}, \cite{Simon2020}, we do not formulate a diffusion potential evolution equation. Instead, we calculate the diffusion potential using Eq. (\ref{Eqn24}), and solve for the mole fraction variable using Eq. (\ref{Eqn26}). This means that we initialize our system with mole fractions instead of diffusion potentials. This is similar to the classical approach of Kim \textit{et al.}\cite{Kim1999}, and is useful because it is more intuitive to set an initial condition based on composition rather than diffusion potential. Particularly, in case of non-dilute and non-ideal alloys since composition cannot be analytically expressed as a function of diffusion potential \cite{Plapp2011, Chatterjee2021}. In Appendix E , we provide additional implementation details and the source code to calculate the diffusion potential using Eq. (\ref{Eqn24}).

For comparison with the VTS scheme, the PRH scheme is reduced by setting all components of the strain jump $\llbracket \boldsymbol{\epsilon} \rrbracket$ tensor to zero, while keeping Eqs. (26)-(29) unchanged.
It should however be noted that due to this the last term in Eq. (\ref{Eqn27}), i.e., the mechanical part of the ``bulk" driving force is different in the VTS scheme compared to the PRH scheme. In the PRH scheme, this driving force is consistent with the sharp-interface formulation (see Eq. (41) in  \cite{Larche1978}), while in the VTS case it is equal to the difference in elastic strain energy. This is the key difference between the two schemes.

\subsection{Parameters and material properties}
The interfacial properties in this model are controlled by three constant parameters: $\kappa$, $m$ and $L_{\phi}$. The first two parameters, $\kappa$ and $m$, in Eq. (\ref{Eqn23}) were calculated from interfacial energy $\sigma$ and interface width $l_{w}$ using the formula given by Kim \textit{et al.} \cite{Kim1999}: $
\kappa  = (3/\alpha)\sigma l_{w}$ and  $m = (6\alpha)(\sigma/l_{w}), $
where $\alpha$ was taken to be $2.94$; because the interface width, $l_{w}$, was defined as the region $0.05 <\phi< 0.95$ \cite{Kim1999}. The interfacial energies used for the $\gamma/\gamma^{\prime}$ and UO$_2$/void systems are given in Table \ref{Table1.1}. Moreover, the kinetic parameter $L_{\phi}$ for a binary A-B alloy was determined by rearranging Eq. (63) given in Ref. \cite{Kim2007} and using $f=(1/3)\sqrt(\frac{m}{2\kappa})$
\begin{align}
\frac{1}{L_{\phi}} = \frac{1}{f}\left(\frac{1}{M_{\phi}} + a_{2}\zeta\sqrt{\frac{\kappa}{2m}}\right),
\end{align}
where $a_{2} = \int_{0}^{1}\left(h(\phi)/\phi\right) d\phi$, $\zeta= \left(X_{B}^{\beta,eq} - X_{B}^{\alpha,eq}\right)^{2}/L_{BB}^{\beta}V_{m}$ and $M_{\phi}$ is interface mobility. Here $X_{B}^{\alpha,eq}$ and $X_{B}^{\beta,eq}$ are the equilibrium mole fractions when the alloy is in the unstressed state. Note that $\zeta$ is a material property and its value is given in Table \ref{Table1.1}. Since the value of $M_{\phi}$ was unavailable for the case of $\gamma/{\gamma^{\prime}}$ case, infinite interface mobility ($1/M_{\phi} \rightarrow 0 $) was assumed. While for the UO$_2$/void system, $M_{\phi}$ was taken to be $1.8926\mathrm{e}{-16}$ m$^{4}$/Js \cite{Greenquist2020}. Moreover, for the Ni-Al $\gamma^{\prime}/\gamma$ alloy, the bulk chemical grand-potentials were calculated by Taylor expansion about the equilibrium diffusion potentials \cite{Plapp2011}. Specifically, 
\begin{align}
\Omega_{m}^{\theta}(\tilde{\mu}_{Al}) = \Omega_{m}^{\theta,eq}(\tilde{\mu}^{eq}_{Al}) - X_{Al}^{\theta,eq}\left(\tilde{\mu}_{Al}-\tilde{\mu}^{eq}_{Al}\right) - \frac{1}{2\Theta_{Al}(\tilde{\mu}_{B}^{eq})}(\tilde{\mu}_{Al} - \tilde{\mu}_{Al}^{eq})^{2},
\label{Eqn25}
\end{align}
where $\theta=\gamma/\gamma^{\prime}$. Since only the difference in molar grand-potentials contributes to the chemical driving force (see Eq. \ref{Eqn27}), the first term in Eq. (\ref{Eqn25}) has no effect on phase transformation. Consequently, the phase mole fraction $X_{Al}^{\theta}(\tilde{\mu}_{Al})$ was calculated using
\begin{align}
X_{Al}^{\theta}(\tilde{\mu}_{Al})  = -\frac{\partial \Omega_{m}^{\theta}}{\partial \mu_{Al}} = \frac{\tilde{\mu}_{Al} - \tilde{\mu}_{Al}^{eq}}{\Theta_{Al}(\tilde{\mu}_{Al}^{eq})} + X_{Al}^{\theta,eq}.
\end{align}
Finally, for the UO$_2$/void alloy, we have used all thermochemical and kinetic properties from the work of Greenquist \textit{et al.} \cite{Greenquist2020} (see Table \ref{Table1.1}). In contrast to the Taylor approach, in their work, the grand-potential densities are worked out by assuming parabolic (Helmholtz) free energy densities $f_{\theta}$. Specifically, since the free energy density is related to molar Gibbs energy by $f_{\theta} = G_{m}^{\theta}/V_{m}$, we assumed a parabolic molar Gibbs energy of the form: $G_{m}^{\theta} = \Theta_{Va}^{\theta}/2(X_{Va}^{\theta} - X_{Va}^{\theta,eq})^{2}$, where $X_{Va}$ is vacancy mole fraction in phase $\theta$. This yields
\begin{align}
\Omega_{m}^{\theta}(\tilde{\mu}_{Va}) &= -\frac{1}{2}\left(\frac{\tilde{\mu}_{Va}^{2}}{\Theta_{Va}^{\theta,eq}}\right) - \mu_{Va} X_{Va}^{\theta,eq},\\
X^{\theta}_{Va}(\tilde{\mu}_{Va})  &= \left(\frac{\tilde{\mu}_{Va}}{\Theta_{Va}^{\theta,eq}}\right) + X_{Va}^{\theta,eq}.
\end{align}
For sake of simplicity, we have also assumed all elastic properties to be constant, and these are listed in Table \ref{Table1.2}.

\begin{table}[!ht]
\begin{center}
\caption{Constant material parameters for the Ni-Al and UO$_2$/void alloy systems. Here, TC stands for ThermoCalc; $\Theta_{B}$ is thermodynamic factor; the superscript $\alpha$ denotes the $\gamma$ and UO$_2$ phases; the superscript $\beta$ denotes $\gamma^{\prime}$ and void phases; the superscript $eq$ indicates that the properties are calculated at the unstressed equilibrium state; the subscript B is the diffusing solute which is Al in $\gamma/\gamma^{\prime}$and Va in UO$_2$/void alloy systems. Note that $\Theta_{B}^{\theta}$ and $V_{m}$ in this work are related to $k_{\theta}$ and $V_{a}$ in Greenquist \textit{et al.} \cite{Greenquist2020} by $\Theta_{B}^{\theta} = k_{\theta}V_{m}$ and $V_{m}= V_{a}N_{A}$, where $N_{A}$ is Avogadro number}
\label{Table1.1}
\begin{tabular}{p{4cm}l l l l}
\hline
 & Ni-Al alloy& Ref. & UO$_2$/void alloy & Ref. \\
\hline
$T$ $[\text{K}]$  &$1473$& - & $1816$ & \cite{Greenquist2020} \\
$\sigma$ [J/m$^{2}$]  & $36.2\mathrm{e}{-3}$&\cite{Sonderegger2009} & $3.2$ &\cite{Greenquist2020} \\
$V_{m}$ [m$^{3}$/mol] &$7.5\mathrm{e}{-5}$& -& $2.46\mathrm{e}{-5}$ &\cite{Greenquist2020}\\

$X_{B}^{\alpha,eq}$&$0.183922$& TC& $3.4252\mathrm{e}{-8}$ & \cite{Greenquist2020}\\
$X_{B}^{\beta,eq}$&$0.230730$& TC& $0.99999$ & \cite{Greenquist2020}\\
$\tilde{\mu}_{B}^{eq}$ [J/mol] & $-1.08531\mathrm{e}{5}$& TC &0&\\
$\Theta_{B}^{\alpha,eq}$ [J/mol] & $3.6937\mathrm{e}{5}$& TC &$6.02537\mathrm{e}{5}$&\cite{Greenquist2020} \\
$\Theta_{B}^{\beta,eq}$ [J/mol] & $2.86035\mathrm{e}{5}$& TC &$6.02537\mathrm{e}{6}$&\cite{Greenquist2020} \\
$L_{BB}^{\alpha, eq}$ [mol m$^{2}$/Js] &$7.1907\mathrm{e}{-19}$& TC& $1.3248\mathrm{e}{-24}$ &  \cite{Greenquist2020} \\
$L_{BB}^{\beta, eq}$ [mol m$^{2}$/Js] &$1.1094\mathrm{e}{-18}$& TC& $1.3248\mathrm{e}{-25}$ &  \cite{Greenquist2020} \\
$\zeta$ [ Js/ m$^{5}$] &$2.6332\mathrm{e}{19}$& - & $3.0632\mathrm{e}{28}$ & -\\
\hline
\end{tabular}
\end{center}
\end{table}

\begin{table}[!ht]
\begin{center}
\caption{Constant elastic properties for the Ni-Al and UO$_2$/void alloy systems. Here $\alpha$ denotes the $\gamma$ and UO$_2$ phases, while $\beta$ denotes $\gamma^{\prime}$ and void phases. The isotropic elastic properties for $\gamma$ and $\gamma^{\prime}$ phases were obtained from the work of Tien and Copley listed in Ref. \cite{Socrate1992}.  Moreover, the anisotropic elastic properties were obtained from Ref.       \cite{Ardell2014} at temperatureT=$1473$ K}
\label{Table1.2}
\begin{tabular}{p{2cm}llll}
\hline
 & $\alpha$ & Ref. &$\beta$& Ref.\\
\hline
Isotropic  & $E =  158$ GPa, & \cite{Socrate1992} &$E=144$ GPa, &\cite{Socrate1992}\\
$\gamma/\gamma^{\prime}$&$\nu = 0.3$ && $\nu= 0.3$&\\
\\
Isotropic  &$E=192$ GPa, &\cite{NIST_web} &$E = 1.92\mathrm{e}{-2}$ GPa,&-\\
UO$_2$/void&$\nu = 0.3$& &$\nu= 0.3$&\\
\\
Anisotropic& $C_{11} = 188.3$ GPa& \cite{Ardell2014} & $C_{11} = 194.37$ GPa& \cite{Ardell2014}\\
$\gamma^{\prime}/\gamma$&$C_{12} = 143.54$ GPa & &$C_{12} = 142.82$ GPa&\\
&$C_{44} = 80.734$ GPa& &$C_{44} = 84.04$ GPa&\\
\hline
\end{tabular}
\end{center}
\end{table}
\section{Results and discussion}
In this section, we discuss the performance of our model by simulating two planar single-particle, two non-planar single-particle and one multi-particle simulation. For simplicity's sake, we have assumed isotropic elastic constants in the first four simulations (see Table \ref{Table1.2}). Moreover, for the single-particle Ni-Al simulations, two scenarios are considered to study the effect of elastic fields on transformation kinetics: i) a dilatational eigenstrain $\boldsymbol{\epsilon}^{\star}  = \begin{psmallmatrix}-0.3\% & 0\\ 0 & -0.3\%\end{psmallmatrix}$ has been assumed in the $\gamma^{\prime}$ phase; and ii) with a zero eigenstrain. Henceforth, we refer to these two scenarios as Case I and Case II, respectively. This eigenstrain has been taken from the work of Tien and Copley listed in Table 2 of Ref. \cite{Socrate1992}.

Similarly, for the single-particle UO$_2$/void system, we have considered two scenarios: i) with applied load and ii) without applied load. We refer to these cases as case III and case IV, respectively.
It must be noted that no eigenstrain has been assumed in either UO$_2$ or void phase for these two cases. As emphasized in the Introduction, this means that the results obtained using the Voigt-Taylor scheme are also applicable for the Khachaturayan scheme since they become identical. For sake of clarity, Table \ref{table:bc} summarizes the assumed eigenstrains and applied boundary conditions for all simulated cases.

\begin{table}[ht]
\caption{Table summarizing the assumed eigenstrains and applied mechanical boundary conditions for all simulated cases.  Here, $\epsilon^{\star}$ is the eigenstrain, $\boldsymbol{u}$ is the displacement vector, $u_{x}$ is the x-component of the displacement, $u_{y}$ is the y-component of displacement, and $l_{c}$ refers to the characteristic simulation length which is a constant. Although not explicitly shown, Case II refers to the Ni-Al simulation with zero eigenstrains in both phases, and Case IV is the UO$_2$-Va simulation without any applied displacements.} 
\centering
\begin{tabular}{l l l }
\hline\hline
Simulation & Eigenstrains $[\text{Phase}]$ &  Boundary conditions  \\  
\hline
\multirow{2}{*}{Planar Ni-Al}&\multirow{2}{*}{$\boldsymbol{\epsilon}^{\star}$  $\left[\gamma^{\prime} \right]= -0.3\%\boldsymbol{1}$}& $\boldsymbol{u}$ (at left boundary) = $\boldsymbol{0}$ \\
 \multirow{2}{*}{(Case I)}&&$\boldsymbol{u}$ (at right boundary) = $\boldsymbol{0}$\\
 &{$\boldsymbol{\epsilon}^{\star}$  $\left[\gamma \right]= \boldsymbol{0}$}&\multirow{2}{*}{$\boldsymbol{u}$ is periodic along y-direction}\\
 &&\\
&& $\boldsymbol{u}$ (at left boundary) = $\boldsymbol{0}$ \\
 \multirow{2}{*}{Planar UO$_2$-Va}&\multirow{2}{*}{$\boldsymbol{\epsilon}^{\star}$  $\left[\text{void} \right]= \boldsymbol{0}$}&${u}_{x}/l_{c}$ (at right boundary) $= 5$\\
  \multirow{2}{*}{(Cases III)}&&${u}_{y}/l_{c}$ (at right boundary) $= -5$\\
 &{$\boldsymbol{\epsilon}^{\star}$  $\left[\text{uo}_2 \right]= \boldsymbol{0}$}&\multirow{1}{*}{$\boldsymbol{u}$ is periodic along y-direction}\\
 &&\\
\multirow{2}{*}{Non-planar Ni-Al}&\multirow{2}{*}{$\boldsymbol{\epsilon}^{\star}$  $\left[\gamma^{\prime} \right]= -0.3\%\boldsymbol{1}$}& $u_{x}$ (at left boundary) = 0 \\
 \multirow{2}{*}{(Case I)}&&$u_{y}$ (at bottom boundary) = 0\\
 &{$\boldsymbol{\epsilon}^{\star}$  $\left[\gamma \right]= \boldsymbol{0}$}&\multirow{2}{*}{traction is zero at outer boundary}\\
 &&\\
\multirow{2}{*}{Non-planar UO$_2$-Va}&\multirow{2}{*}{$\boldsymbol{\epsilon}^{\star}$  $\left[\text{void} \right]= \boldsymbol{0}$}& $u_{x}$ (at left boundary) = 0 \\
 \multirow{2}{*}{(Case III)}&&$u_{y}$ (at bottom boundary) = 0\\
 &{$\boldsymbol{\epsilon}^{\star}$  $\left[\text{uo}_2 \right]= \boldsymbol{0}$}&\multirow{1}{*}{$u_{x}$ (at outer boundary) = $0.1\%x$}\\
 &&\multirow{1}{*}{$u_{y}$ (at outer boundary) = $0.1\%y$}\\
 \\
\multirow{2}{*}{Multi-particle Ni-Al}&\multirow{2}{*}{$\boldsymbol{\epsilon}^{\star}$  $\left[\gamma^{\prime} \right]= -0.3\%\boldsymbol{1}$}& \multirow{4}{*}{$\boldsymbol{u}$ is periodic along both x and y} \\
 \multirow{2}{*}{(Case I)} &&\\
 &{$\boldsymbol{\epsilon}^{\star}$  $\left[\gamma \right]= \boldsymbol{0}$}&\multirow{2}{*}{}\\
 &&\\
\hline
\end{tabular}
\label{table:bc}
\end{table}
 
 \clearpage
\subsection{Planar Ni-Al $\gamma^{\prime}/\gamma$ simulation}
\begin{figure}[!ht]
\begin{center}
\begin{subfigure}{0.6\textwidth}
\includegraphics[keepaspectratio,width=\linewidth]{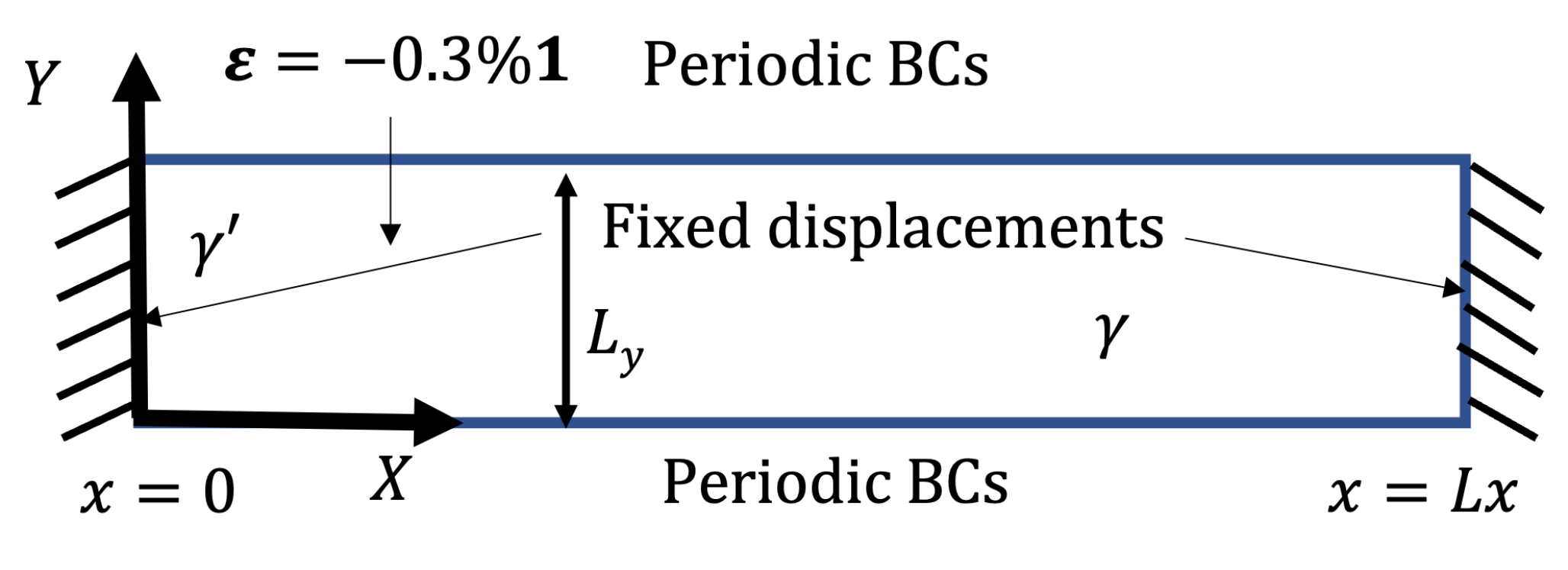}
\caption{}
\label{Fig.0new}
\end{subfigure}
\end{center}
\begin{center}
\begin{subfigure}{1.0\textwidth}
\includegraphics[trim=0 0 60 0, clip,keepaspectratio,width=\linewidth]{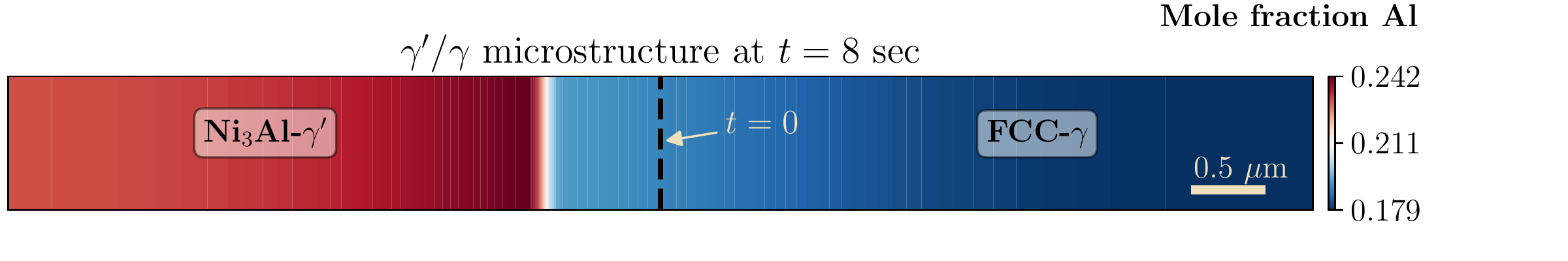}
\caption{}
\label{Fig.0a}
\end{subfigure}
\end{center}
\begin{subfigure}{0.50\textwidth}
\includegraphics[keepaspectratio,width=\linewidth]{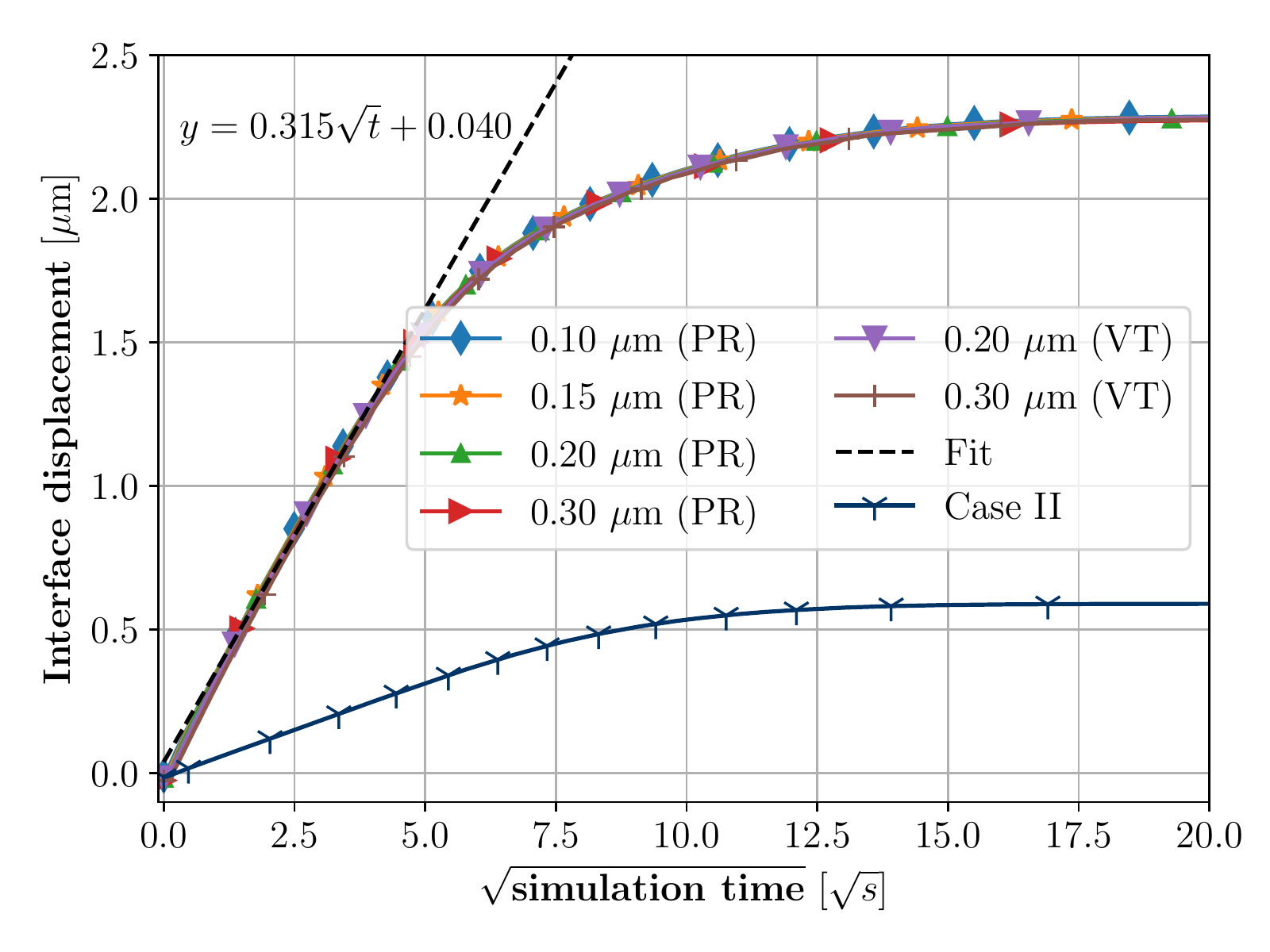}
\caption{}
\label{Fig.1a}
\end{subfigure}
\begin{subfigure}{0.50\textwidth}
\includegraphics[keepaspectratio,width=\linewidth]{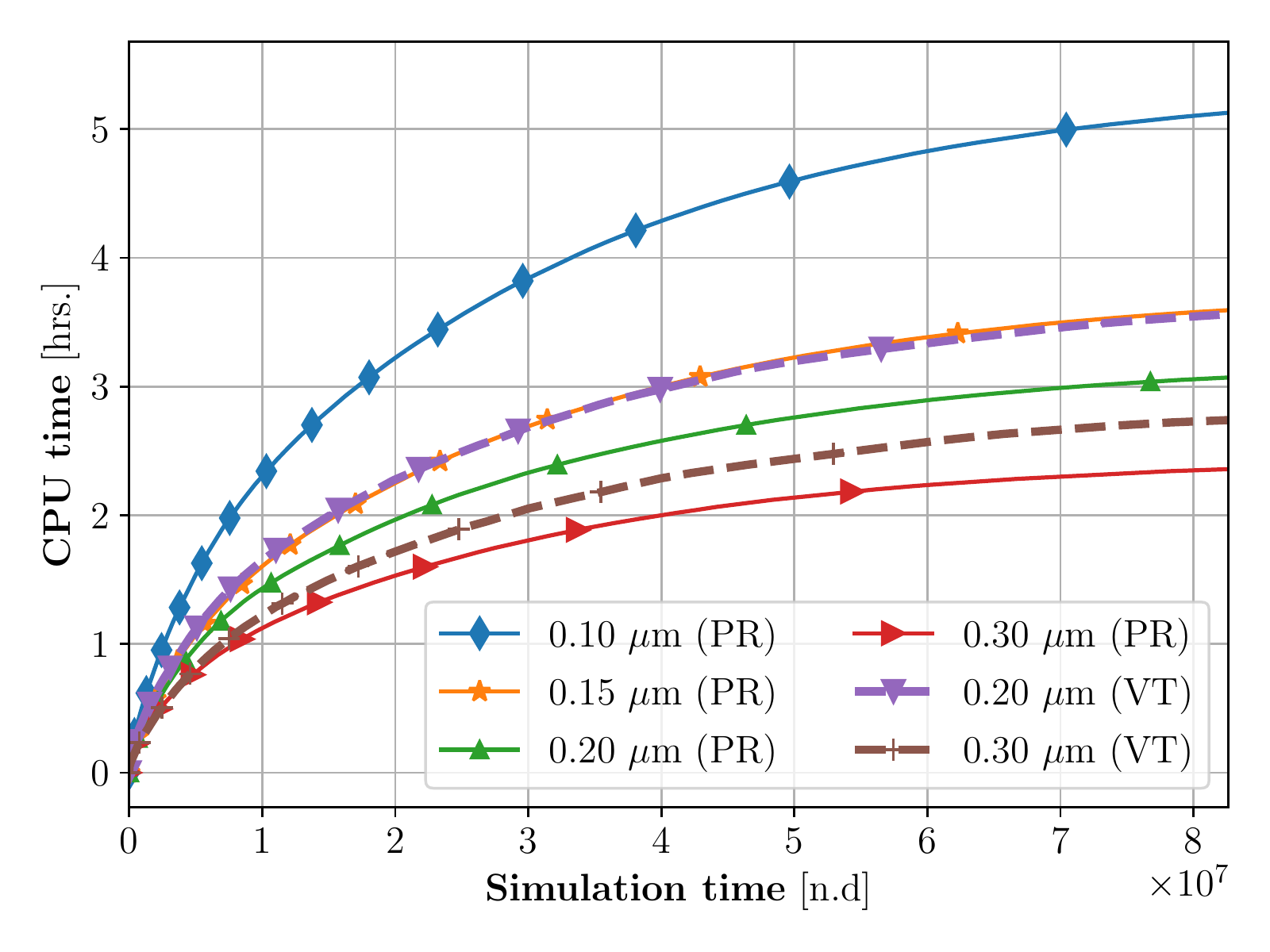}
\caption{}
\label{Fig.1b}
\end{subfigure}
\caption{a) Schematic showing the imposed boundary conditions and eigenstrains in the Ni-Al system. b) Simulated Al-mole fraction field in an elastically stressed $\gamma^{\prime}/\gamma$ diffusion couple ($10 \times 0.33$ $\mu$m$^{2}$) at time $t=8$ sec. b) Variation of $\gamma^{\prime}/\gamma$ interface displacement ($y$) as a function of the square root of simulation time ($t$) for four different interface widths ($l_{w} = 0.10$ $\mu$m to $0.30$ $\mu$m) simulated using the partial rank-one (PR) homogenization scheme. This displacement was calculated by first tracking the position of the phase-field variable $\phi=0.5$  and then subtracting it from its initial position. Calculated interface displacement using the Voigt-Taylor (VT) homogenization scheme for two different interface widths ($0.20$ $\mu$m and $0.30$ $\mu$m) are also superimposed on Fig.\ref{Fig.1a}. Subsequently, the data is fitted to a parabolic growth law for $t < 25 $s and the fitted curve is superimposed on Fig. \ref{Fig.1a}. The interface displacement with zero eigenstrain (\textcolor{black}{case II}) is also superimposed on Fig.\ref{Fig.1b}. For each simulated case, c) the variation of CPU time with non-dimensional simulation time. }
\end{figure}

We first selected an elastically heterogeneous and isotropic  Ni-Al $\gamma/\gamma^{\prime}$ alloy at $1473$ K. The interface is assumed to be planar, and periodic boundary conditions have been enforced along the y-direction (Fig.\ref{Fig.0new}). While the mechanical displacements ($\boldsymbol{u}$) are fixed, and zero Neumann boundary conditions are enforced on the phase-field ($\phi$) and mole fraction variables along the x-direction. That is for all y and $x=\left\{0, L_{x}\right\}$ (see Table {\ref{table:bc}} \& Fig. \ref{Fig.0new})
\textcolor{black}{
\begin{align}
\boldsymbol{u}(x,y) &= \boldsymbol{0},\\
\boldsymbol{n}\cdot\nabla \phi(x, y) &= 0,\label{BCI}\\
\boldsymbol{n}\cdot \nabla \tilde{{\mu}}(x,y) &= 0\label{BCII}.
\end{align}
}
\textcolor{black}{Here $\boldsymbol{n}$ is the outward unit normal to the external surface and $\tilde{\mu}$ is the diffusion potential of Al. Although not apparent from Fig. \ref{Fig.0new}, it bears emphasis that the longitudinal dimension $L_{x}$ in this simulation is 30 times that of the lateral dimension $L_{y}$. Moreover, two scenarios have been considered: i) with a dilatational eigenstrain in the $\gamma^{\prime}$ phase (case I); ii) no eigenstrains (case II). It should be emphasized that in case II, the transformation is driven solely by chemical driving forces. But in case I, both chemical (Eq. \ref{EqnRV33}) and mechanical (Eq. \ref{EqnRV34}) driving forces are present at the interface.}

Our simulations show that the $\gamma$ phase grows at the expense of $\gamma^{\prime}$ phase with increasing time (Fig. \ref{Fig.0a}).  For analysis, we focussed on the effect of two relevant model assumptions on simulation results: first is the choice of interface width, and second is the scheme of homogenization. For this reason, we compared the partial rank-one (PR) homogenization scheme with the Voigt-Taylor (VT) homogenization scheme for different interface width choices.

We find that the variation of interface position with time remains unaffected by our choice of interface width when simulated using both PR and VT schemes (Fig.\ref{Fig.1a}). \textcolor{black}{For case I, the $\gamma$ phase follows a parabolic growth law for time $t<25$s.} As the system reaches towards the equilibrium state, the growth of the $\gamma$ phase slows down. \textcolor{black}{We also observe a parabolic growth behaviour for case II (Fig. \ref{Fig.1a}). However, we find that the coherent $\gamma^{\prime}/\gamma$ phase boundary grows much faster in case I as compared to case II.}

\begin{figure}
\begin{subfigure}{0.50\textwidth}
\includegraphics[keepaspectratio,width=\linewidth]{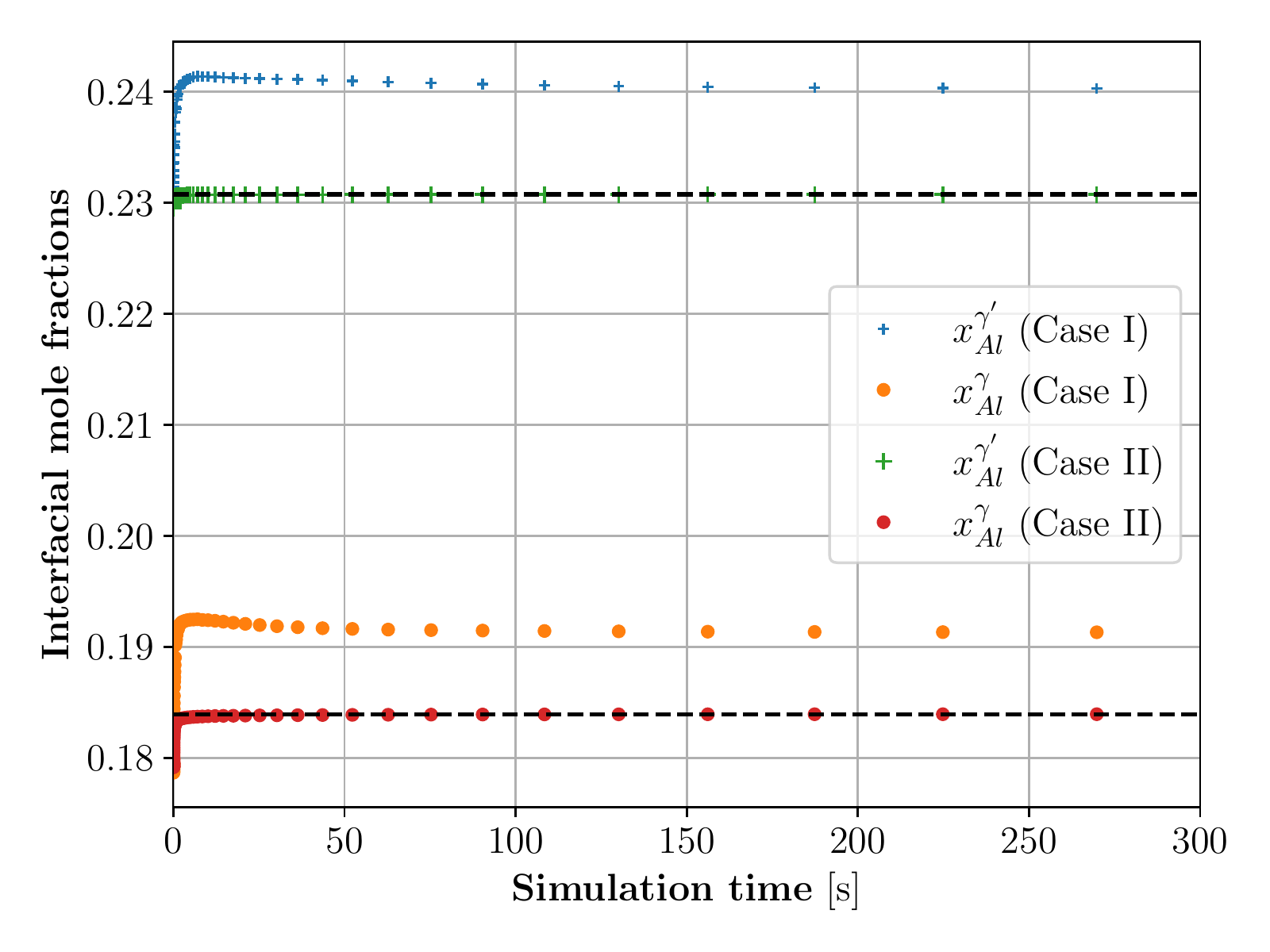}
\caption{}
\label{Fig.1c}
\end{subfigure}
\begin{subfigure}{0.50\textwidth}
\includegraphics[keepaspectratio,width=\linewidth]{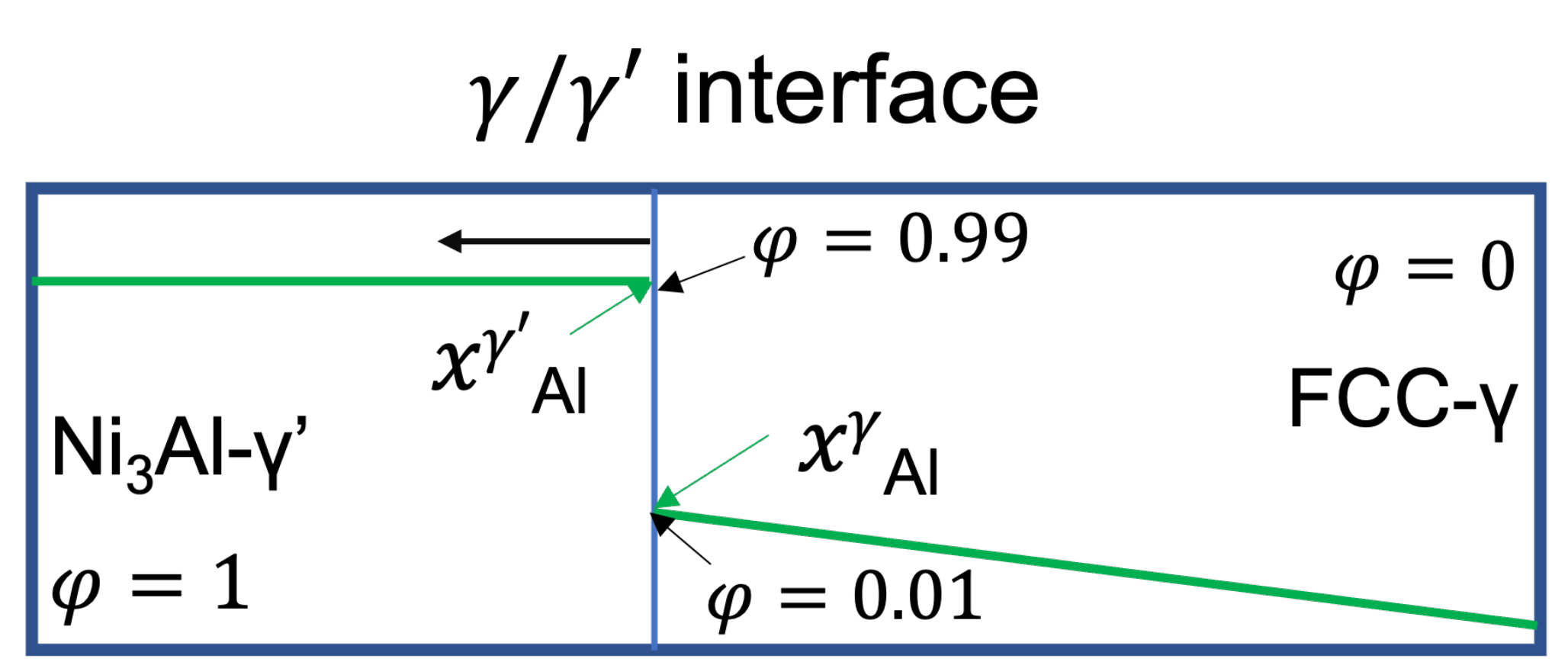}
\caption{}
\label{Fig.1d}
\end{subfigure}
\caption{\color{black}a) Variation of interfacial Al mole fractions in the $\gamma$ and $\gamma^{\prime}$ side of the interface as a function of time for the case of the planar interface. b) Schematic depicting the interfacial concentrations and the system geometry. Notice that the interfacial concentrations in both phases are higher in case I (with eigenstrains) as compared to case II (without eigenstrains). The dotted lines show the equilibrium concentrations in both phases, and $\phi$ represents the phase-field variable.}
\end{figure}

\textcolor{black}{This behaviour can be attributed to the shift in local interfacial concentrations due to the presence of elastic fields. Johnson \cite{Johnson1987} analytically calculated this shift in the case of a sharp-interface model. To verify this, we track the temporal evolution of the Al mole fraction on the $\gamma^{\prime} (\gamma)$ side of the interface by assigning the phase-field variable to be $\phi=0.99 (\phi=0.01)$ (Fig.{\ref{Fig.1d}}). It is evident from Fig.\ref{Fig.1c} that the interfacial concentrations in case II are higher compared to case I and are clearly due to the presence of elastic fields. Also, notice that the interfacial concentrations in case II deviate significantly from the equilibrium values (dotted line)}.

Interestingly, we also find that the PR scheme shows marginally better convergence than the VT scheme (Fig.\ref{Fig.1b}). For a given value of simulation time and interface width, the PR scheme always outperforms the VT scheme in terms of CPU time (Fig.\ref{Fig.1b}). Moreover, a trivial observation is that the CPU time decreases with increasing interface width (Fig.\ref{Fig.1b}). This shows the advantage of controlling the interface width in a phase-field model without loss of simulation accuracy.

\begin{figure}[ht]
\begin{subfigure}{0.50\textwidth}
\includegraphics[keepaspectratio,width=\linewidth]{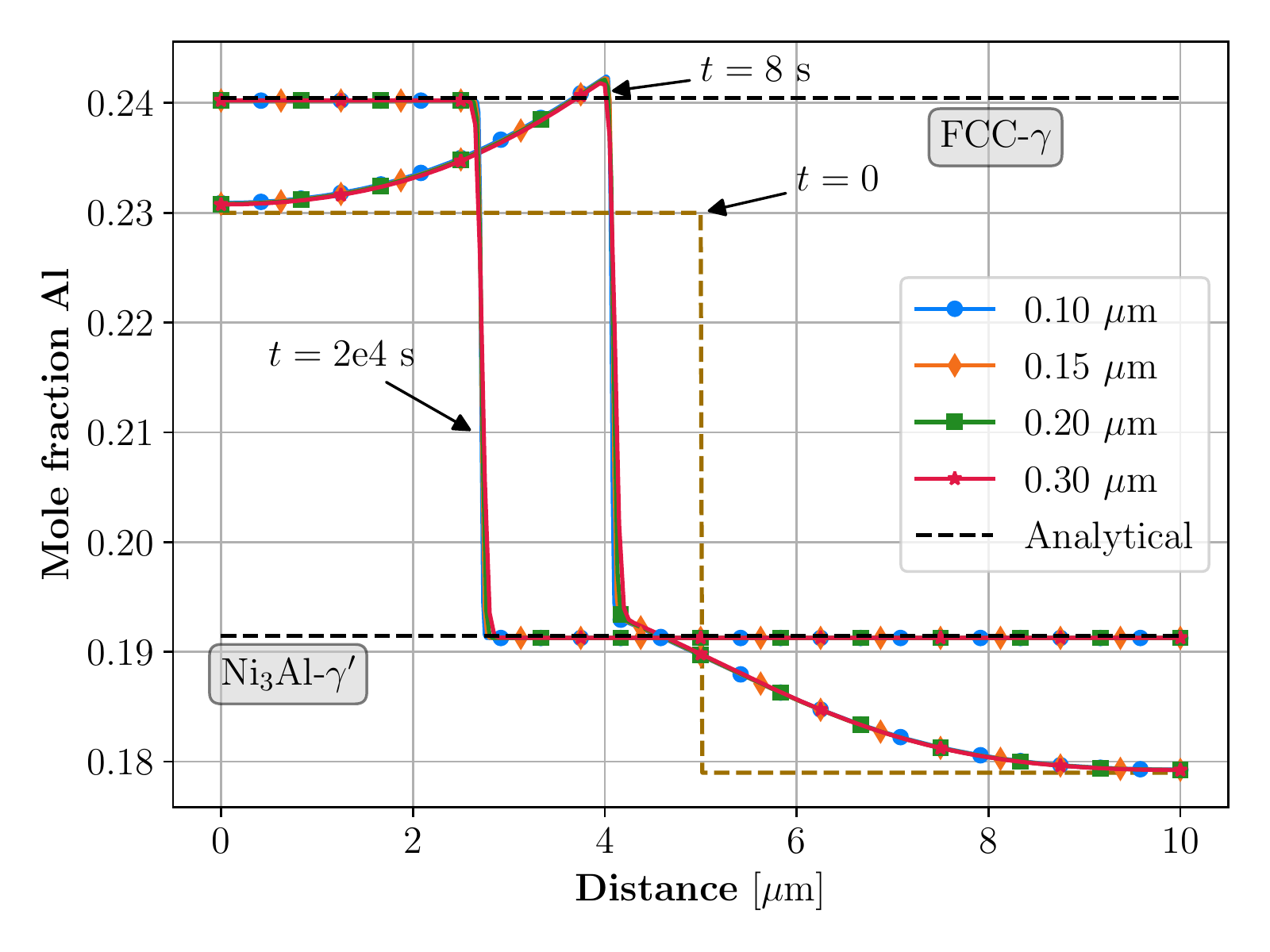}
\caption{}
\label{Fig.2a}
\end{subfigure}
\begin{subfigure}{0.50\textwidth}
\includegraphics[keepaspectratio,width=\linewidth]{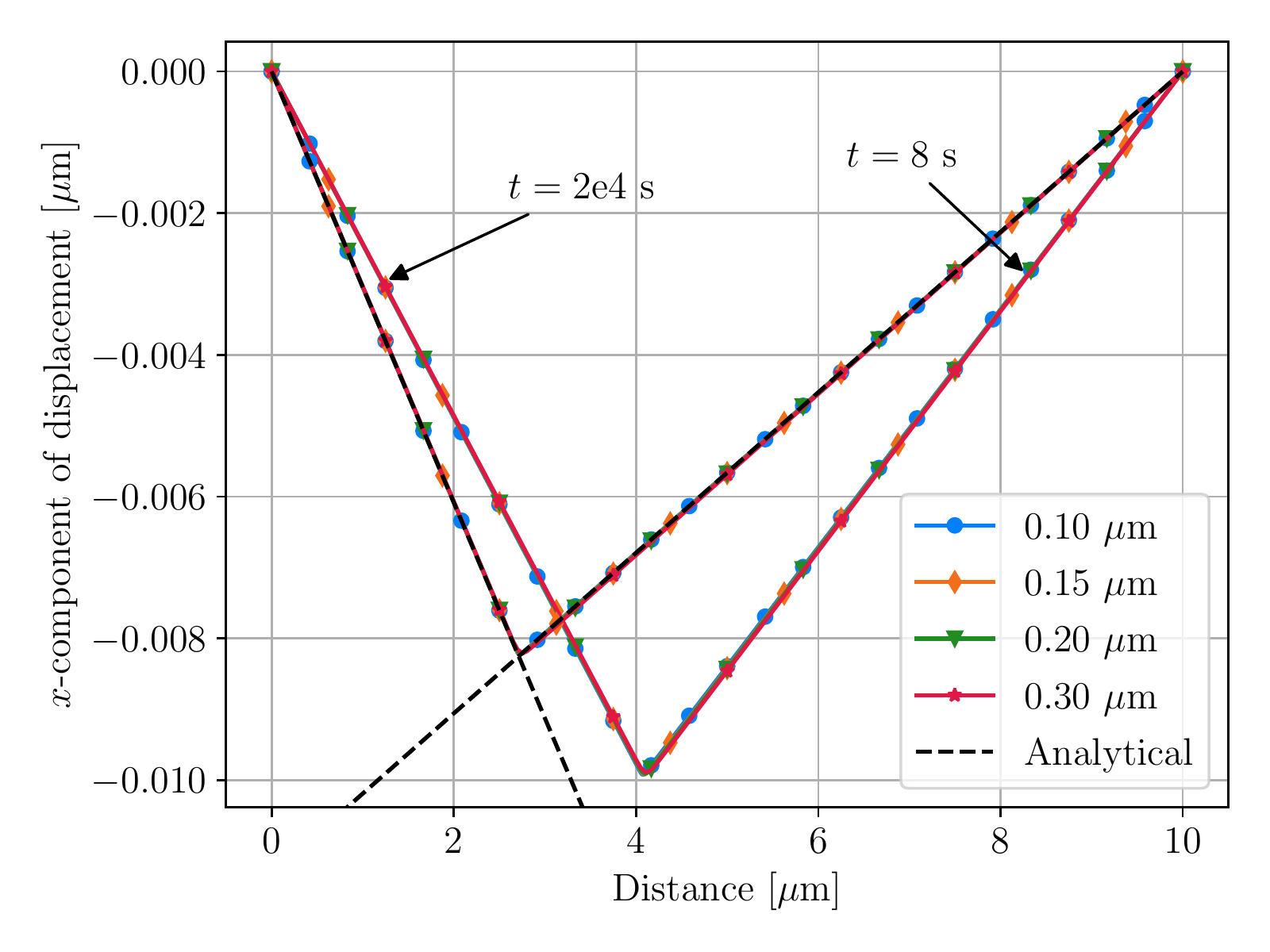}
\caption{}
\label{Fig.2b}
\end{subfigure}
\begin{subfigure}{0.50\textwidth}
\includegraphics[keepaspectratio,width=\linewidth]{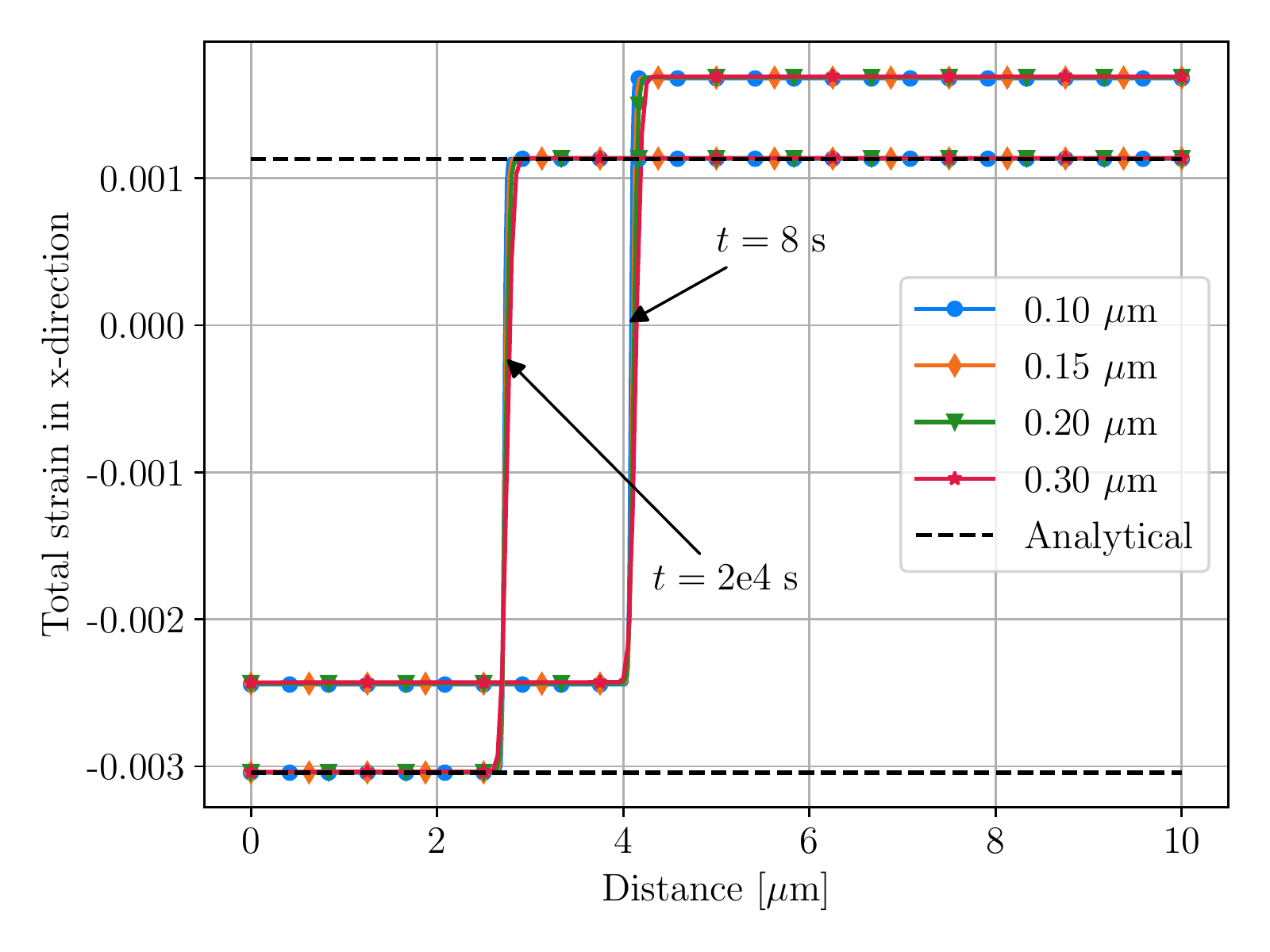}
\caption{}
\label{Fig.2c}
\end{subfigure}
\begin{subfigure}{0.50\textwidth}
\includegraphics[keepaspectratio,width=\linewidth]{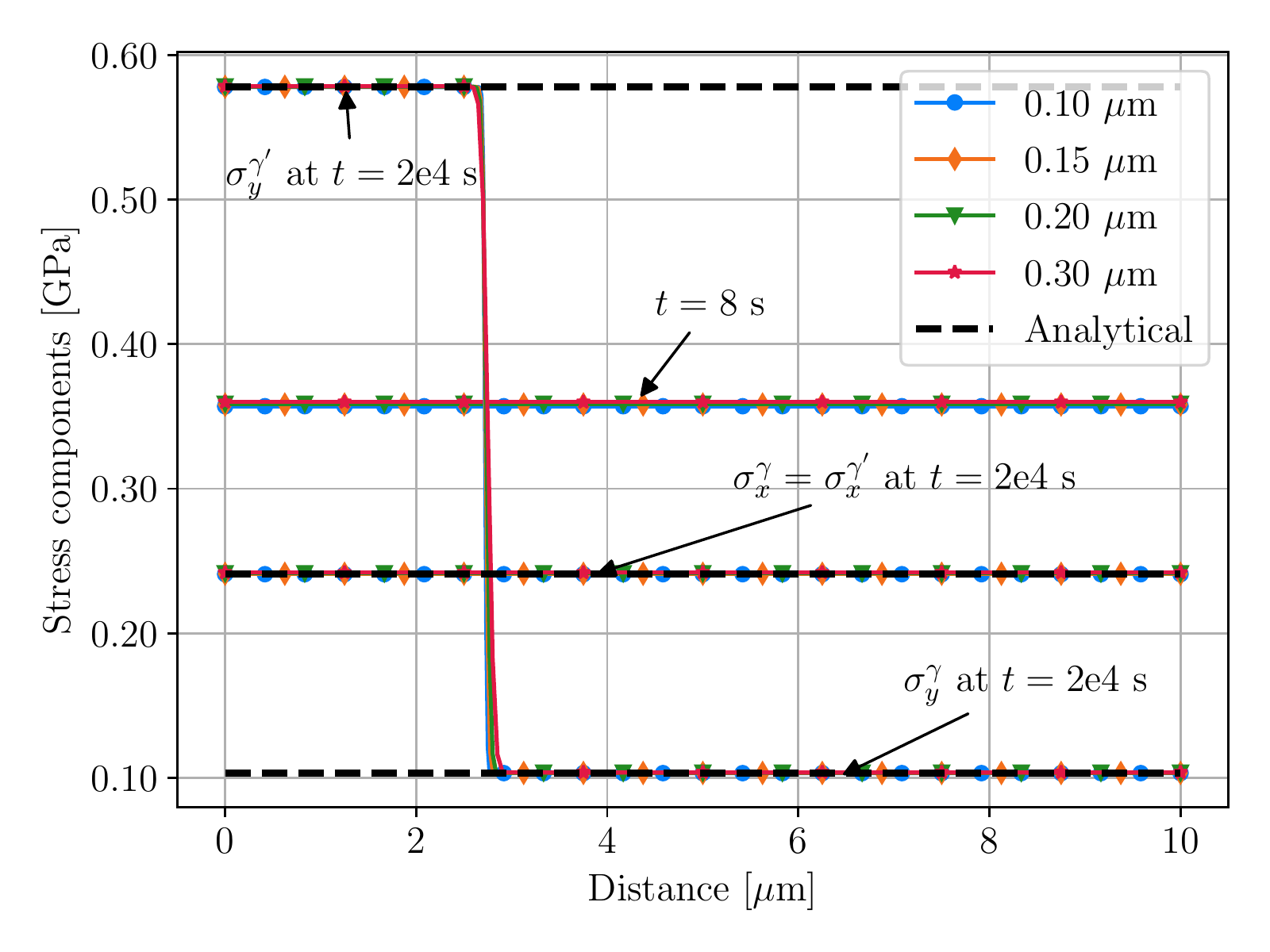}
\caption{}
\label{Fig.2d}
\end{subfigure}
\caption{Comparison of a) Al-mole fraction profiles; b)  x-component of displacement field; c) total strain; and d) non-zero stress components as functions of spatial coordinates for four different interface widths at time $t=8$ s and $t=2\mathrm{e}{4}$s. In Figs. \ref{Fig.2b}-\ref{Fig.2d}, the dotted black lines indicate the analytically calculated elastic fields \textcolor{black}{for the special case of plane stress} (Appendix D).\textcolor{black}{The black dotted lines in Fig.\ref{Fig.2a} are the analytically calculated equilibrium Al mole fractions for the case of non-zero eigenstrains in the $\gamma^{\prime}$ phase.}}
\label{Fig.2}
\end{figure}

We then compared our simulated composition and elastic fields using the PR scheme with the analytically obtained elastic fields \textcolor{black}{for the special case of plane stress} (see Appendix D.1). To this end, we used the calculated interface position (Fig.\ref{Fig.1a}) as an input in our analytical calculations, since the size of the inclusion is needed in the analytical result. Once the elastic fields were known, the interfacial equilibrium compositions were determined from the generalized Gibbs-Thomson equation (Appendix D). This comparison for different interface widths is shown in Fig.\ref{Fig.2}. We find that the growth of $\gamma$ phase is driven by the diffusion of Al from $\gamma$ to $\gamma^{\prime}$ (Fig.\ref{Fig.2a}). Moreover, we find that the $x$-component of the displacement field is maximum at the interface, and its slope decreases with increasing simulation time (Fig. \ref{Fig.2b}). This is reflected in the total strain field normal to the interface (Fig. \ref{Fig.2c}). It is also apparent that this strain is discontinuous at the interface. Due to the system geometry and boundary conditions, both the normal and the shear strains tangential to the interface are zero. Because of the eigenstrain, however, the tangential stress along the y-direction is non-zero and discontinuous (Fig.\ref{Fig.2d}).  These simulations also show that the calculated elastic fields are independent of interface width in this case.

\subsection{Planar UO$_2$-Vacancy simulation}
\begin{figure}[!ht]
\begin{center}
\begin{subfigure}{0.6\textwidth}
\includegraphics[keepaspectratio,width=\linewidth]{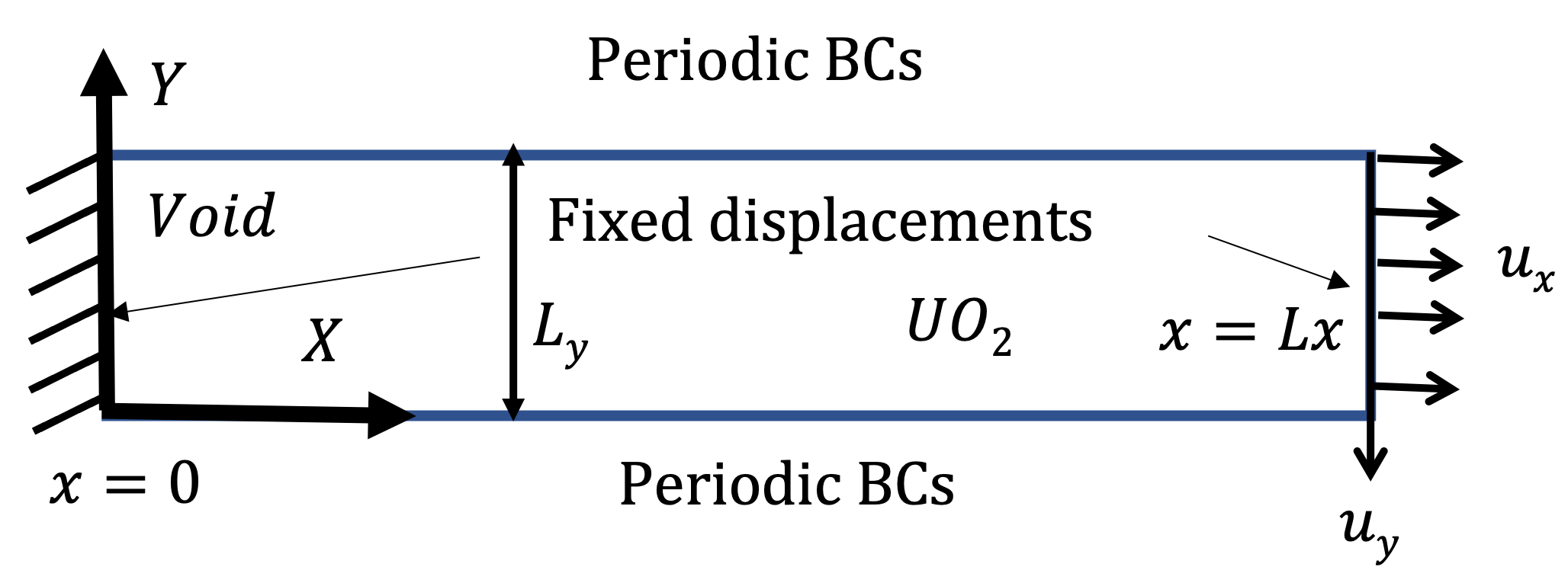}
\caption{}
\label{Fig.3new}
\end{subfigure}
\end{center}
\begin{center}
\begin{subfigure}{0.99\textwidth}
\includegraphics[trim=0 0 50 0, clip,keepaspectratio,width=\linewidth]{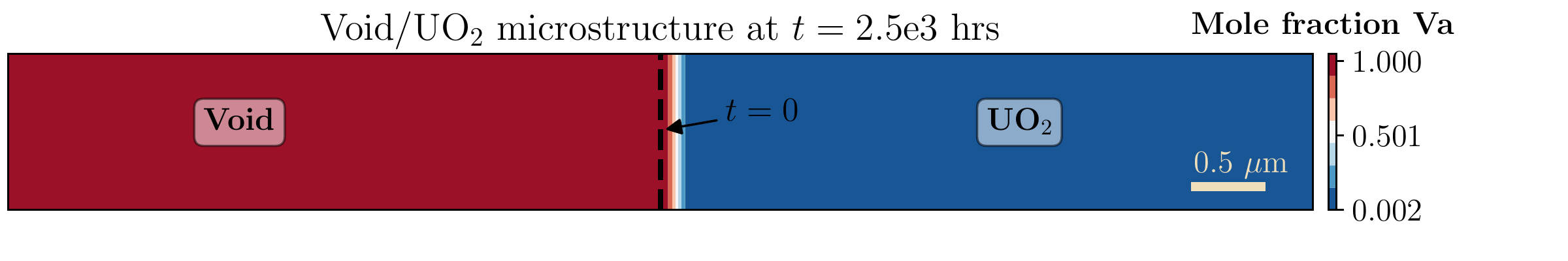}
\caption{}
\label{Fig.3a}
\end{subfigure}
\end{center}
\begin{subfigure}{0.50\textwidth}
\includegraphics[keepaspectratio,width=\linewidth]{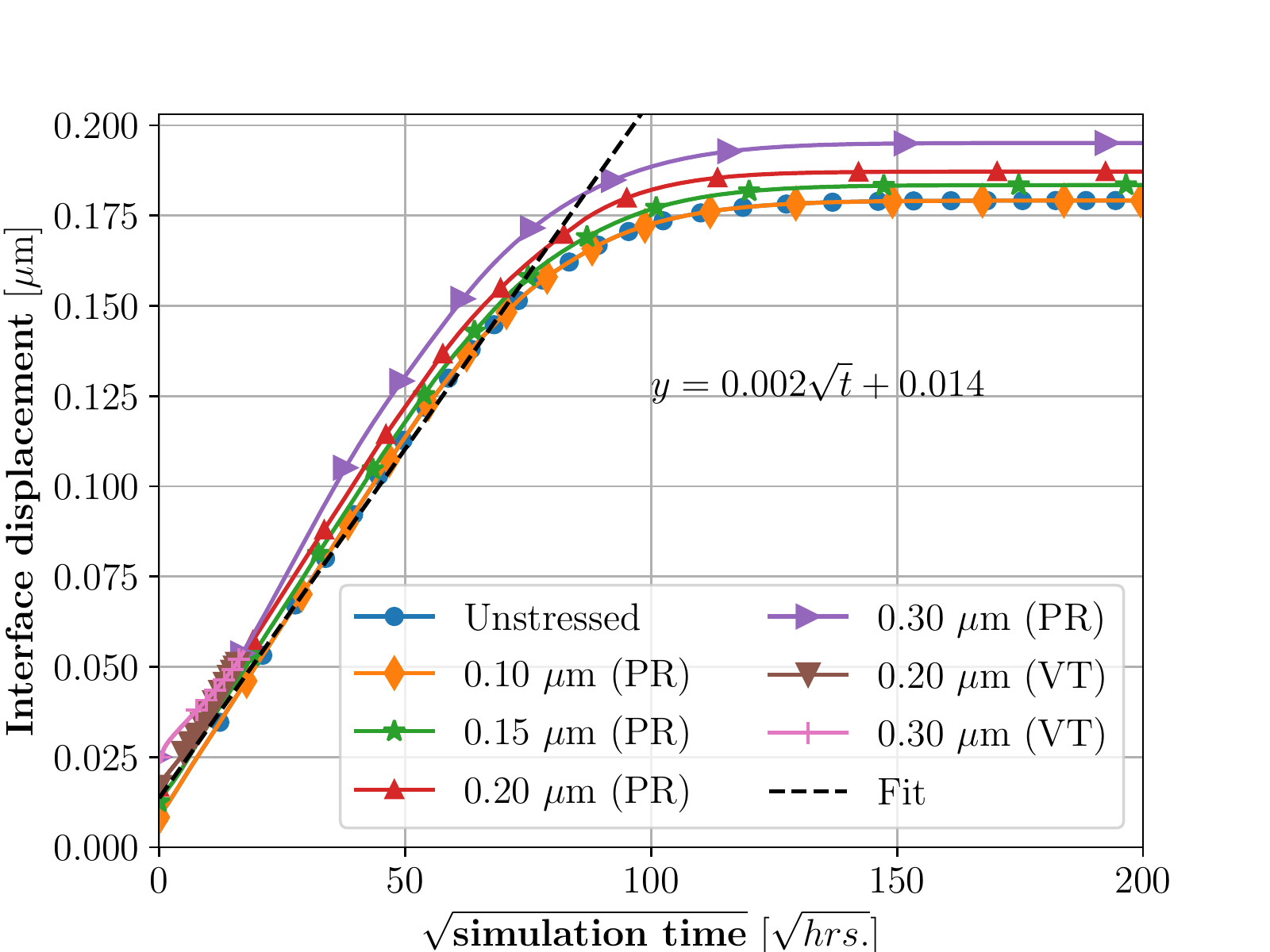}
\caption{}
\label{Fig.3b}
\end{subfigure}
\begin{subfigure}{0.50\textwidth}
\includegraphics[keepaspectratio,width=\linewidth]{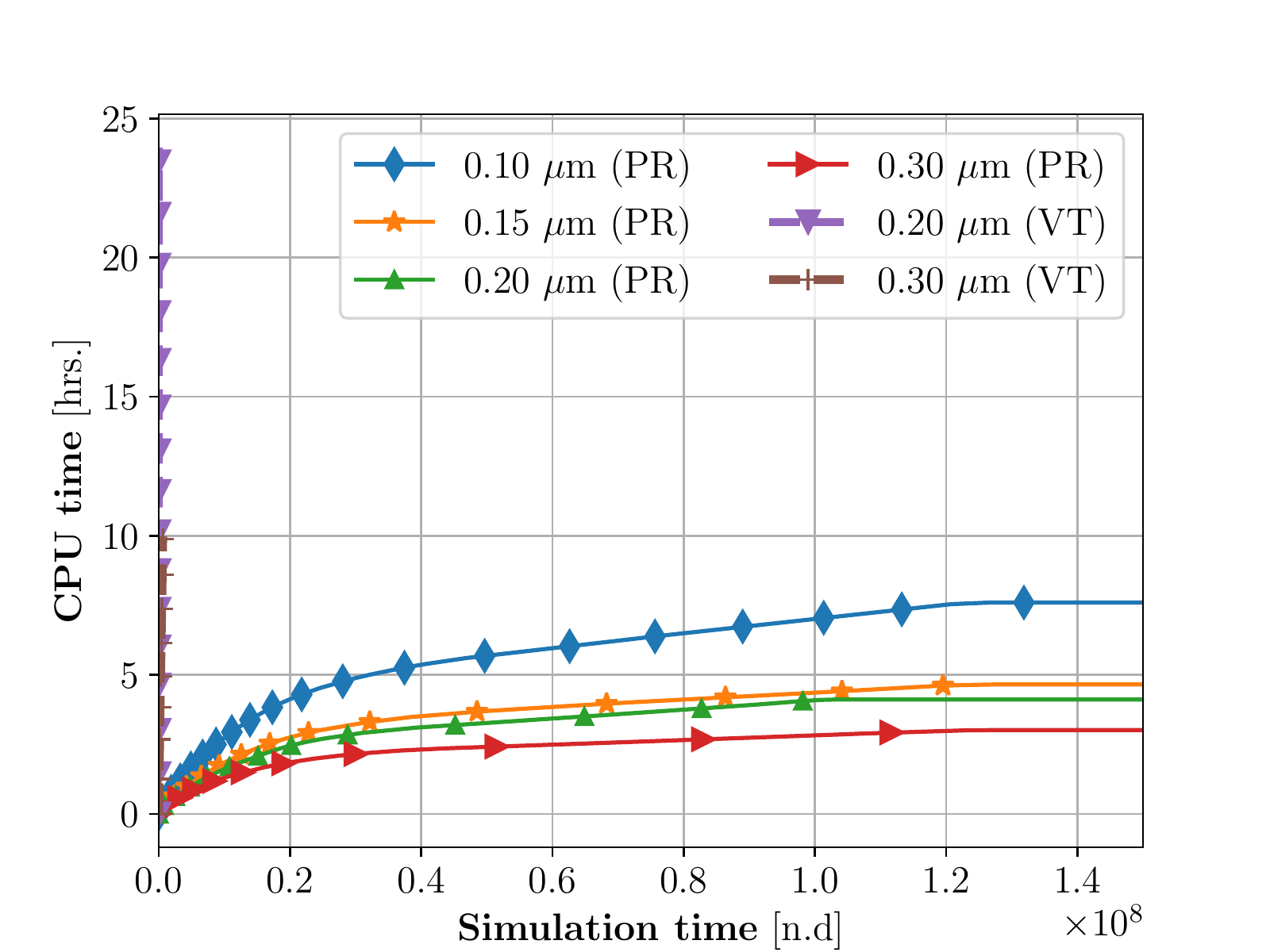}
\caption{}
\label{Fig.3c}
\end{subfigure}
\caption{a) Schematic showing the imposed boundary conditions on displacements. b) Simulated vacancy mole fraction field in an elastically stressed Void/UO$_2$ diffusion couple ($10 \times 0.33$ $\mu$m$^{2}$) at time $t=2.5$$\mathrm{e}${4} hrs. b) The interface displacement ($y$) as a function of the square root of simulation time ($t$) for different interface widths simulated using both the partial rank-one (PR) homogenization scheme and the Voigt-Taylor (VT) homogenization schemes. The unstressed case is also superimposed on Fig.\ref{Fig.1b}. For each simulated case, c) the variation of CPU time with non-dimensional simulation time. Notice that the interface displacement is very small for the cases using the VT scheme compared to the PR scheme.}
\end{figure}

Next, we chose a UO$_2$/void alloy \textcolor{black}{where Young's moduli show higher phase contrast compared to the Ni-Al case, but the system has no eigenstrains. Similar to the planar Ni-Al case, we have assumed periodic boundary conditions along the lateral direction. However, along the longitudinal direction, we have enforced the following Dirichlet boundary conditions on displacements (see Table  {\ref{table:bc}} and Fig.\ref{Fig.3new})}:
\textcolor{black}{
\begin{align}
\boldsymbol{u} (0,y) &= 0 ,\\
u_{x}(L_{x}, y)/l_{c} &= 5 ,\\
u_{y}(L_{x}, y)/l_{c}&= -5.
\end{align}
}
\textcolor{black}{Here $l_{c}=1.133$ nm is the characteristic length of the system, $u_{x}$ and  $u_{y}$ are the x and y components of the displacement vector $\boldsymbol{u}$, respectively. Similar to our previous case, we have enforced zero Neumann boundary conditions on phase-field, and mole fraction variables (see Eqs. \ref{BCI} \& \ref{BCII}) along the left and right boundaries, and the longitudinal dimension is $30$ times that of the lateral dimension. It is also important to note that there are no eigenstrains in this system. Consequently, the elastic stresses are due to the imposed right boundary conditions on displacements.}

\textcolor{black}{Ideally, Young's modulus of the void phase should be vanishingly small. We, however, assume the Young's modulus of the void phase to be $10^{-4}$ times the UO$_2$ phase (see Table \ref{Table1.2}). This choice was based on two practical reasons. First, we referred to the work of Gururajan and Abinandanan \cite{Guru2007}. Second, we found that for the non-planar case (discussed in section $3.4$), convergence to the equilibrium state was achieved only when the ratio of young's moduli $E_{void}/E_{uo_2} \leq 10^{-4}$. Although for the planar cases, we found convergence even with a ratio of young's moduli of the order of $10^{-7}$  using the PR scheme.} 

\begin{figure}[!ht]
\begin{subfigure}{0.50\textwidth}
\includegraphics[keepaspectratio,width=\linewidth]{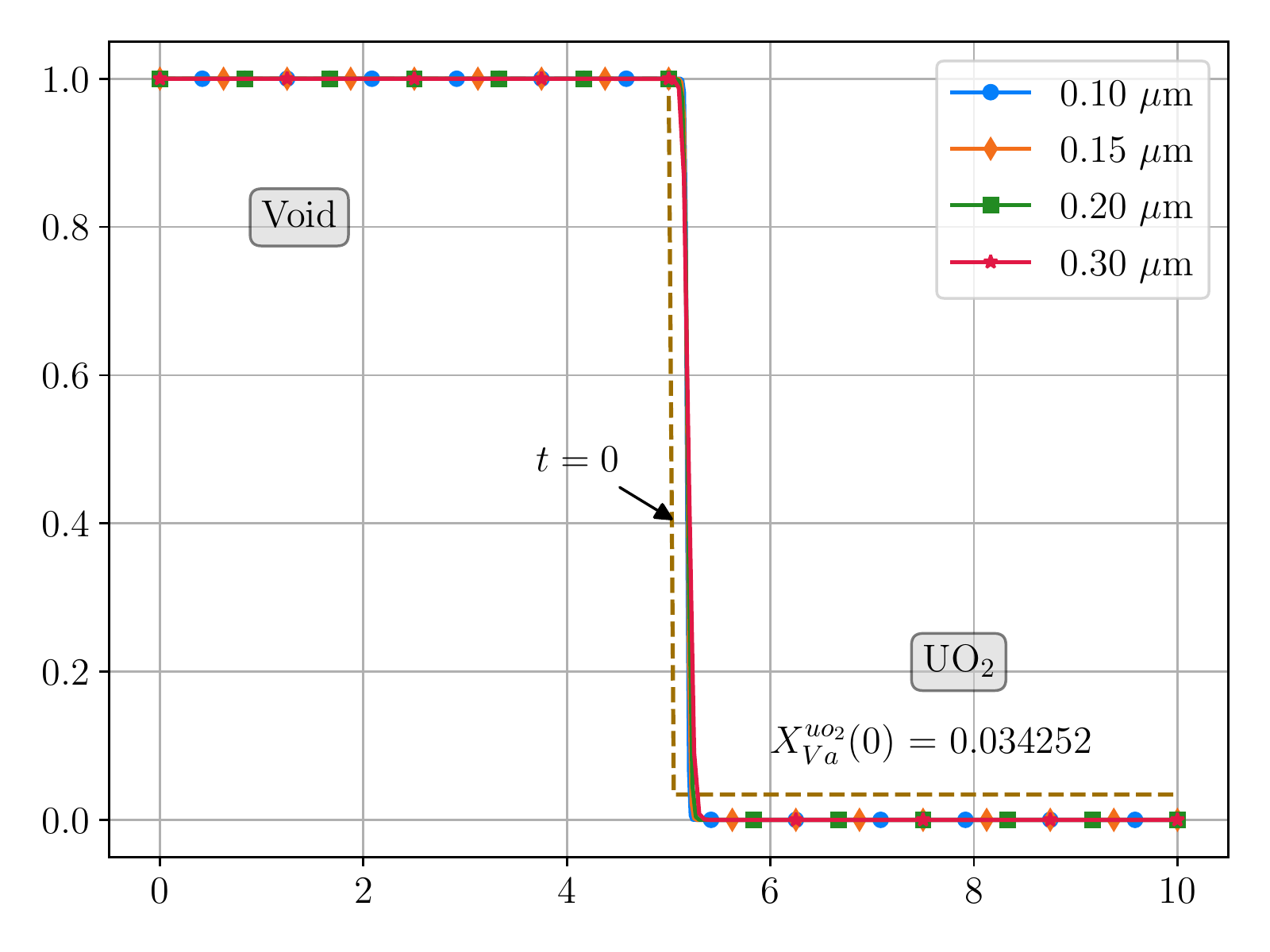}
\caption{}
\label{Fig.4a}
\end{subfigure}
\begin{subfigure}{0.50\textwidth}
\includegraphics[keepaspectratio,width=\linewidth]{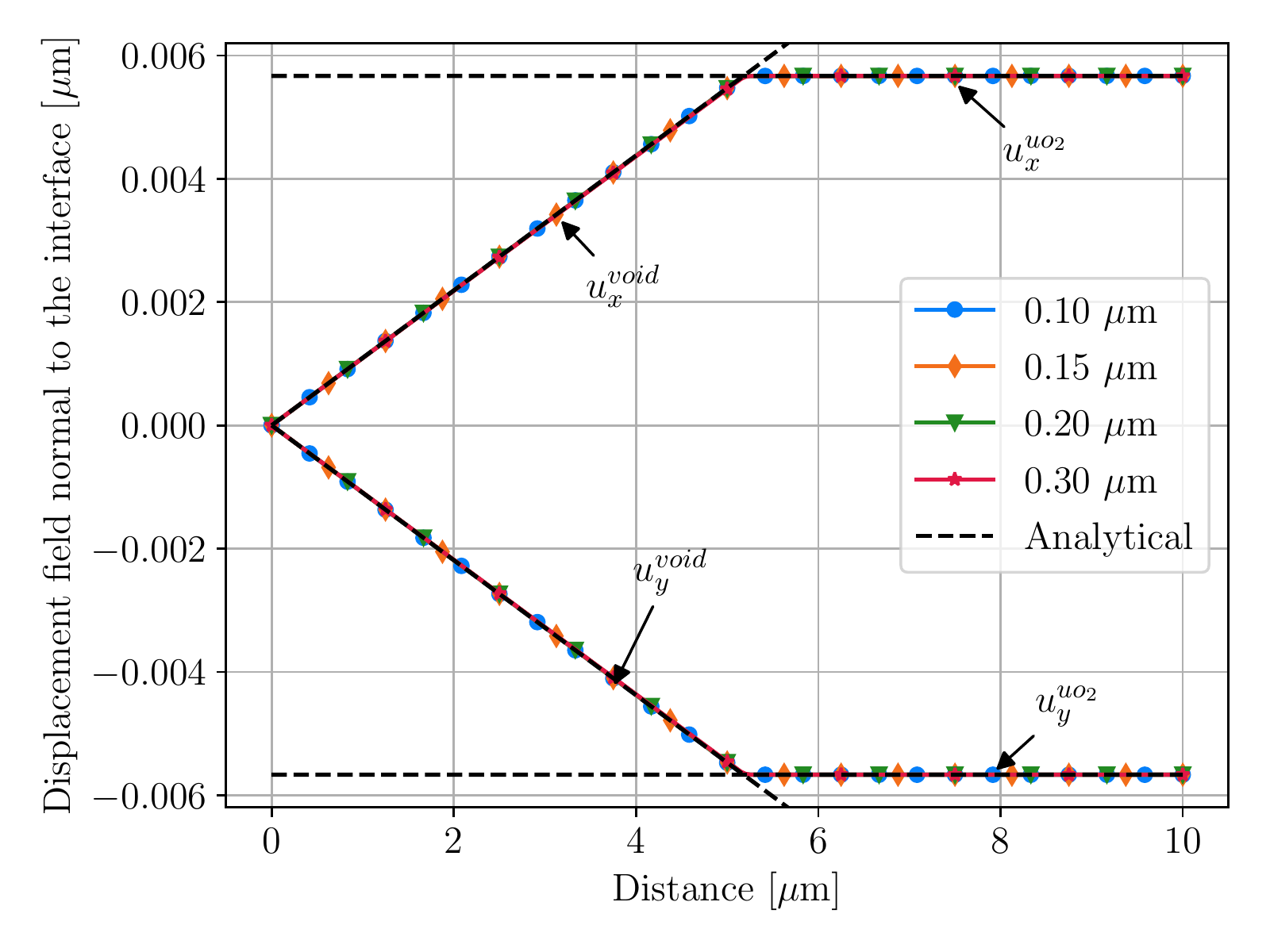}
\caption{}
\label{Fig.4b}
\end{subfigure}
\begin{subfigure}{0.50\textwidth}
\includegraphics[keepaspectratio,width=\linewidth]{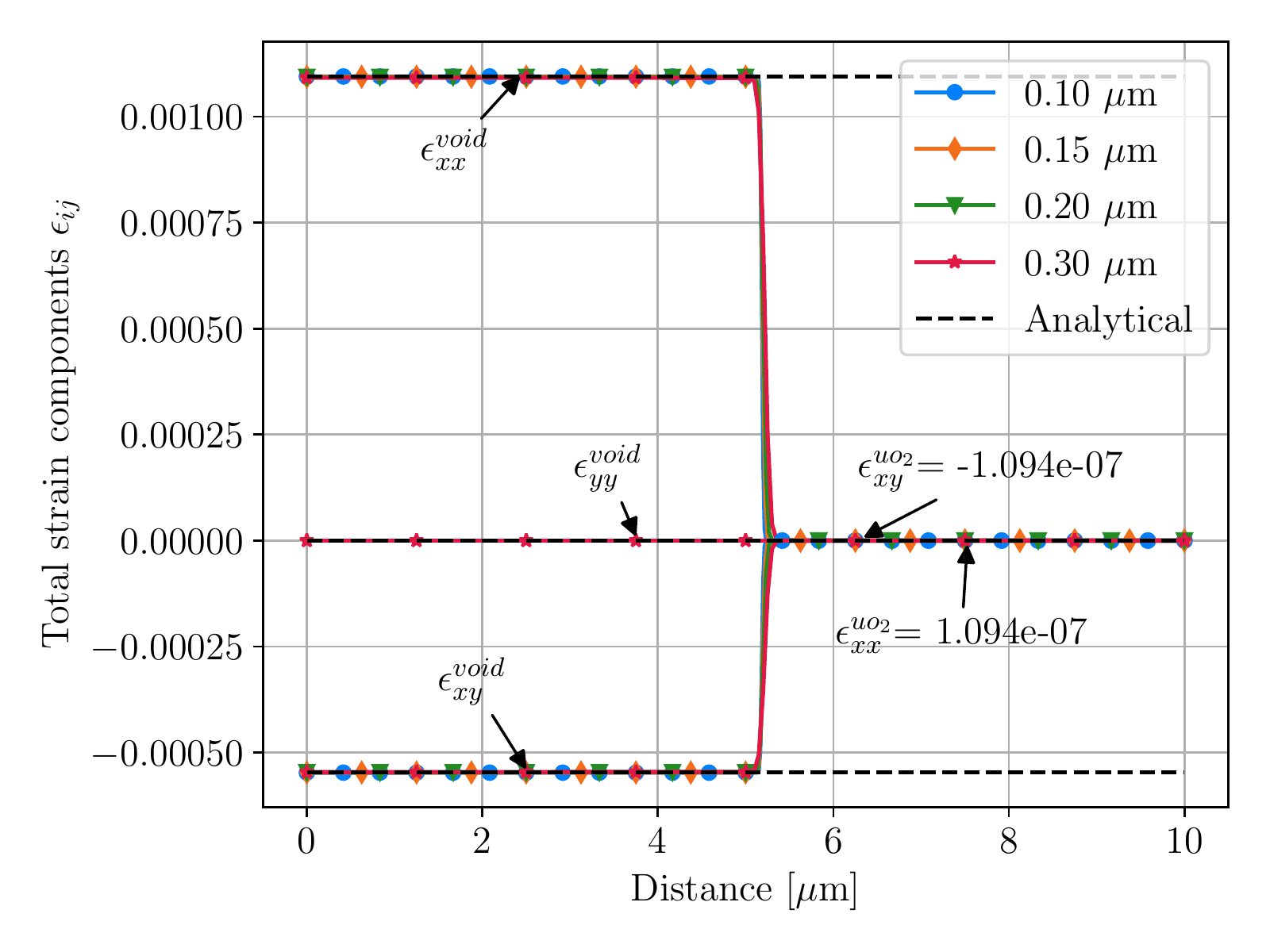}
\caption{}
\label{Fig.4c}
\end{subfigure}
\begin{subfigure}{0.50\textwidth}
\includegraphics[keepaspectratio,width=\linewidth]{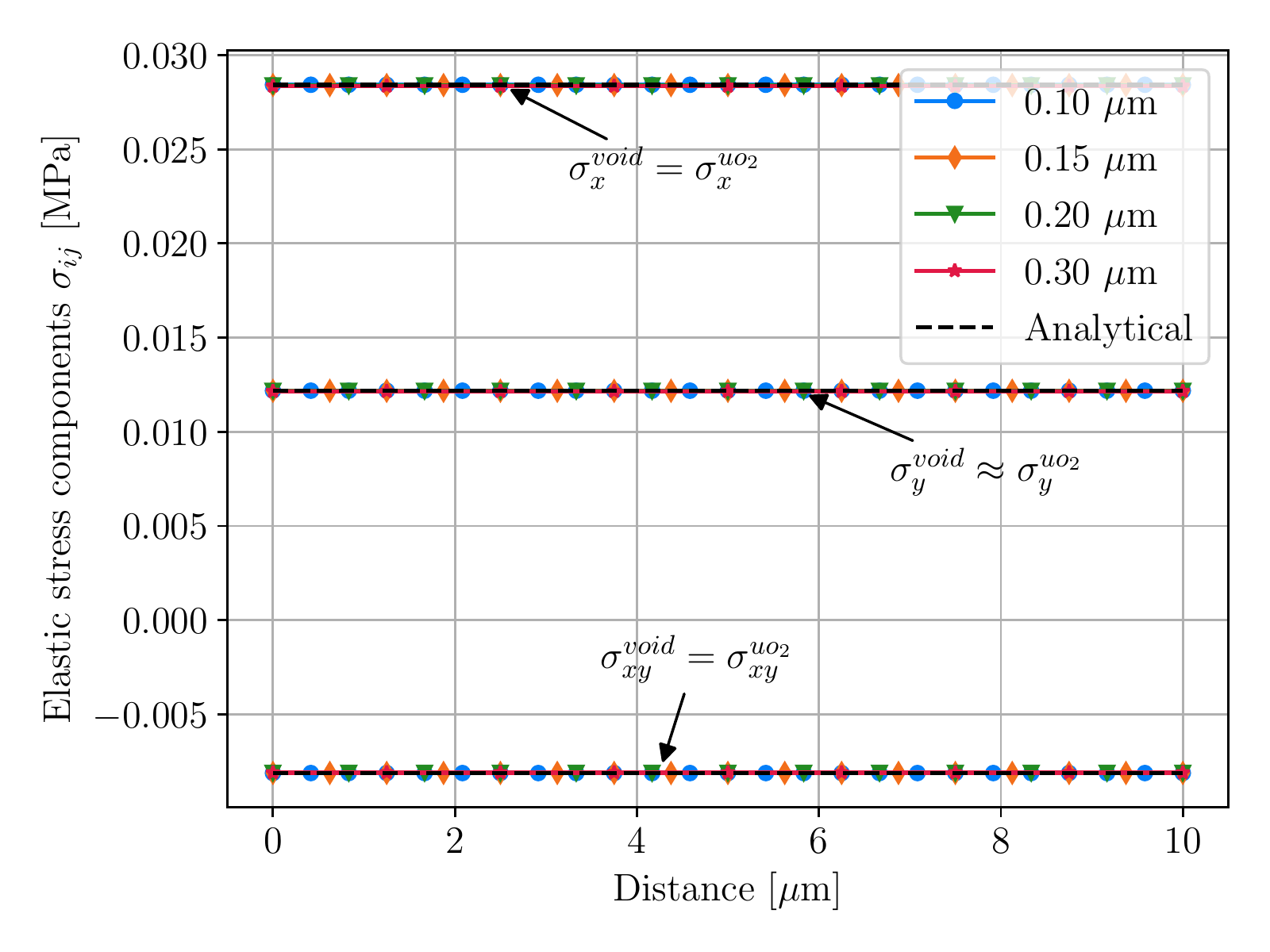}
\caption{}
\label{Fig.4d}
\end{subfigure}
\caption{Comparison of a) vacancy mole fraction profiles; b) both x and y-components of displacement field; c) total strain components; and d) stress components as functions of spatial coordinates for four different interface widths at time $t=8$ s and $t=2\mathrm{e}{4}$s. \textcolor{black}{For the special case of plane stress, the analytically calculated elastic fields (see Appendix D for derivation) in both phases are indicated as dotted black lines in Figs.\ref{Fig.4b}-\ref{Fig.4d}}. The initial vacancy concentration is shown in Fig.\ref{Fig.4a}.}
\label{Fig.4}
\end{figure}

In contrast to the Ni-Al alloy, we find that the simulated interface displacement using the PR scheme depends slightly on the interface position in this alloy (Fig.\ref{Fig.3b}). It is possible to explain this by measuring the strain jump normal to the interface. At the equilibrium state, our calculations show that the strain jump, in this case, is nearly $3.716$ times that of the $\gamma^{\prime}/\gamma$ case (compare Figs. \ref{Fig.2c} \& \ref{Fig.4c}).

\textcolor{black}{The reason is the high phase contrast with respect to elastic moduli. Specifically, the ratio of Young's moduli, in this case, is nearly $9.1\mathrm{e}{5}$ times that of the Ni-Al $\gamma^{\prime}/\gamma$ case (Table \ref{Table1.2})}. Surprisingly, we find that this simulation was excruciatingly slow when the VT scheme was employed. Specifically, the interface migrated by $\approx 0.050$ $\mu$m after a CPU time of 24 hrs with an interface width of $0.20$ $\mu$m using the VT scheme (Fig.\ref{Fig.3c}). On the other hand, this simulation was completed in less than $5$ hrs. using the PR scheme for the same interface width (Fig.\ref{Fig.3c}). We observed similar results when a higher interface width was assumed. In our opinion, this is due to the difference in \textit{local} driving force in the two schemes. \textcolor{black}{This shows the advantage of the PR scheme over the VT scheme for systems exhibiting higher phase contrast with respect to Young's moduli.} Another contrasting observation compared to the Ni-Al case is that the interface position for case IV and case III does not differ significantly. The reason for this is the very low modulus of the void phase that leads to almost negligible stress within the system (Fig.\ref{Fig.4d}). 

We then compared the simulated and the analytically calculated elastic fields \textcolor{black}{for the special case of plane stress}. We find that our simulated elastic fields show reasonable quantitative agreement with the analytical results (Fig.\ref{Fig.4}). Since the net interface displacement, in this case, was very small ($\approx 0.2$ $\mu$m), we plotted the elastic fields only at the final time step. In direct contrast to the Ni-Al case, the displacement field in the y-direction is non-zero in this alloy (Fig.\ref{Fig.4b}). This is due to the imposed downward displacement at the right boundary. Consequently, this mechanical displacement engenders shear strains within the system  (Fig.\ref{Fig.4c}), which are discontinuous at the interface. Note that, compared to the void phase, the deformation in the UO$_2$ phase is negligible (Fig.\ref{Fig.4c}). As mentioned before, due to the very low modulus of the void phase, the stresses within the system are negligible (Fig.\ref{Fig.4d}). Nevertheless, we find that the quantitative accuracy of the simulated vacancy field and the elastic fields remain unaltered with changes in interface width.

\subsection{Non-planar Ni-Al $\gamma^{\prime}/\gamma$ simulation}
\begin{figure}[ht]
\begin{subfigure}{0.5\textwidth}
\includegraphics[trim=0 7 0 7,clip, keepaspectratio,width=\linewidth]{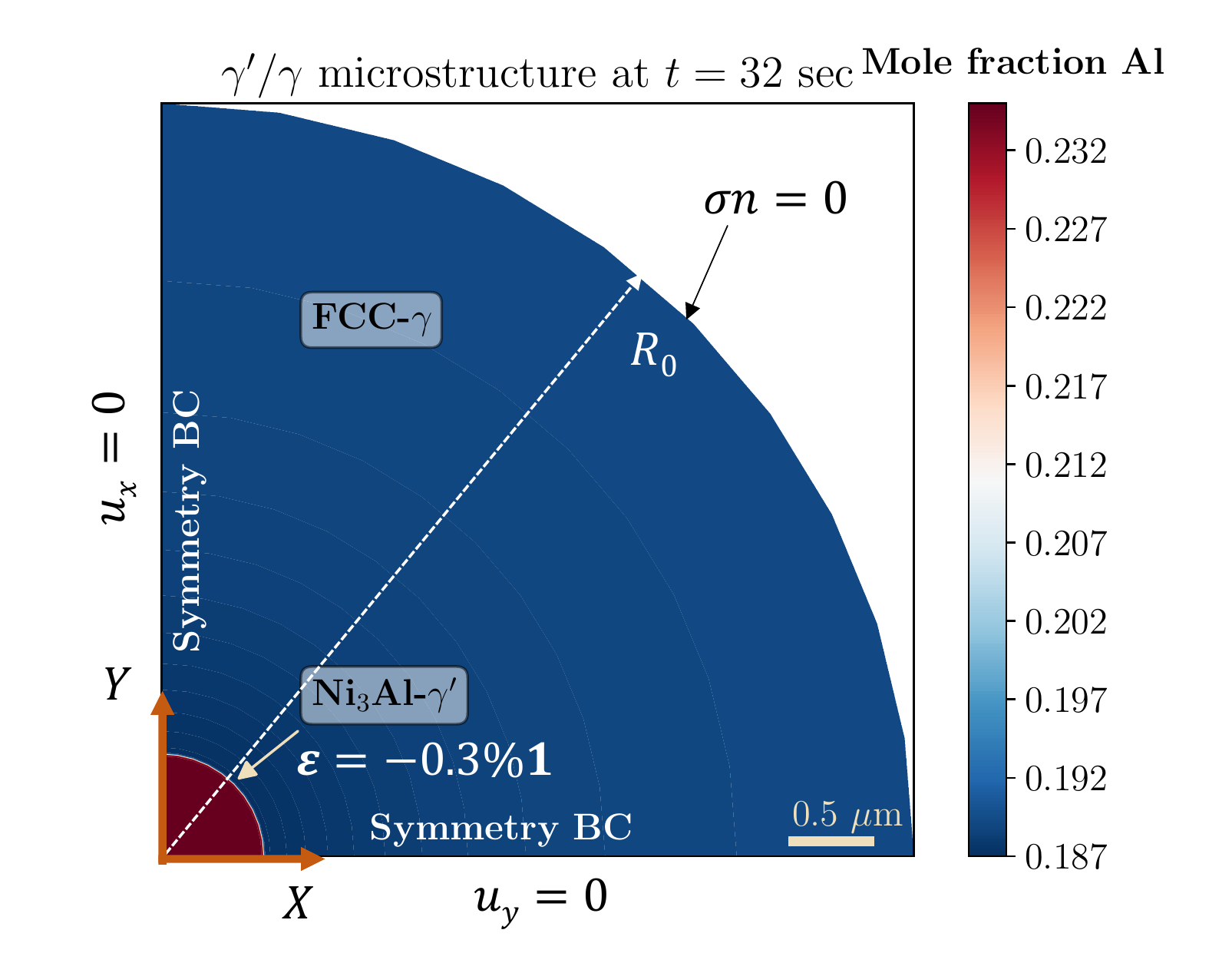}
\caption{}
\label{Fig.5a}
\end{subfigure}
\begin{subfigure}{0.5\textwidth}
\includegraphics[keepaspectratio,width=\linewidth]{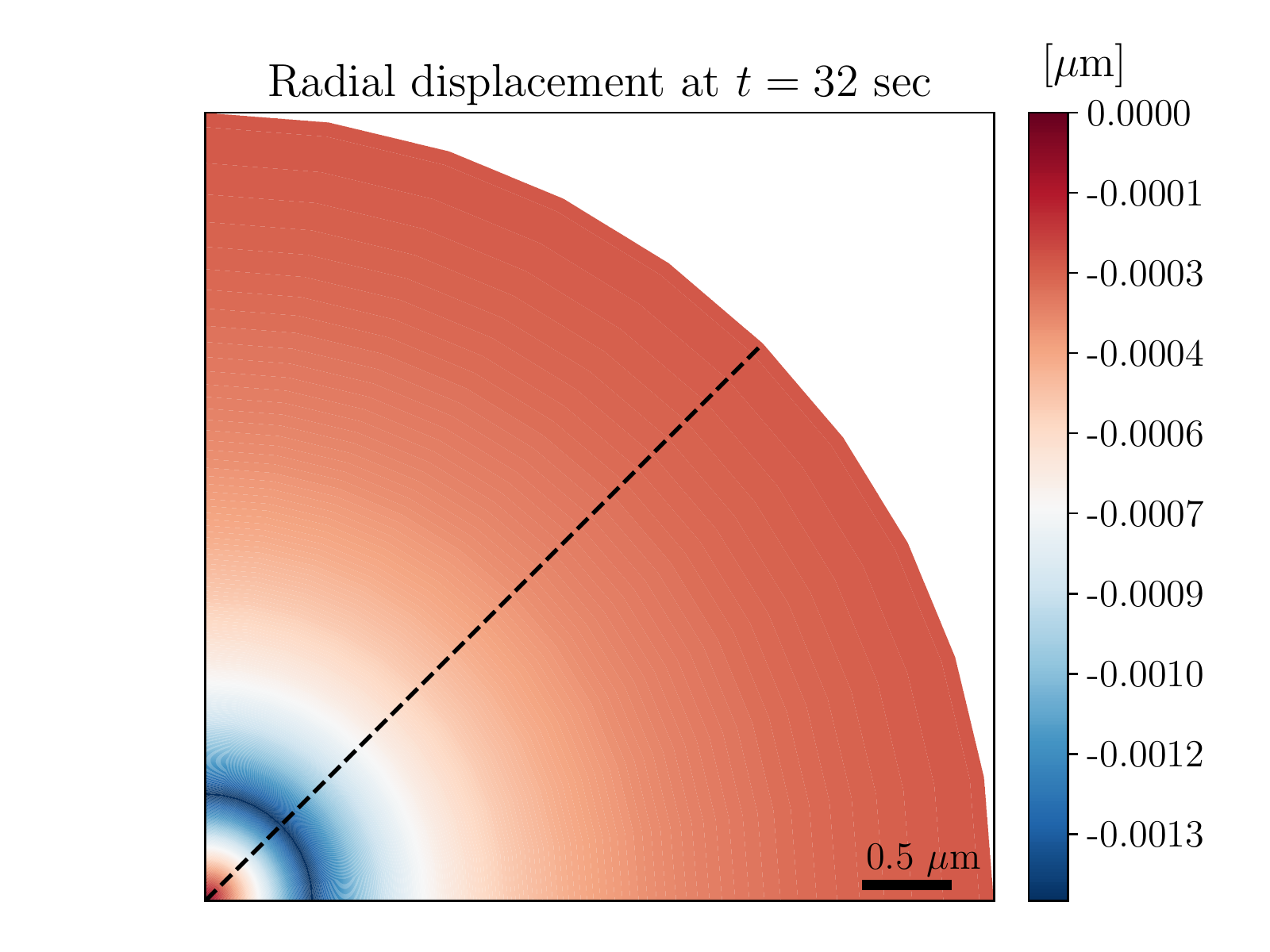}
\caption{}
\label{Fig.5b}
\end{subfigure}
\begin{subfigure}{0.50\textwidth}
\includegraphics[keepaspectratio,width=\linewidth]{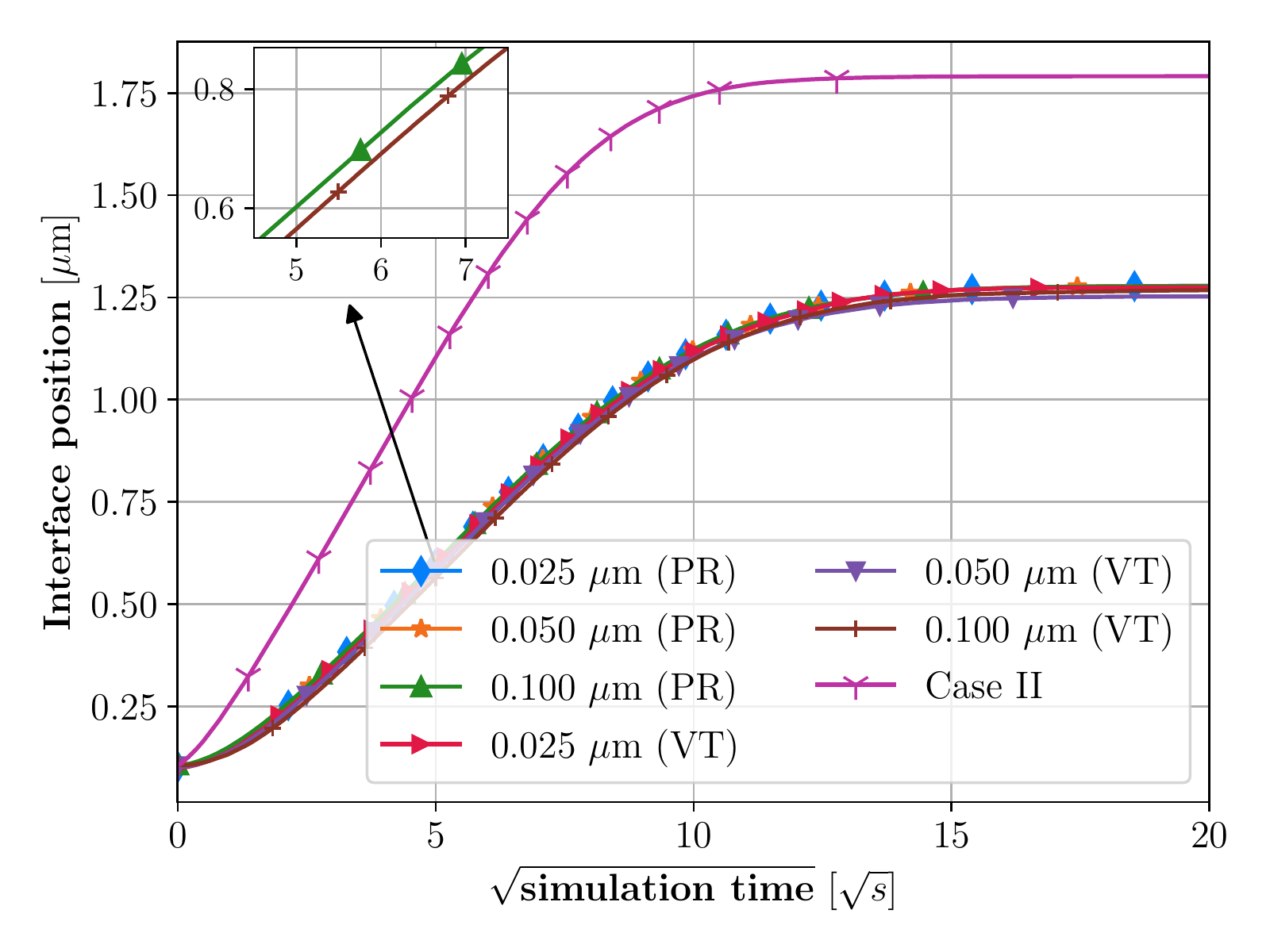}
\caption{}
\label{Fig.5c}
\end{subfigure}
\begin{subfigure}{0.50\textwidth}
\includegraphics[keepaspectratio,width=\linewidth]{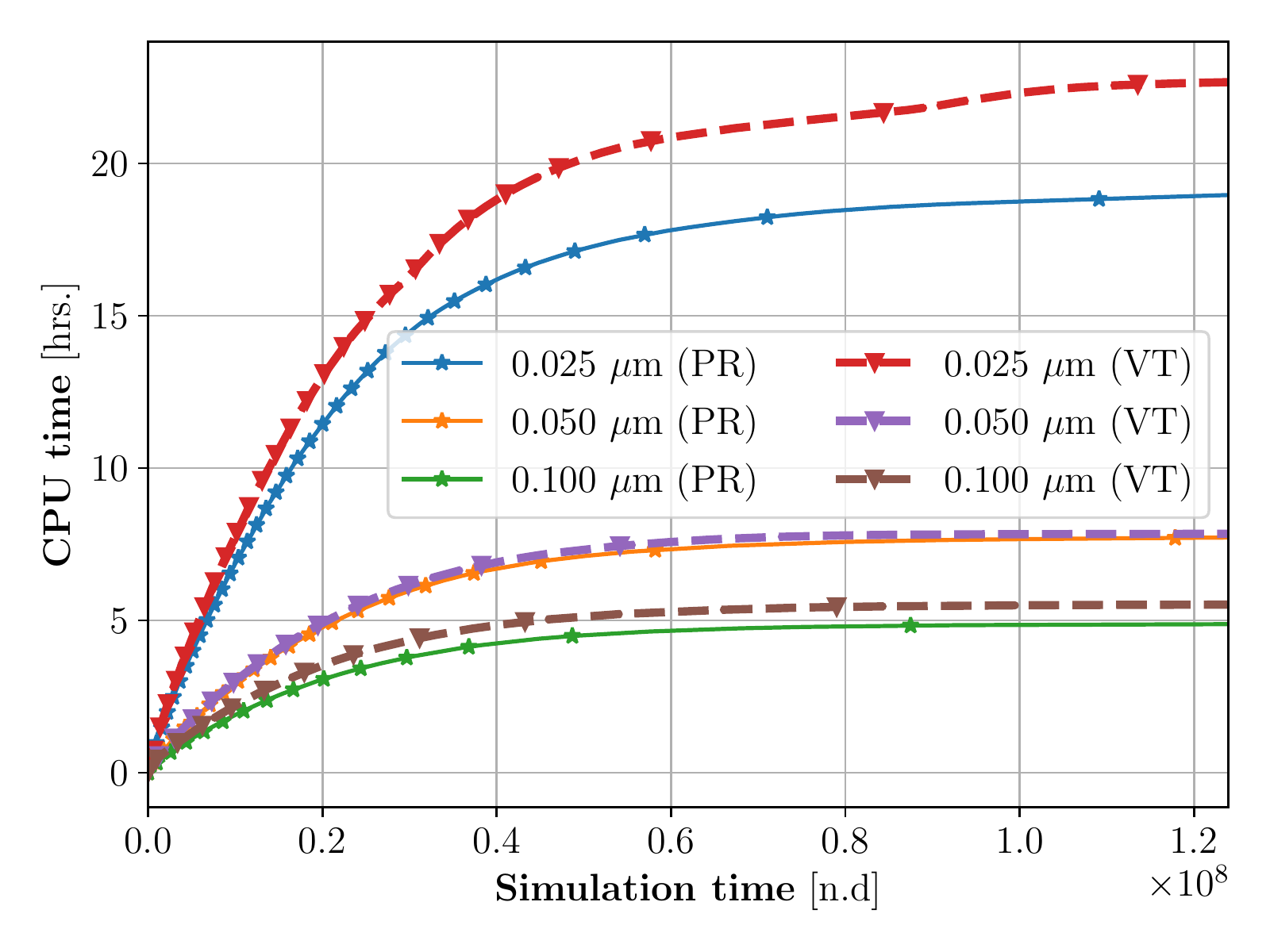}
\caption{}
\label{Fig.5d}
\end{subfigure}
\caption{ Simulated a) Al-mole fraction and b) radial strain fields in an elastically stressed $\gamma^{\prime}/\gamma$ alloy of radius $5$ $\mu$m. Symmetry boundary conditions are imposed at the left and bottom edges of the domain. c) Comparison of  $\gamma^{\prime}/\gamma$ interface position as a function of the square root of simulation time for three different interface widths ($0.025$ $\mu$m to $0.10$ $\mu$m), simulated using both partial rank-one (PR) homogenization and Voigt-Taylor (VT) homogenization. d) The CPU time as a function of non-dimensional simulation time for all simulated cases.  The dotted line in Fig. \ref{Fig.5b} indicate the radial distance along which the field quantities are evaluated in Fig.\ref{Fig.6}.}
\end{figure}

As a third example, we chose a Ni-Al alloy in which the $\gamma^{\prime}/\gamma$ phase boundary is non-planar. \textcolor{black}{Specifically, we have considered a circular shaped $\gamma^{\prime}$ precipitate embedded at the center of  a circular shaped $\gamma$ matrix. Due to the symmetry of the problem, we have simulated only a quarter of the domain. Moreover, zero Neumann boundary conditions have been assumed with respect to phase-field and mole fraction variables at all boundaries (see Eqs. \ref{BCI} \& \ref{BCII}). Further, symmetric boundary conditions have been imposed on displacements at the left and bottom boundaries, while the outer boundary is traction free (see Table \ref{table:bc} and Fig. \ref{Fig.5a}). This yields}
\textcolor{black}{
\begin{align}
u_{x}(0,y) &= 0,\label{EqnResult7}\\
u_{y}(x,0) &= 0,\label{EqnResult8}\\
\boldsymbol{\sigma}\boldsymbol{n}(x,y) &=0\quad \forall \quad x^{2} + y^{2} = R_{0}^{2}
\end{align}
}
\textcolor{black}{Here $\boldsymbol{\sigma}$ is the stress and $R_{0} = 5 \mu$m is the outermost radius of the $\gamma$ matrix. Again, we have considered two scenarios—with dilatational eigenstrains in the $\gamma^{\prime}$ phase (case I) and without any eigenstrains (case II). It is noteworthy that we have performed these simulations in a Cartesian reference frame. For analysis, the simulated results were then transformed into polar coordinates for visualization and comparison with analytical results.}

\textcolor{black}{In contrast to our planar Ni-Al case, the initial conditions in this simulation are such that the $\gamma^{\prime}$ phase grows at the expense of the $\gamma$ phase. Due to system geometry and isotropic elastic properties, the simulated elastic fields are radially symmetric. To show this, the radial $u_{r}$ and tangential $u_{t}$ displacements are determined from the $x$ and $y$ displacements using}
\textcolor{black}{
\begin{align}
\begin{Bmatrix}
u_{r}\\
u_{t}
\end{Bmatrix}
=
\begin{bmatrix}
\cos\theta & \sin\theta\\
-\sin\theta & \cos\theta
\end{bmatrix}
\begin{Bmatrix}
u_{x}\\
u_{y}
\end{Bmatrix},
\label{EqnCT}
\end{align}
}
\textcolor{black}{
where $\theta= \tan^{-1}(y/x)$. Here, $y$ and $x$ are the coordinates in the Cartesian frame.
In Fig.\ref{Fig.5b}, we have plotted the radial displacement at time $t=32$ s. Moreover, we found that the tangential displacement is vanishing throughout the domain in this case.}

\textcolor{black}{Next, we focussed on the effect of interface width and the homogenization assumption on the simulation results.} Our simulations show that the variation of interface position with increasing time is independent of the interface width choice when simulated using the partial rank-one homogenization scheme (Fig.\ref{Fig.5c}). However, using the VT scheme, we find that this variation is slightly dependent on the interface width (see inset in Fig.\ref{Fig.5c}). Moreover, we also find that the computation time to run the simulation using the PR scheme is always less than using the VT scheme (Fig.\ref{Fig.5d}). \textcolor{black}{It is interesting to observe that case I is kinetically slower compared to case II, which contrasts with the planar case during which the reverse was true}. This is possibly due to the shift in the local equilibrium compositions at the interface due to the curvature driving force, as given by the generalized Gibbs-Thomson condition.

\begin{figure}
\begin{subfigure}{0.5\textwidth}
\includegraphics[keepaspectratio,width=\linewidth]{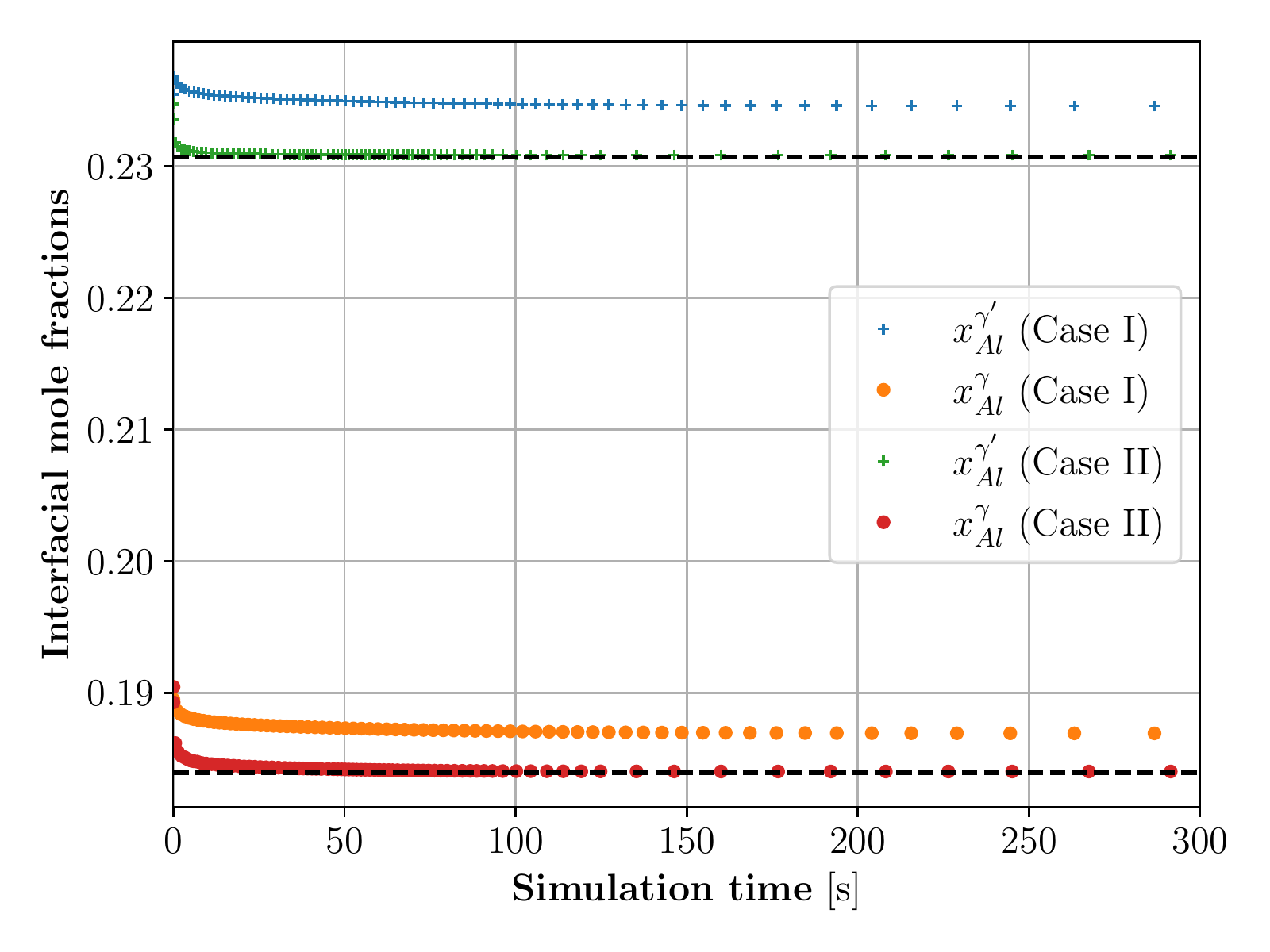}
\caption{}
\label{Fig.5e}
\end{subfigure}
\begin{subfigure}{0.45\textwidth}
\includegraphics[keepaspectratio,width=\linewidth]{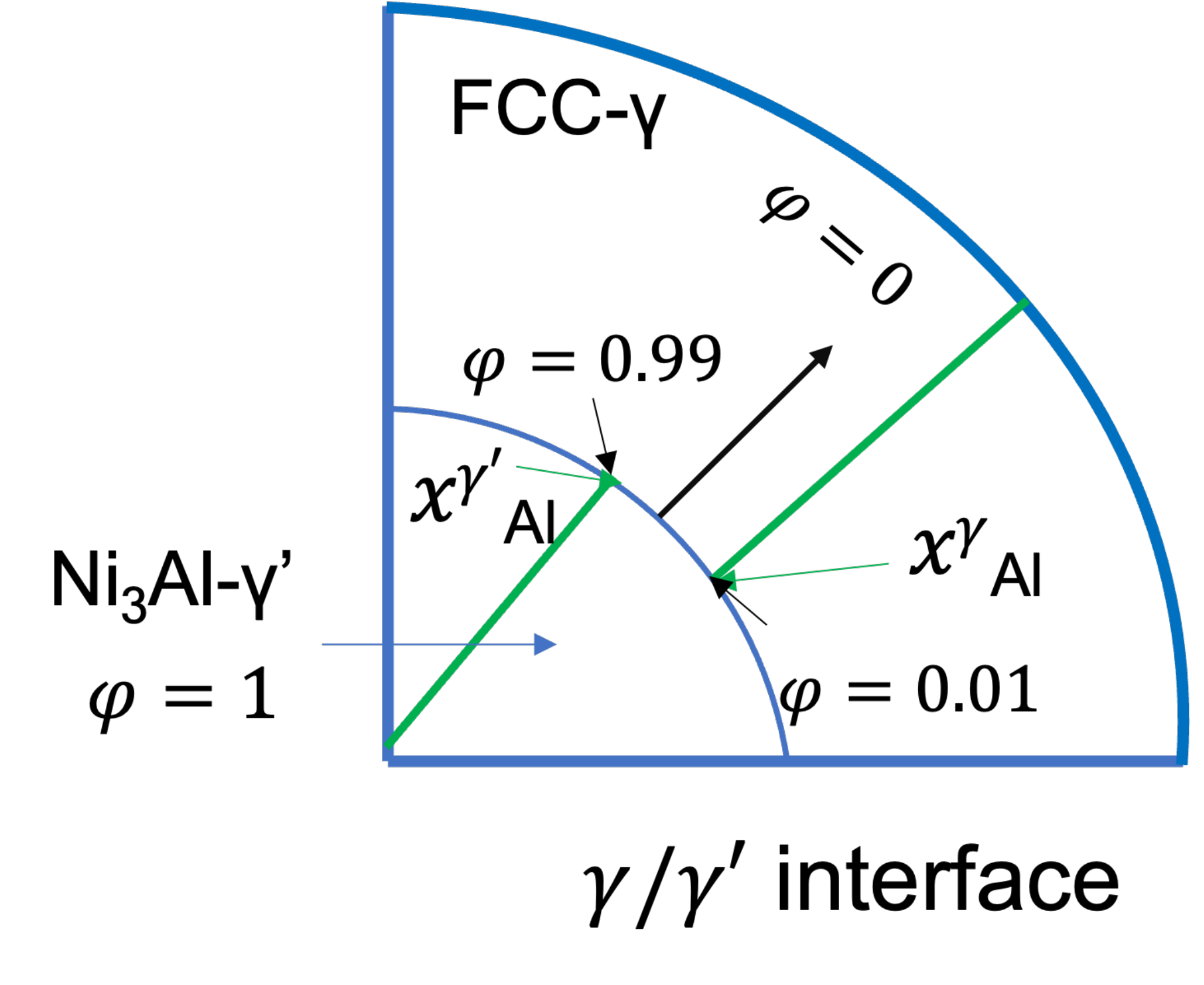}
\caption{}
\label{Fig.5f}
\end{subfigure}
\caption{\color{black}a) Variation of interfacial Al mole fractions in the $\gamma$ and $\gamma^{\prime}$ side of the interface as a function of time for the case of the non-planar interface. b) Schematic depicting the interfacial concentrations and the system geometry. Notice that the interfacial concentrations in both phases are higher in case I (with eigenstrains) as compared to case II (without eigenstrains). The dotted lines show the equilibrium concentrations in both phases, and $\phi$ represents the phase-field variable.}
\end{figure}

\textcolor{black}{To verify this, we again determine the interfacial concentrations as a function of simulation time. Moreover, we follow the same procedure as described in the planar case to determine these concentrations. This is shown schematically in Fig. \ref{Fig.5f}. In this case, we also find that the interfacial concentrations in case II are higher compared to case I (Fig. \ref{Fig.5e}). However, the magnitude of shift from the equilibrium values is lower in the non-planar $\gamma/\gamma^{\prime}$  case as compared to the planar $\gamma/\gamma^{\prime}$ case due to curvature effects.}

\begin{figure}[!ht]
\begin{subfigure}{0.5\textwidth}
\includegraphics[keepaspectratio,width=\linewidth]{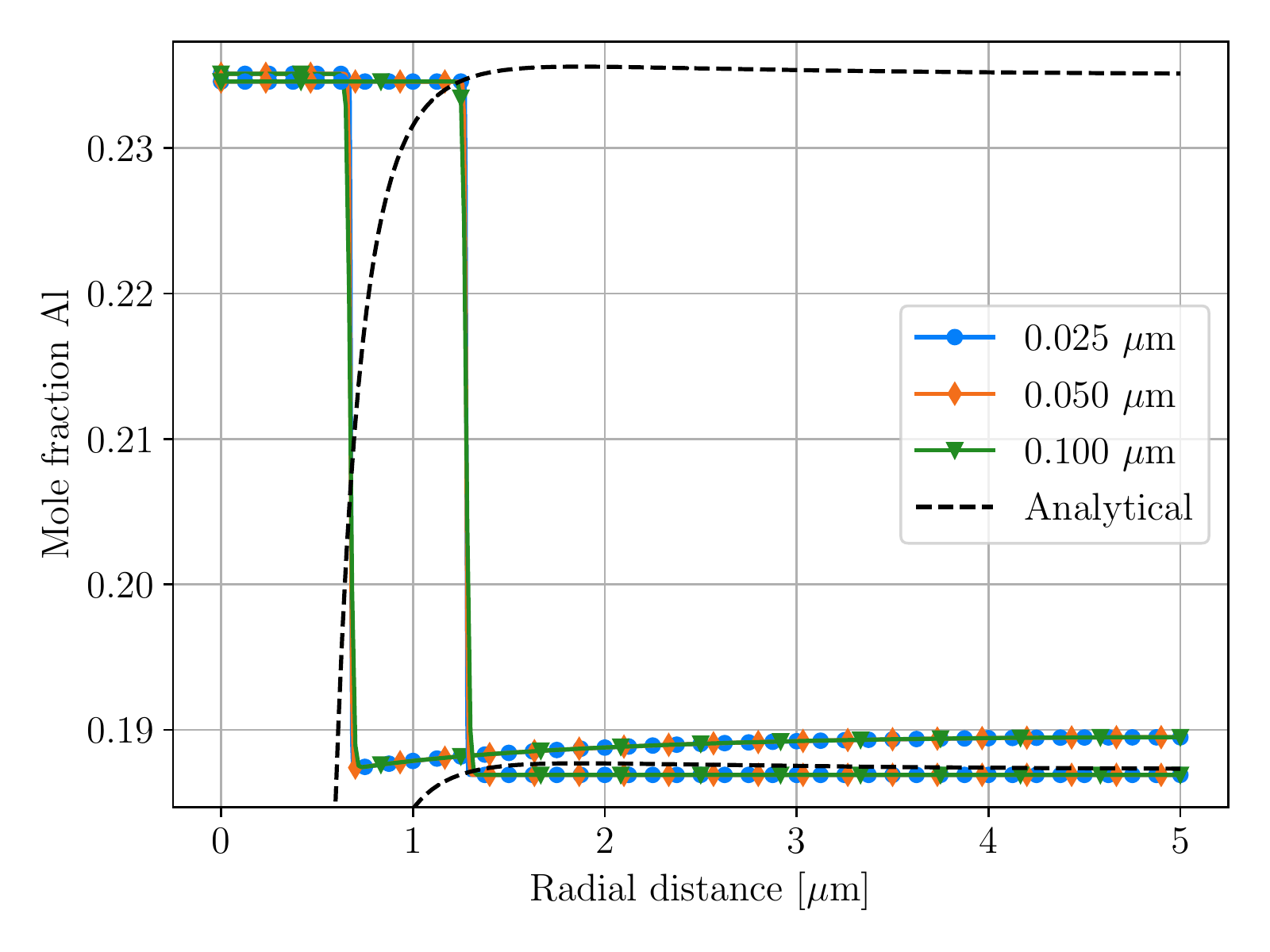}
\caption{}
\label{Fig.6a}
\end{subfigure}
\begin{subfigure}{0.5\textwidth}
\includegraphics[keepaspectratio,width=\linewidth]{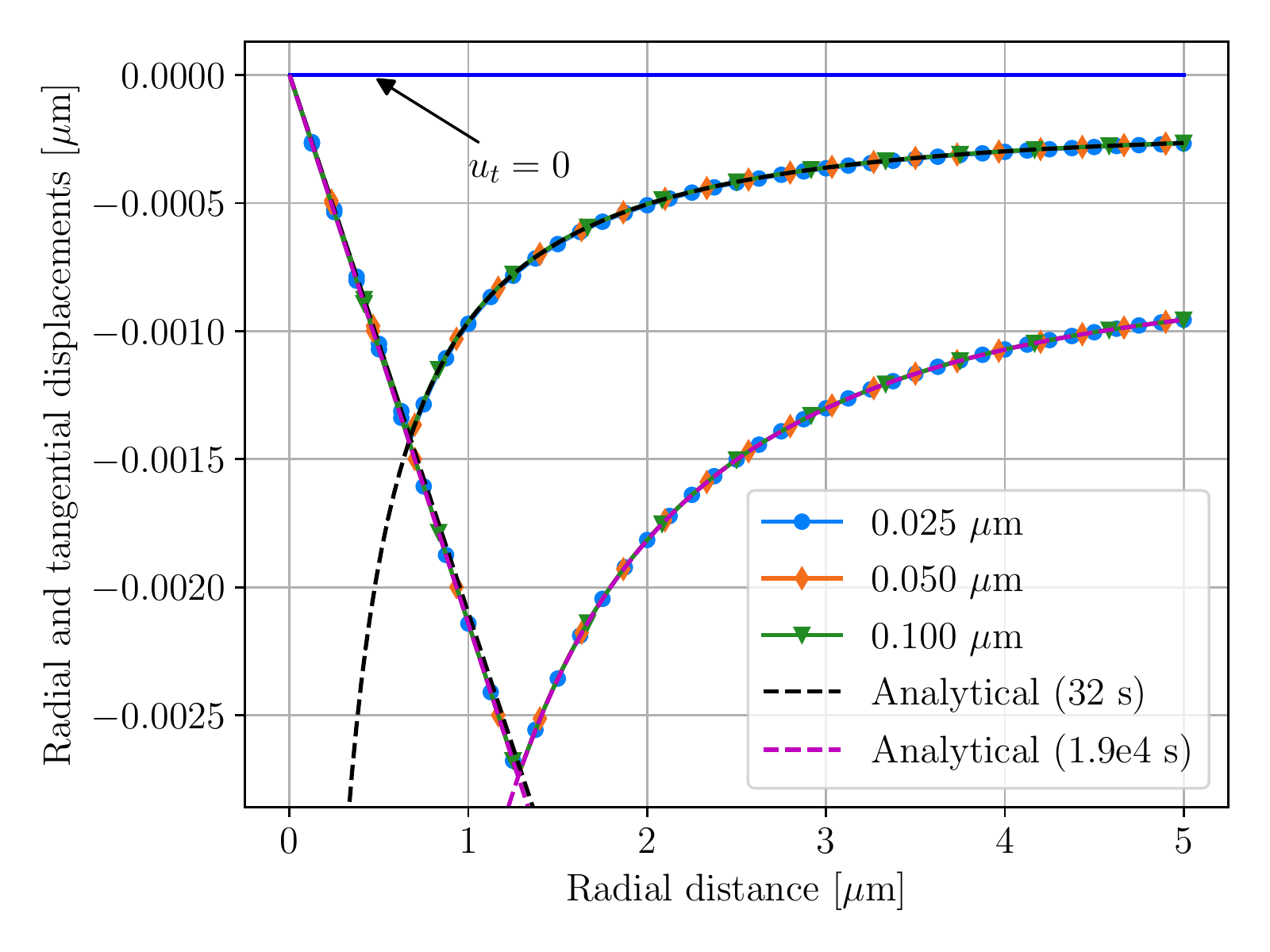}
\caption{}
\label{Fig.6b}
\end{subfigure}
\begin{subfigure}{0.5\textwidth}
\includegraphics[keepaspectratio,width=\linewidth]{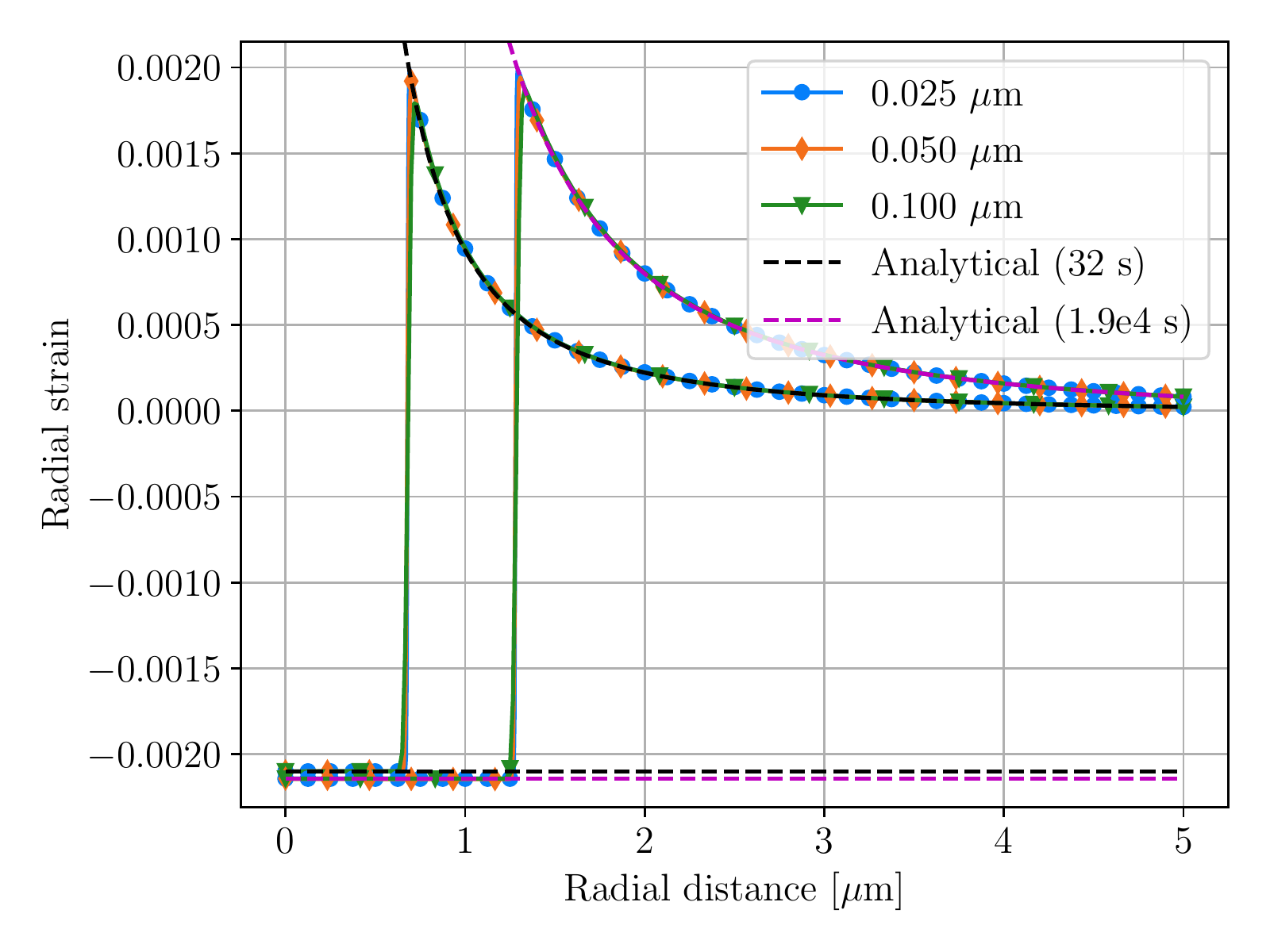}
\caption{}
\label{Fig.6c}
\end{subfigure}
\begin{subfigure}{0.5\textwidth}
\includegraphics[keepaspectratio,width=\linewidth]{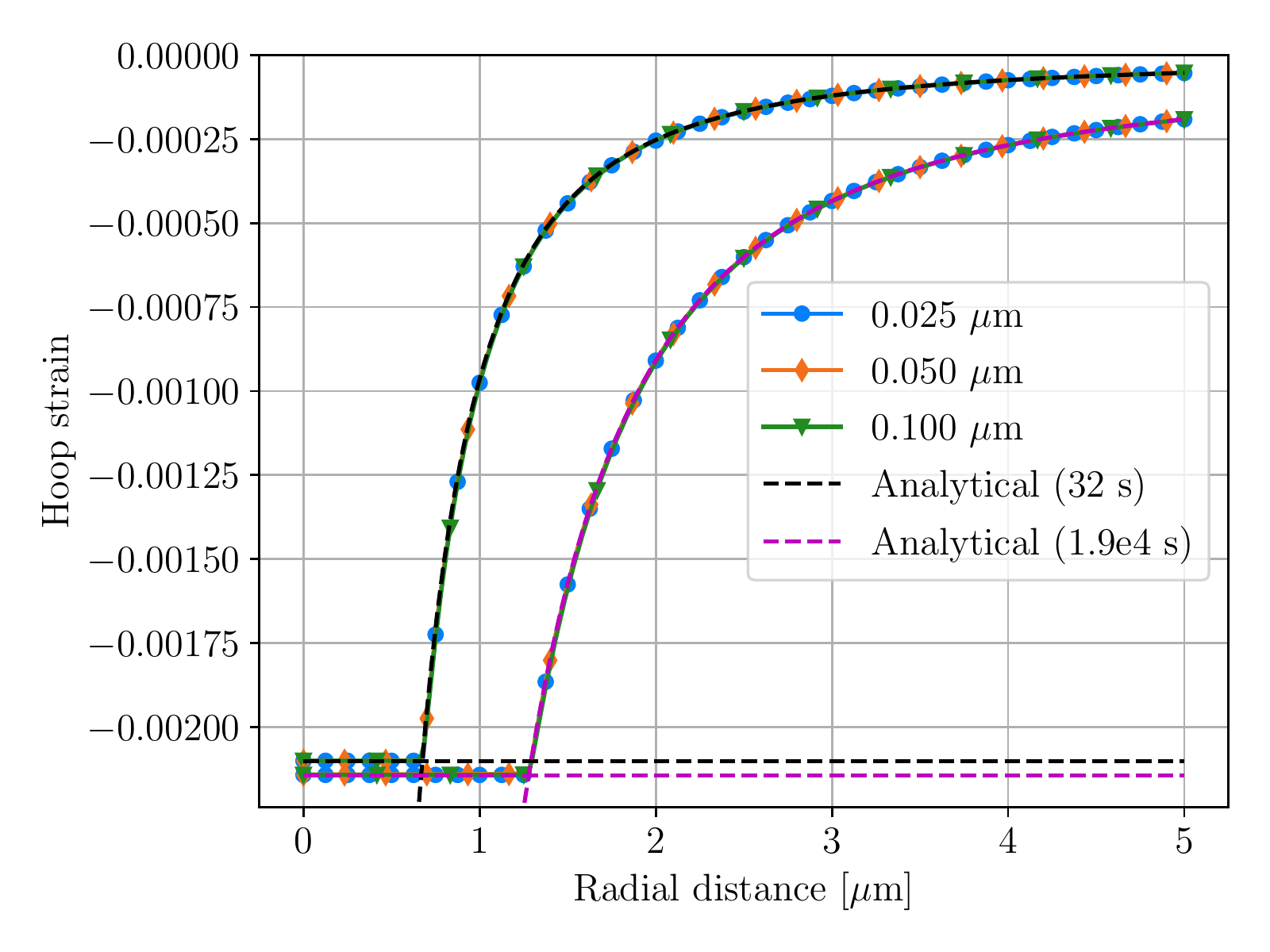}
\caption{}
\label{Fig.6d}
\end{subfigure}
\begin{subfigure}{0.5\textwidth}
\includegraphics[keepaspectratio,width=\linewidth]{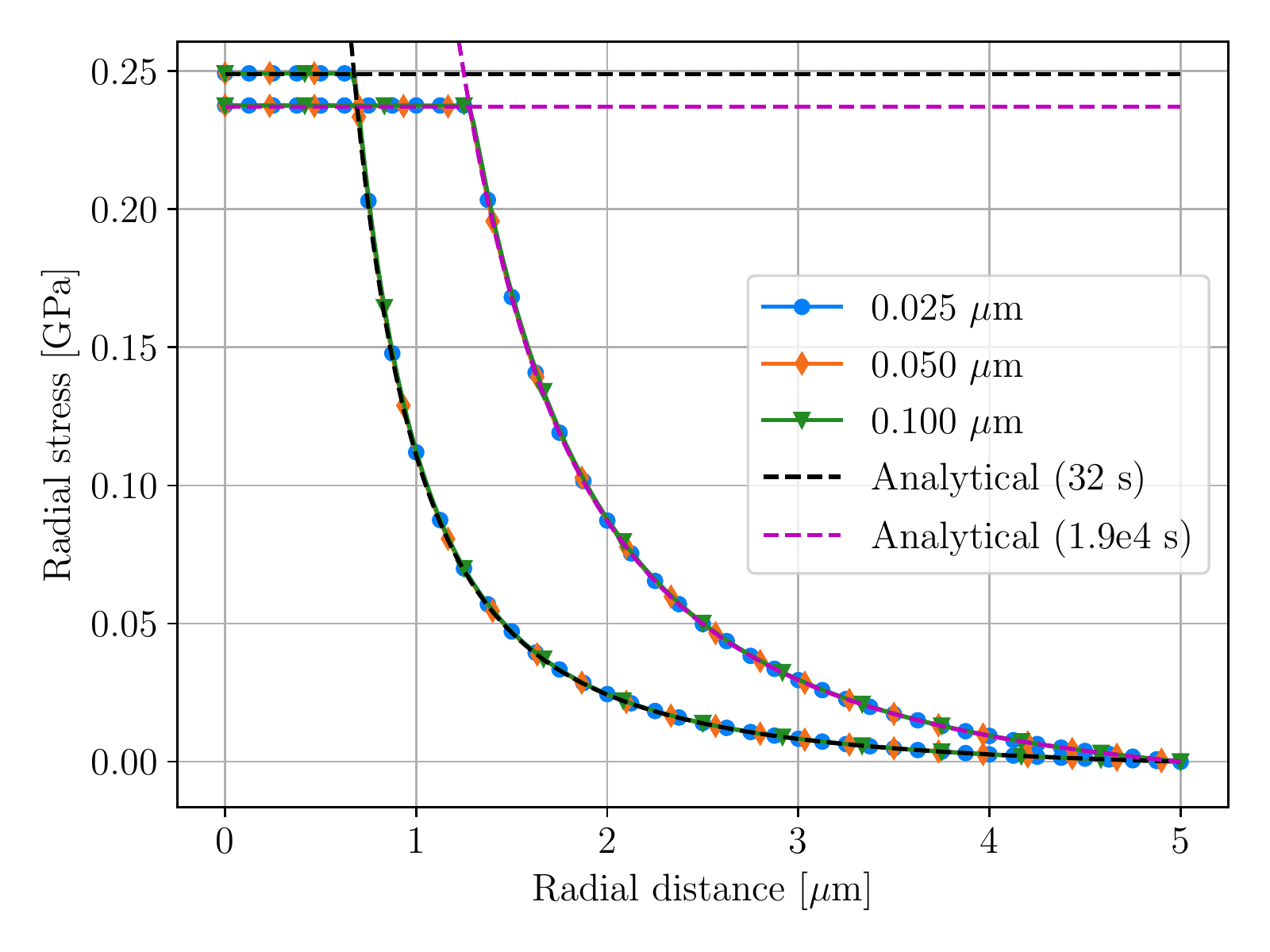}
\caption{}
\label{Fig.6e}
\end{subfigure}
\begin{subfigure}{0.5\textwidth}
\includegraphics[keepaspectratio,width=\linewidth]{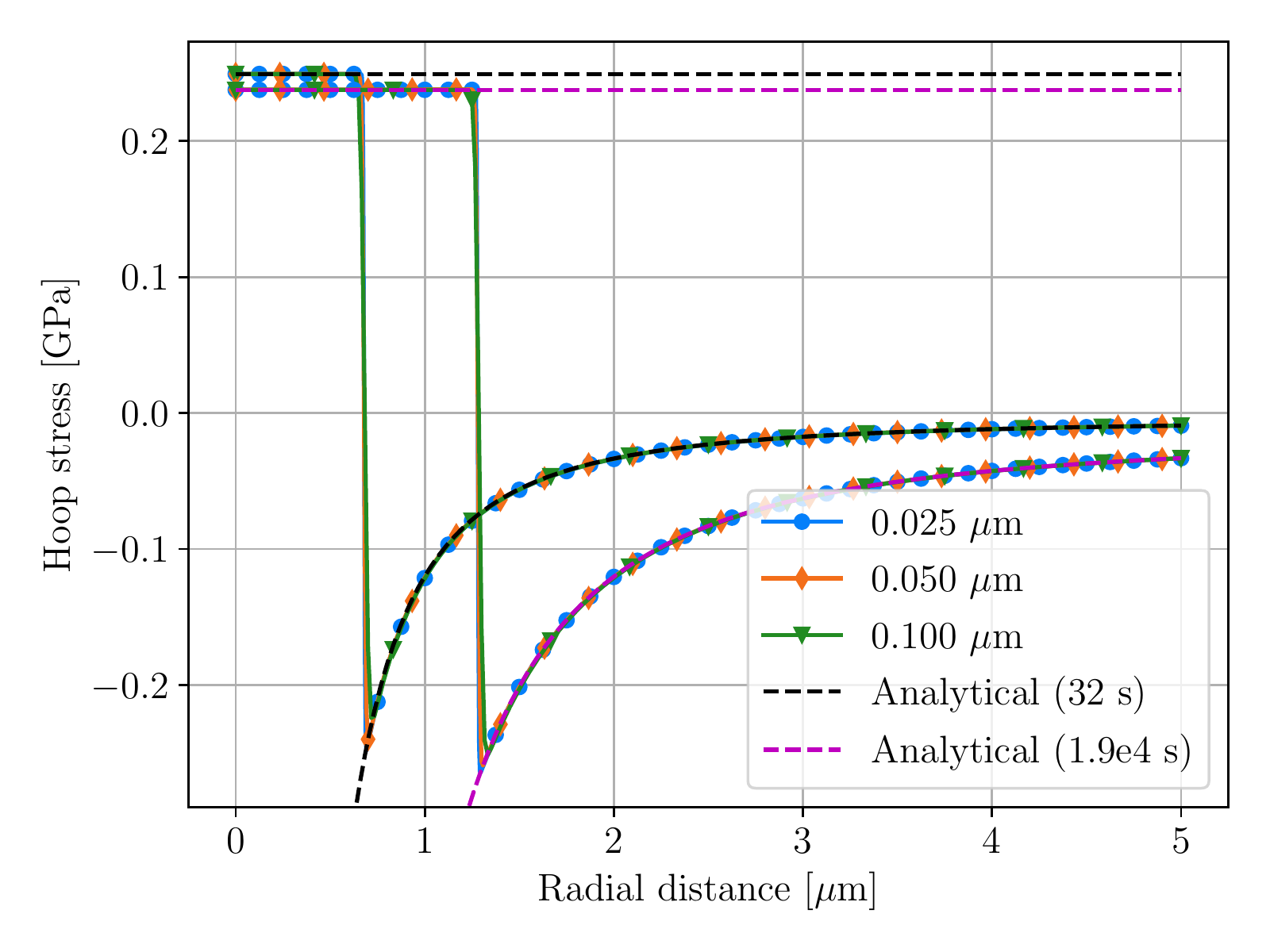}
\caption{}
\label{Fig.6f}
\end{subfigure}
\caption{Comparison of the simulated a) Al-mole fraction field; b) radial and tangential displacements; c) radial strain; d) hoop strain; e) radial stress and f) hoop stress at two different time steps and for three different widths with analytical solutions. \textcolor{black}{The dotted lines indicate the analytically obtained solution for the special case of plane stress (Appendix D).} Notice that the superimposed analytical solutions depend on the interface position.}
\label{Fig.6}
\end{figure}

Moreover, we find that the simulated elastic fields (shown in polar coordinate frame) are in quantitative agreement with analytical solutions \textcolor{black}{obtained assuming plane stress conditions} (Fig.{\ref{Fig.6}}). Fig.\ref{Fig.6b} shows that the radial displacement field is maximum near the interface and increases with time. While the tangential displacement within both the precipitate and matrix phases is zero. Notice that the radial displacement field within the $\gamma^{\prime}$ phase is linear as a function of radial distance. This indicates that both the radial and hoop strains are uniform within the precipitate phase with increasing growth (Figs.\ref{Fig.6c} and \ref{Fig.6d}). However, in the matrix phase, the variation of radial and hoop strains with radial distance is non-linear. Consequently, the radial and hoop stresses are also uniform within the precipitate phase (Figs.\ref{Fig.6c} and \ref{Fig.6d}). It is clear from this comparison that the simulated elastic fields using the partial rank-one homogenization scheme are independent of the interface width choice within the bulk phases but show slight deviations in the interfacial region.

\subsection{Non-planar UO$_2$-Vacancy simulation}
\begin{figure}[!ht]
\begin{subfigure}{0.5\textwidth}
\includegraphics[trim=0 7 0 7,clip,keepaspectratio,width=\linewidth]{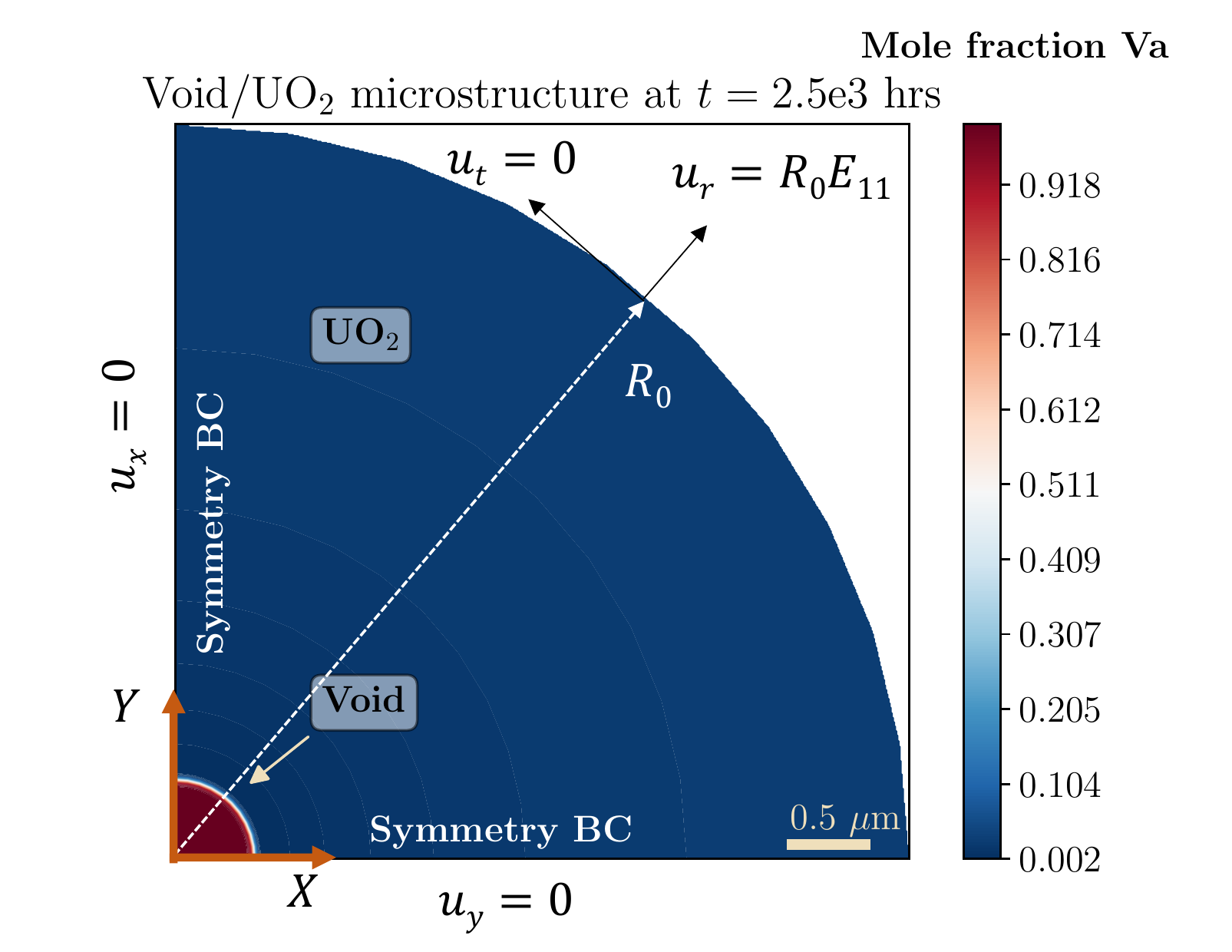}
\caption{}
\label{Fig.7a}
\end{subfigure}
\begin{subfigure}{0.5\textwidth}
\includegraphics[keepaspectratio,width=\linewidth]{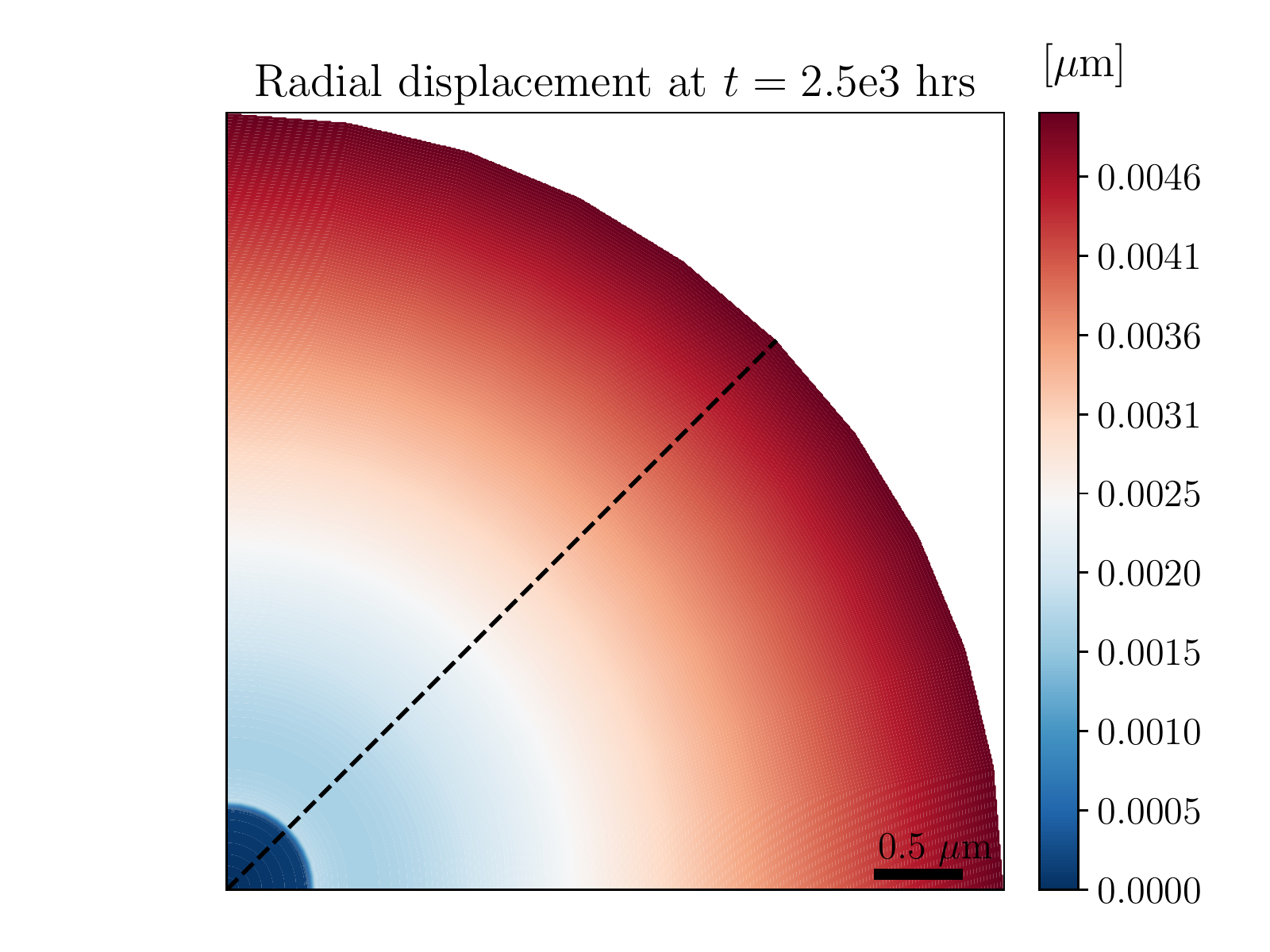}
\caption{}
\label{Fig.7b}
\end{subfigure}
\begin{subfigure}{0.5\textwidth}
\includegraphics[keepaspectratio,width=\linewidth]{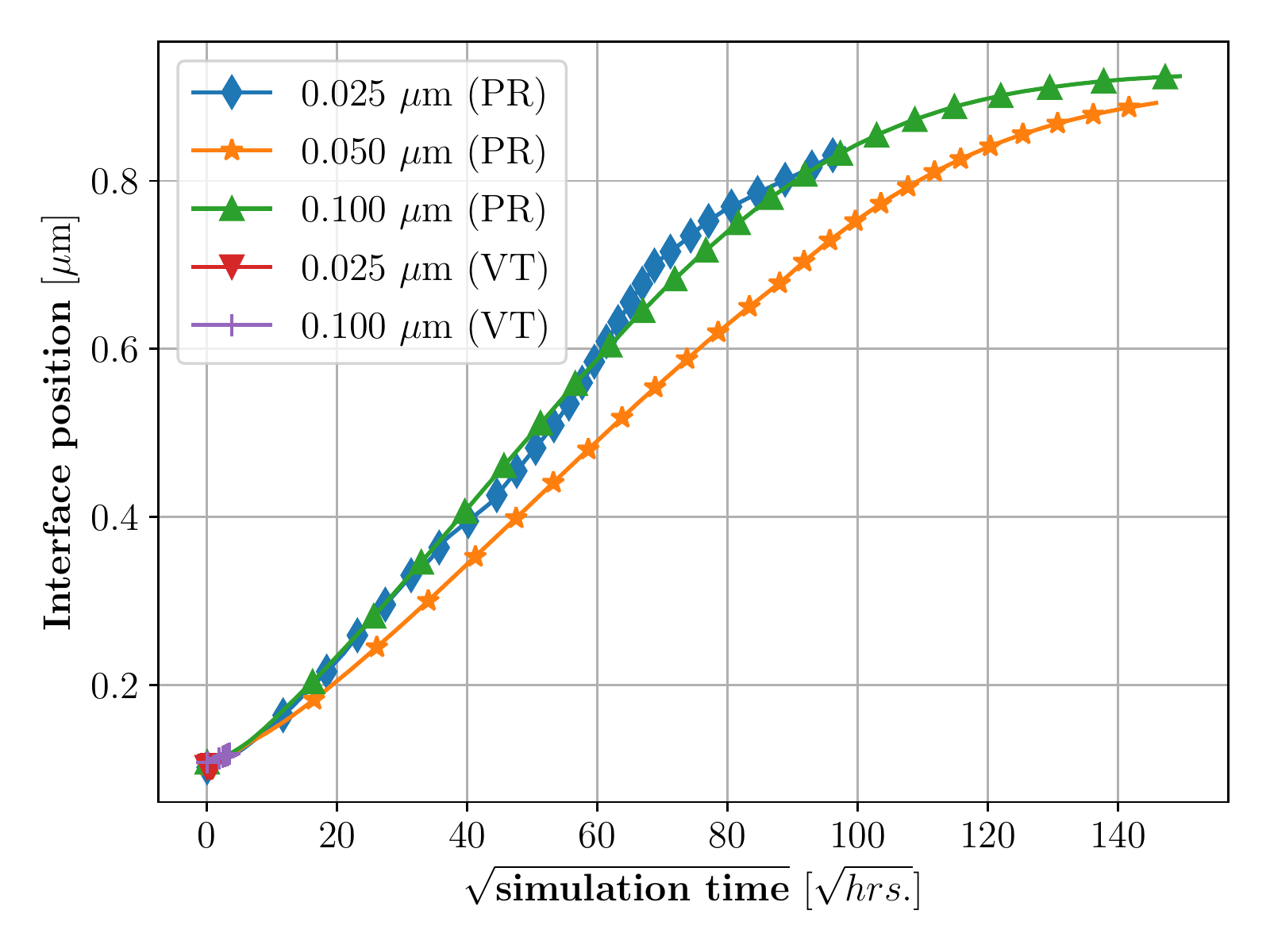}
\caption{}
\label{Fig.7c}
\end{subfigure}
\begin{subfigure}{0.5\textwidth}
\includegraphics[keepaspectratio,width=\linewidth]{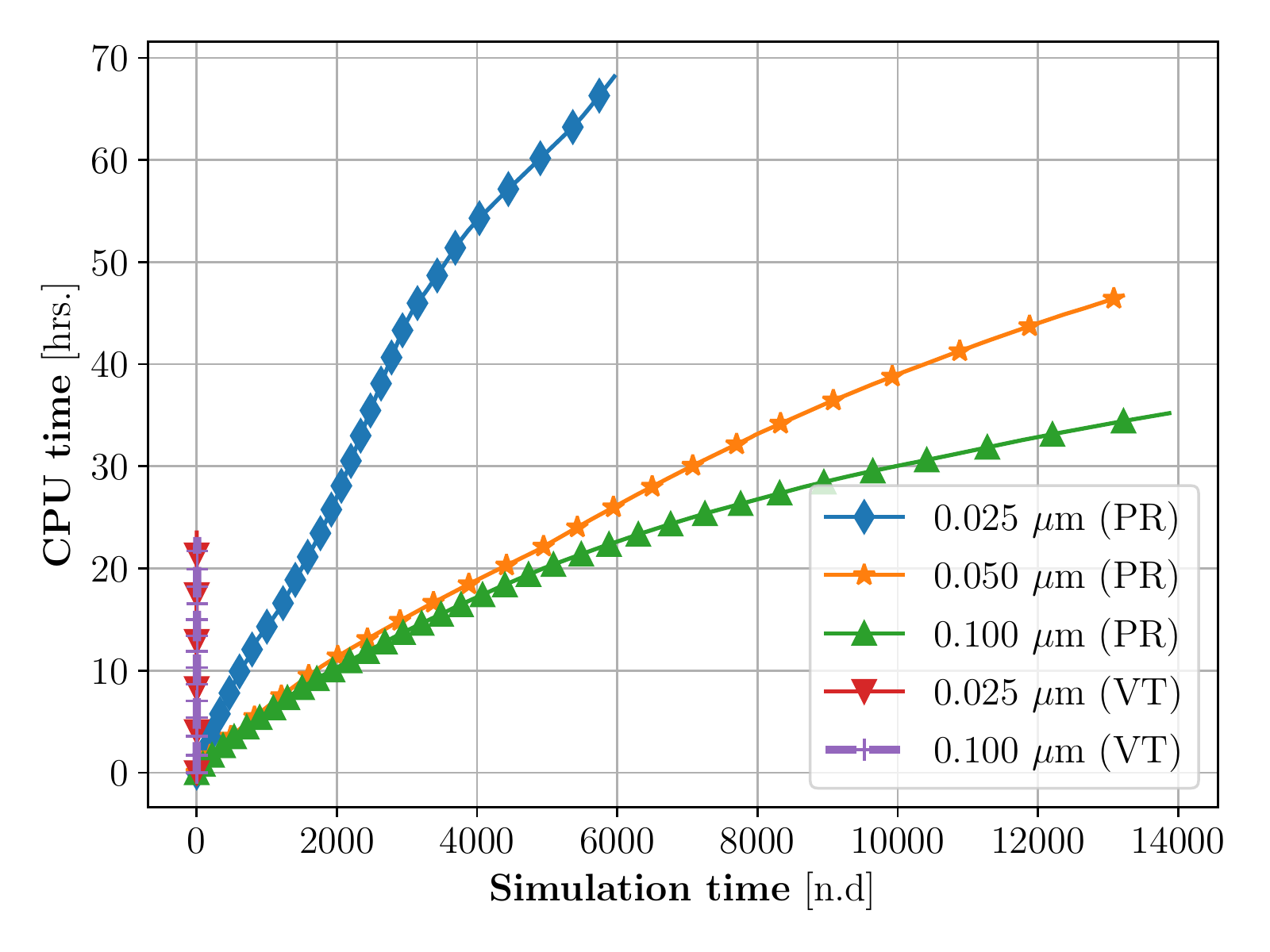}
\caption{}
\label{Fig.7d}
\end{subfigure}
\caption{Simulated a) vacancy mole fraction, and b) radial displacement for the UO$_2$/Void alloy system. For a range of interface width ($0.025$ $\mu$m-$0.10$ $\mu$m), c) the variation of interface position as a function of the square root of time, and d) CPU time as a function of non-dimensional simulation time using the partial rank-one (PR) scheme. Both interface position and the CPU time taken when the Voigt-Taylor (VT) scheme is employed are superimposed on Figs. \ref{Fig.7c} \& \ref{Fig.7d}, respectively. Notice that the time taken using the VT scheme is significantly higher compared to the PR scheme for the same simulation time. The dotted line in Fig. \ref{Fig.7b} indicate the radial distance along which the field quantities are evaluated in Fig.\ref{Fig.8}.}
\end{figure}

Next, we selected a non-planar UO$_2$/void system. \textcolor{black}{In this simulation, we have assumed a circular-shaped void embedded at the center of a circular-shaped UO$_2$ matrix. Due to symmetry reasons, we have simulated only a quarter of the circular domain. Moreover, the boundary conditions on the phase-field and mole fraction variables are identical to the non-planar Ni-Al case. However, the boundary conditions on displacements are different at the outer-most boundary (see Table \ref{table:bc} and Fig. \ref{Fig.7a}). Specifically,}
\textcolor{black}{
\begin{align}
u_{x}(x,y) &= E_{11}x, \quad \forall \quad x^{2} + y^{2} = R_{0}^{2}, \\
u_{y}(x,y)  &= E_{22}y, \quad \forall \quad x^{2} + y^{2} = R_{0}^{2}, 
\end{align}
}
\textcolor{black}{where for simplicity we have taken $E_{11} = E_{22} = 0.001$ and $R_{0} = 5 \mu$m is the outermost radius of the UO$_2$ matrix. As a consequence of this fixed boundary condition, it can be shown using Eq. (\ref{EqnCT}) that the imposed tangential displacement is zero, but the radial displacement is equal to $R_{0}E_{11}=0.005 \mu$m at the outer boundary (see Figs. \ref{Fig.7a}-\ref{Fig.7b}). Since the system has no imposed eigenstrains, the fixed boundary condition at the outermost boundary is the source of elastic stresses. Moreover, at the left and bottom boundaries, the displacement boundary conditions are identical to the non-planar Ni-Al case (Eqs. \ref{EqnResult7}-\ref{EqnResult8}).}

For the same interface width range as in the Ni-Al non-planar case, we find that the variation of interface position as a function of the square root of time depends on the interface width using the PR scheme (Fig.\ref{Fig.7c}). Because of the strong elastic heterogeneity ($E_{void}/E_{uo_2}=1\mathrm{e}{-4}$) this dependence on interface width is observed. However, similar to the planar UO$_2$/void case, the VT scheme nearly fails to simulate the given alloy system (Fig.\ref{Fig.7c}). This is mainly because the VT scheme has poor convergence compared to the PR scheme (Fig.\ref{Fig.7d}). As mentioned before, this poor convergence is not due to our implementation since the governing equations remain the same but the thermodynamic driving force changes. To gain a quantitative understanding, we calculate the average time step,  $\Delta t_{avg}$, that is, the ratio of total time by a number of time steps, for both schemes. Concretely, we find that the $\Delta t_{avg}^{pr}$ using the PR scheme is nearly $693.27$ times higher compared to the $\Delta t_{avg}^{vt}$ using the VT scheme for an interface width value of $0.025$ $\mu$m.

\begin{figure}[!ht]
\begin{subfigure}{0.5\textwidth}
\includegraphics[keepaspectratio,width=\linewidth]{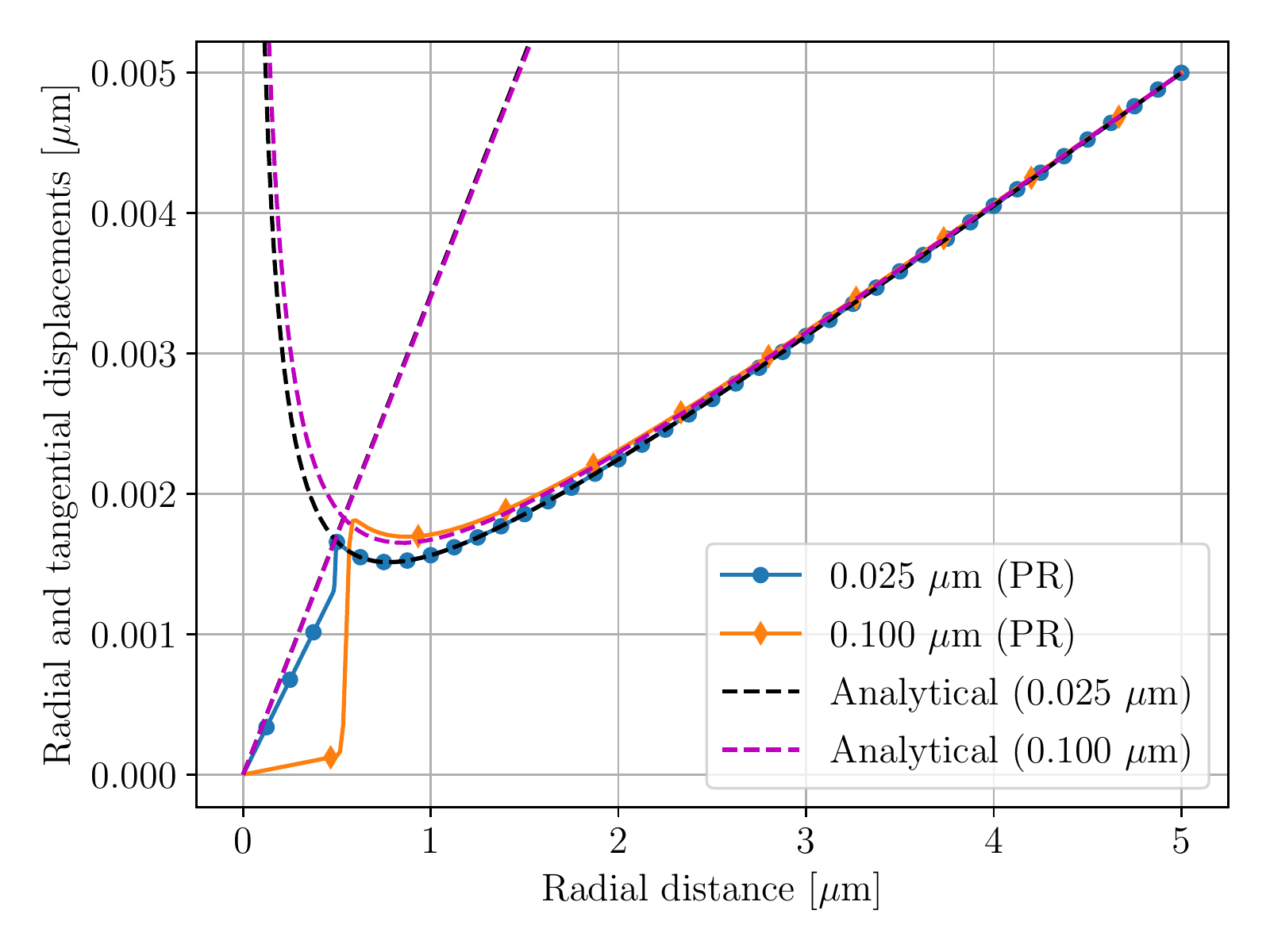}
\caption{}
\label{Fig.8a}
\end{subfigure}
\begin{subfigure}{0.5\textwidth}
\includegraphics[keepaspectratio,width=\linewidth]{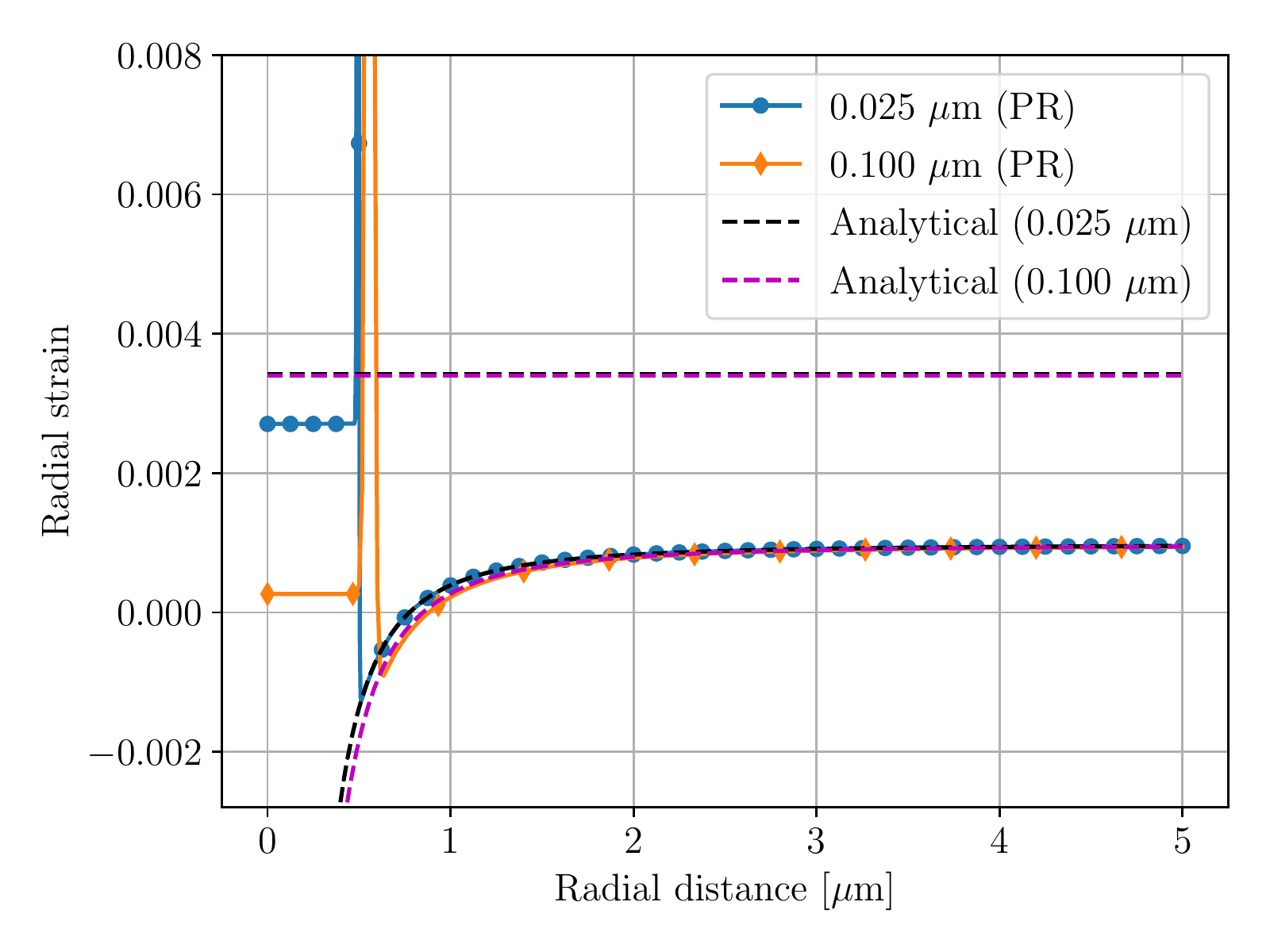}
\caption{}
\label{Fig.8b}
\end{subfigure}
\caption{Comparison of a) radial displacement and b) radial strain with analytical solutions as a function of radial distance for two different interface widths using the partial rank-one (PR) homogenization scheme. The values within brackets indicate the interface width. Since the analytical solution requires the interface position, which was slightly different in both cases (see Fig.\ref{Fig.7c}), this value was mentioned specifically for this case.}
\label{Fig.8}
\end{figure}

Also, in contrast to the planar UO$_2$/void case, we find that the bulk elastic fields within the void phase depend on the interface position and deviate from the analytical solution (Fig.\ref{Fig.8}). \textcolor{black}{This analytical solution has been obtained assuming plane stress conditions.} We find that the radial displacement field in the void phase is lower compared to the analytically predicted value for an interface width of $0.1$ $\mu$m  (Fig.\ref{Fig.8a}). Moreover, the continuity of radial displacement is not satisfied near the interfacial region, in contrast to analytical predictions (Fig.\ref{Fig.8a}). However, as the interface width is decreased to $0.025$ $\mu$m, the simulated radial displacement field gets closer to the analytical solution (Fig.\ref{Fig.8a}). Also, notice that the interface position slightly varies with a change in interface width, and thus the analytical solution in the UO$_2$ phase shifts slightly near the interface. This trend is also reflected in the variation of radial strain as a function of distance (Fig.\ref{Fig.8b}). Although the bulk radial strain is qualitatively similar to the analytical solution in the void phase, we find significant quantitative disagreement with the analytical solution for both interface widths. In the bulk UO$_2$ phase, however, quantitative agreement between simulated and analytical solutions is obtained. We believe this is because of the extremely low modulus of the void phase compared to the UO$_2$ phase. Nevertheless, this shows that for strongly heterogeneous alloys, the PR scheme should be preferred over the VT scheme, provided that the interface width value is taken to be less than (1/4) of the initial particle size.

\subsection{Multiparticle $\gamma^{\prime}/\gamma$ simulation}
Finally, we simulated a multiparticle elastically heterogeneous and anisotropic system using the partial rank-one (PR) scheme. \textcolor{black}{In this case, we have imposed periodic boundary conditions on all variables along both $x$ and $y$ directions. We initialize the system with a random distribution of 160 circular-shaped precipitates with a mean radius of $0.1$ $\mu$m. An inbuilt adaptive mesh refinement capability within MOOSE was employed to reduce the computational costs of running simulations in a uniformly generated mesh.}

\begin{figure}[!ht]
\begin{subfigure}{0.5\textwidth}
\includegraphics[keepaspectratio,width=\linewidth]{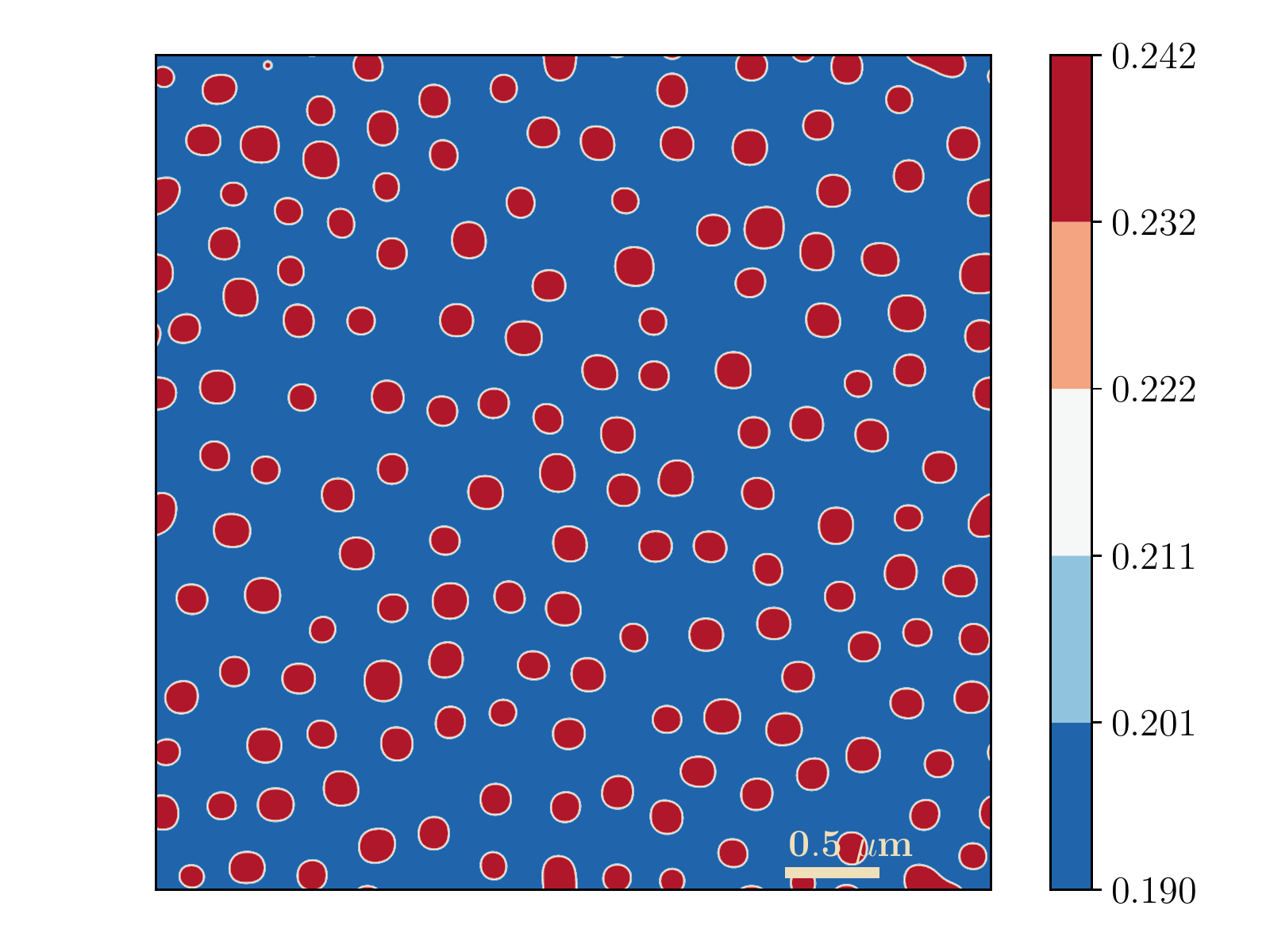}
\caption{}
\label{Fig.9a}
\end{subfigure}
\begin{subfigure}{0.5\textwidth}
\includegraphics[keepaspectratio,width=\linewidth]{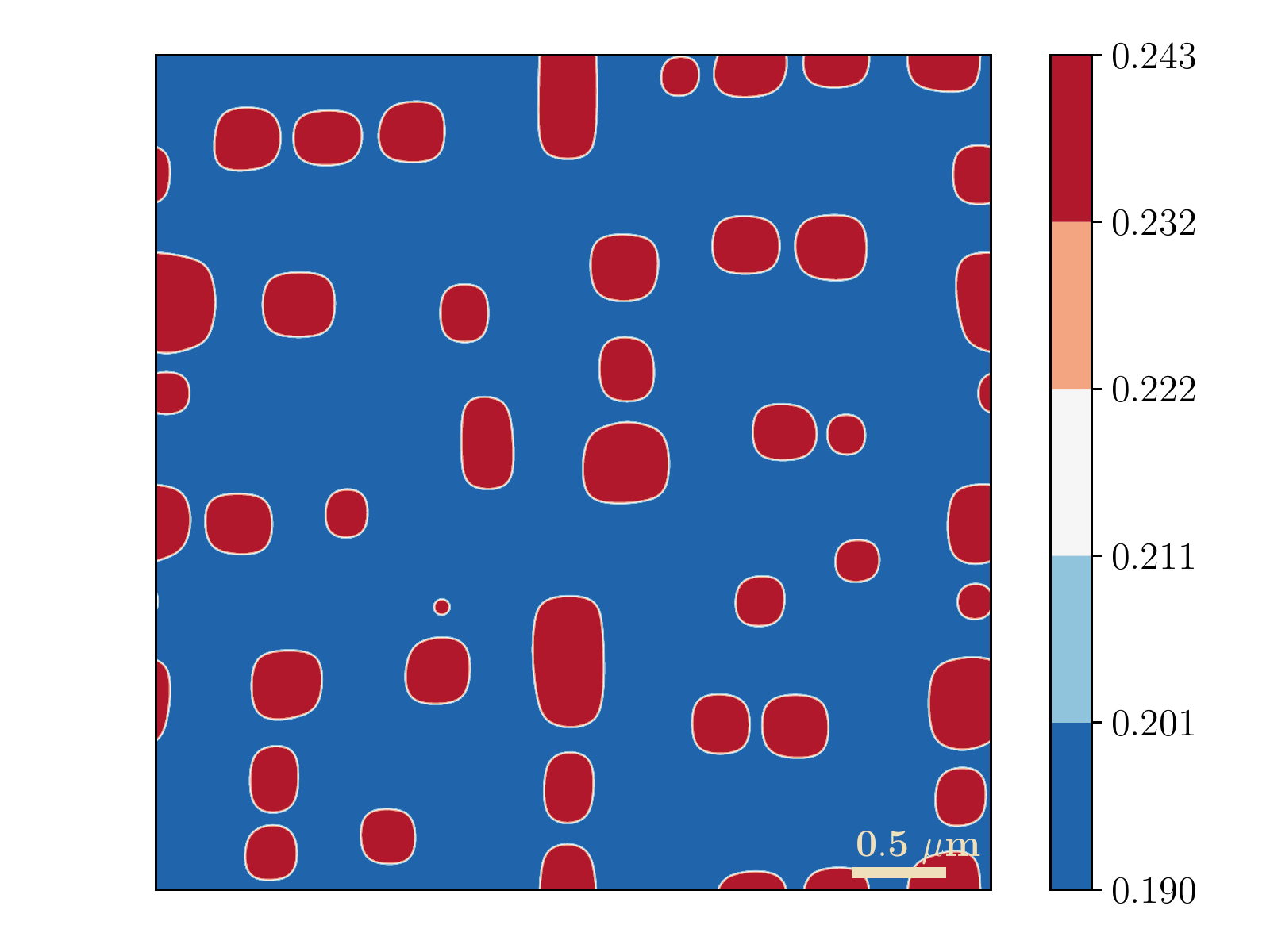}
\caption{}
\label{Fig.9b}
\end{subfigure}
\begin{subfigure}{0.5\textwidth}
\includegraphics[keepaspectratio,width=\linewidth]{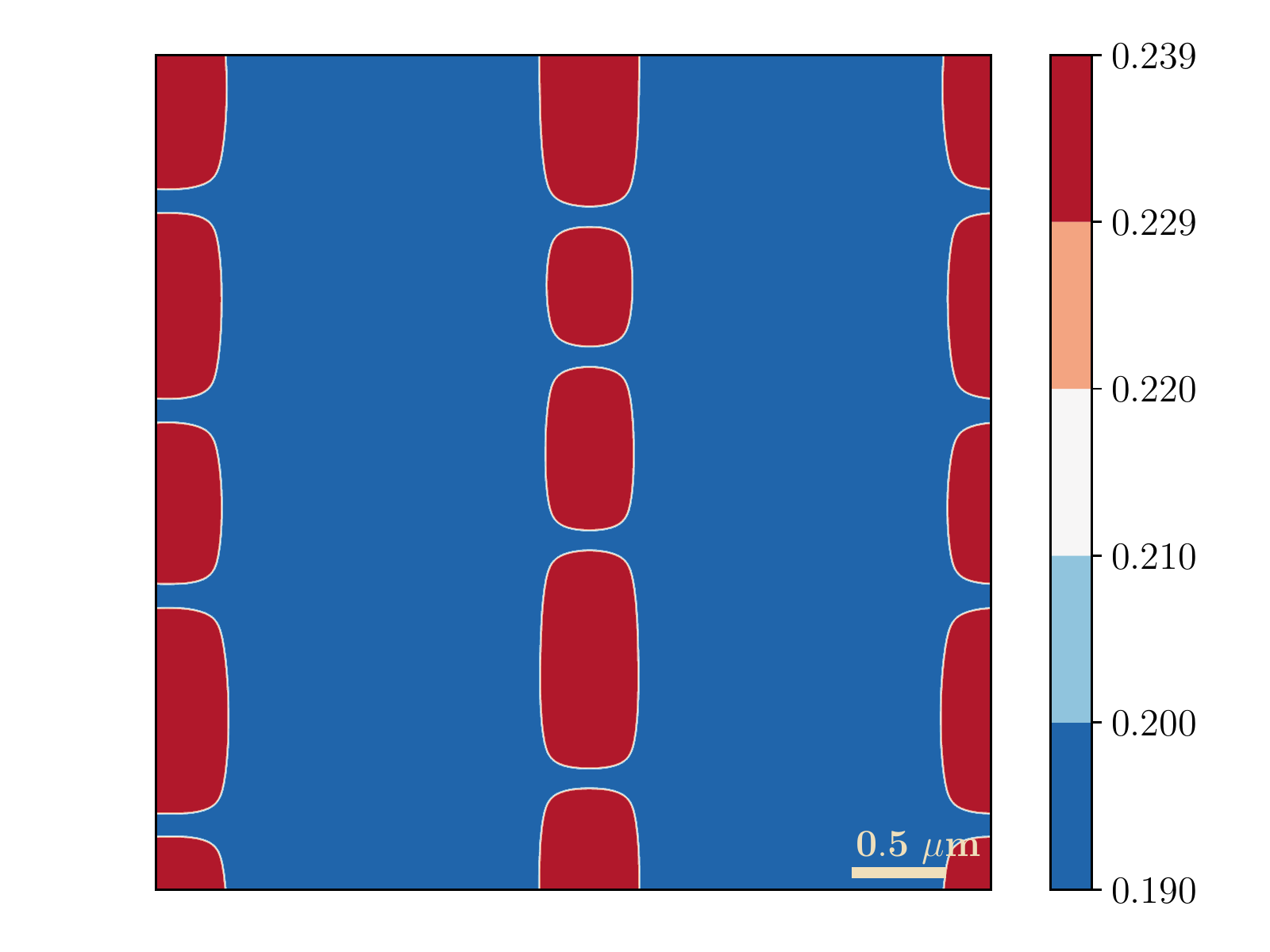}
\caption{}
\label{Fig.9c}
\end{subfigure}
\begin{subfigure}{0.5\textwidth}
\includegraphics[keepaspectratio,width=\linewidth]{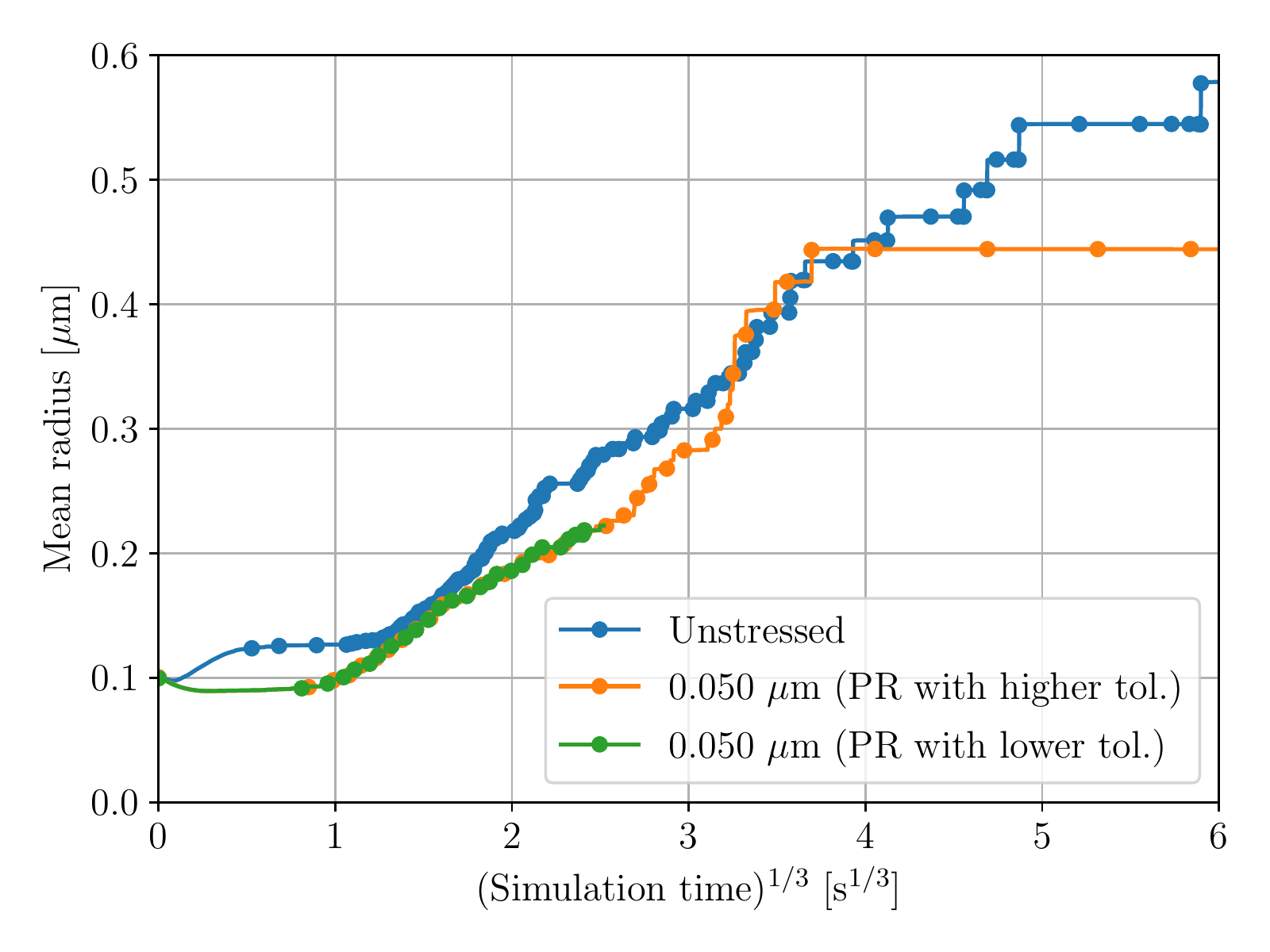}
\caption{}
\label{Fig.9d}
\end{subfigure}
\caption{Simulated Al mole fraction field in a Ni-Al $\gamma^{\prime}/\gamma$ alloy at temperature $T = 1473$ K and time $t$ equal to $0.2186$ s (a); 8.7789 s (b); and 135.3717 s (c). For the same alloy, d) variation of mean radius ($\sqrt{A/\pi}$) as a function of the cube root of simulation time for the unstressed (blue) and stressed cases (orange). Another coarsening simulation of the same alloy with lower absolute and relative tolerances (green) that did not converge to the equilibrium state is superimposed on Fig. \ref{Fig.9d}. The domain size is $5\times 5$ $\mu$m$^{2}$, and periodic boundary conditions were enforced.}
\end{figure}

When the effect of elastic strain energy is small compared to the interfacial energy, the microstructure has a uniform distribution of precipitates (Fig.\ref{Fig.9a}). However, with increasing time, we find that the precipitate shape becomes increasingly cuboidal  (Figs. \ref{Fig.9b} and \ref{Fig.9c}). Moreover, the $\gamma^{\prime}$ precipitates show preferential alignment along the elastically soft $\left<10\right>$ and $\left<01\right>$ directions. This preference is mainly due to the elastically anisotropic interactions between the $\gamma^{\prime}$ precipitates. Moreover, the variation of the mean radius with the cube root of simulation time is shown in Fig.\ref{Fig.9d}. \textcolor{black}{We find that the coarsening kinetics in case I is marginally slower compared to case II (Fig. \ref{Fig.9d}).} 

It must be pointed out that, with the exception of this simulation, we have used a relative non-linear tolerance of $1\mathrm{e}{-8}$ and an absolute non-linear tolerance of $1\mathrm{e}{-10}$  to obtain convergence in all four previous simulations. The reason for this is our simulation did not converge after time $t = 15.93$ s when the tolerance values were set equal to the single-particle simulations (Fig. \ref{Fig.9d}). Therefore, we increased both these tolerances by a factor of $100$. Consequently, the simulation converged to the equilibrium state (Fig. \ref{Fig.9d}). Moreover, a comparison of the coarsening kinetics shows that the accuracy of the simulation results remains unaffected despite this increase in tolerance values (Fig.\ref{Fig.9d}). 

\textcolor{black}{Lastly, to show the effect of composition on elastic fields, we have simulated the same multiparticle system by considering composition-dependent eigenstrains instead of constant eigenstrains. Since in a grand-potential model, the diffusion potential is the independent variable, this dependence arises indirectly through the \textit{phase} compositions. Concretely, by assuming Vegard's law \cite{Wu2001}, the eigenstrains in the $\gamma^{\prime}$ phase $\boldsymbol{\epsilon}^{\star\gamma^{\prime}}$ can be written as}
\begin{align}
\color{black}
\boldsymbol{\epsilon}^{\star\gamma^{\prime}}\left(\boldsymbol{\tilde{\mu}}\right) = \epsilon \boldsymbol{1}\left[X_{Al}^{\gamma^{\prime}}\left(\boldsymbol{\mu}\right) - X_{Al}^{0}\right],
\label{ResultEq2}
\end{align}
\textcolor{black}{
where $\epsilon=-0.3\%$, $\boldsymbol{1}$ is the identity tensor, $X_{Al}^{\gamma^{\prime}}$ is the Al mole fraction of $\gamma^{\prime}$ phase and $X_{Al}^{0}$ is the overall Al mole fraction. For our case, this value was determined to be $0.19$.}

\begin{figure}
\begin{center}
\includegraphics[trim=70 300 70 70, clip,keepaspectratio,width=\textwidth]{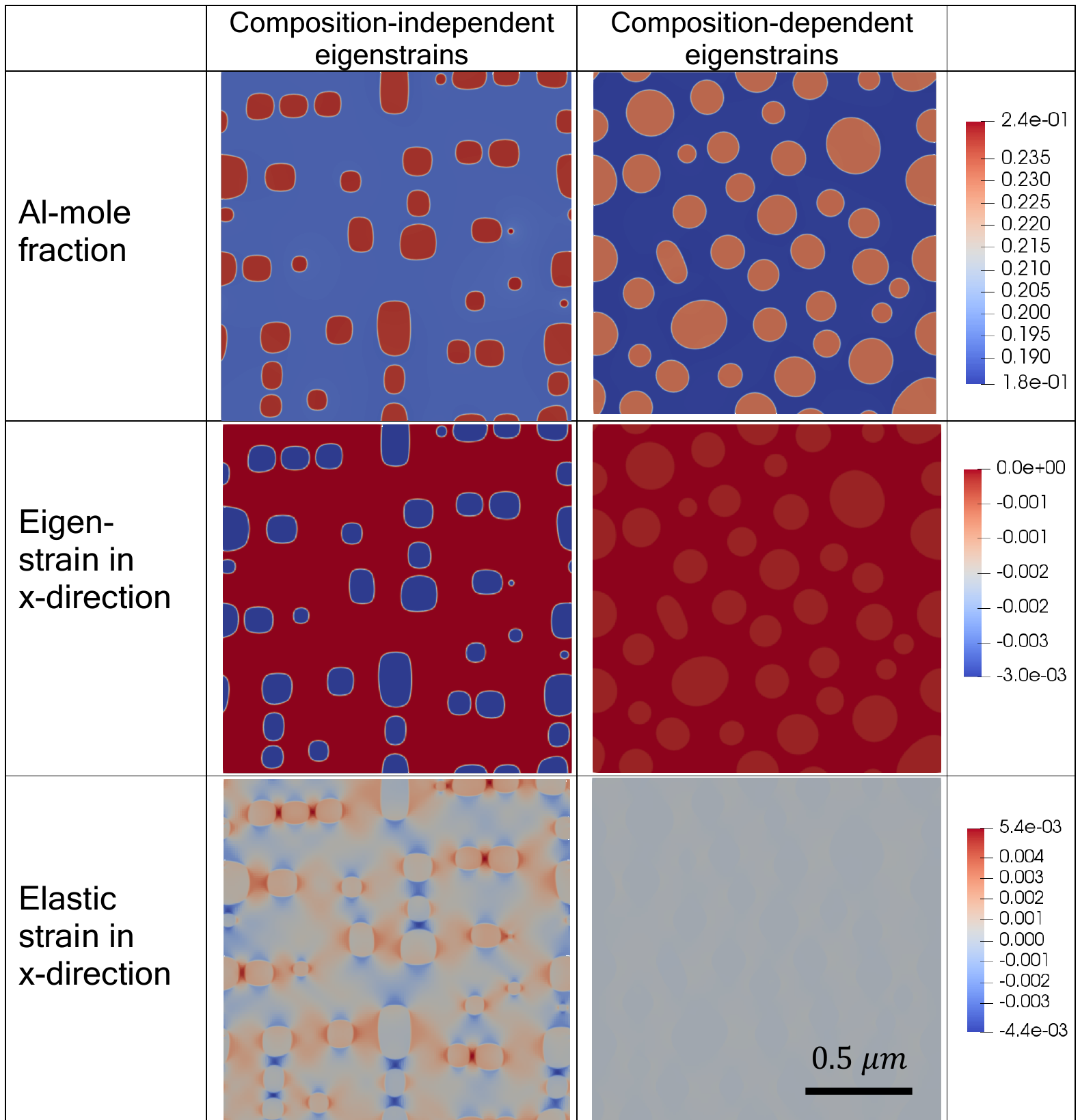}
\caption{\color{black}Comparison of Al-mole fraction fields, eigenstrains and elastic strains in the x-direction for the case of composition-independent eigenstrains and composition-dependent eigenstrains at time $t=10.58$ s. Notice that due to the dependence of eigenstrains on local Al-mole fraction, their effective value within the $\gamma^{\prime}$ phase is much lower compared to the composition-independent case. Consequently, the microstructure is more isotropic in the former case as compared to the latter case.}
\label{Fig.10}
\end{center}
\end{figure}

\textcolor{black}{
Fig.\ref{Fig.10} compares the simulated microstructures in both these cases at time $t=10.58$ s. Notice that as a consequence of Eq (\ref{ResultEq2}), the effective eigenstrains within the $\gamma^{\prime}$ phases are much lower compared to the constant case (see the second row of Fig.\ref{Fig.10}). For this reason, we find that the microstructure, in this case, is relatively more isotropic and randomly distributed compared to the previous case. Also, the elastic strains are negligible as compared to the constant eigenstrain case (Fig.\ref{Fig.10}). It should be emphasized that despite coherency strains, the $\gamma^{\prime}$ morphology, in this case, is isotropic. Moreover, this expected dependence of microstructure on the magnitude of eigenstrain has also been demonstrated experimentally for the case of Ni-Al-Mo alloy \cite{Fahrmann1995}. The possible dependence of compositions on elastic constants can be similarly modelled using Eq. (\ref{Eqn16}). This demonstrates that our model can consider composition-dependent elastic properties through the \textit{phase} compositions, and simulate its subsequent role in microstructure evolution.}

\clearpage

\section{Conclusions}
In this work, we developed a two-phase multicomponent phase-field model for elastically heterogeneous two-phase solids based on a partial rank-one homogenization scheme. The uniqueness of this model is that both mechanical compatibilities and chemical equilibrium is ensured within the interfacial region. It was numerically solved using the fully parallelized MOOSE (Multiphysics Object-Oriented Simulation Environment) package. The current implementation is, therefore, capable of simulating elastically heterogeneous and anisotropic alloy systems. Its performance was demonstrated for two model binary systems—$\gamma^{\prime}/\gamma$ and UO$_2$/void—at length scales much larger than the physical interface width.

We found that the partial rank-one homogenization (PR) scheme converged better compared to the Voigt-Taylor (VT) scheme for both the planar and the non-planar Ni-Al $\gamma^{\prime}/\gamma$ simulations. It was also found that using the former scheme, the variation of interface position with time was unaltered with increasing interface width in both these cases. However, when the latter scheme was employed, slight variation in $\gamma^{\prime}/\gamma$ interface position was observed with increasing interface width for the non-planar case. In both these cases, the simulated bulk elastic fields based on the PR scheme showed excellent quantitative agreement with analytical solutions at different time steps and were found to be independent of interface width choice. 

We interestingly found that the PR scheme should be preferred over the VT scheme, particularly for the UO$_2$/void simulations. This is mainly because of the extremely poor convergence of the VT-based simulations compared to the PR-based simulations for the same interface width range as in the $\gamma^{\prime}/\gamma$ alloy. We attribute this to the strong elastic heterogeneity in this system compared to the $\gamma^{\prime}/\gamma$ case. Precisely, the ratio of Young's modulus of the particle to that of Young's modulus of the matrix phase, in this case, was $9.1\mathrm{e}{5}$ times higher than the Ni-Al $\gamma^{\prime}/\gamma$ case. Despite this strong heterogeneity, we found that the PR-based simulations reached the equilibrium state and showed good quantitative agreement with analytical solutions in the planar case. For the non-planar case, however, this quantitative agreement only holds true as long as the interface width is taken smaller than one quarter of the initial particle size.

Finally, we simulated the coarsening of $\gamma^{\prime}$ precipitates in an elastically anisotropic Ni-Al alloy using the PR scheme. As expected in a coherently stressed system, the microstructure showed preferential alignment along the elastically soft directions. It was also observed that the coarsening kinetics was marginally slower in the stressed case compared to the unstressed case.

\section{Acknowledgement}
This work was supported by the European Research Council (ERC) under the European Union’s Horizon 2020 research and innovation program (INTERDIFFUSION, grant agreement no. $714754$). The computational resources and services used in this work were provided by the VSC (Flemish Supercomputer Center), funded by the Research Foundation - Flanders (FWO) and the Flemish Government - department EWI. \textcolor{black}{S.C thanks Dr. Michael Tonks at University of Florida, for allowing to use the supercomputer facility at University of Florida during the review process.}
\clearpage
\renewcommand{\theequation}{A.\arabic{equation}}
\setcounter{equation}{0}
\section*{Appendix A}
\label{Appendix_A}
\section*{Derivation of jump vector $\boldsymbol{a}$ and its derivatives}
Here, we analytically derive expressions for the jump vector $\boldsymbol{a}$ and its derivatives with respect to $\phi$ and $\boldsymbol{\epsilon}$ for anisotropic two-phase solids. These derivates are then used to calculate the Jacobian terms in Appendices B and C.

To determine an expression for $\boldsymbol{a}$, we start by substituting Eq. (\ref{Eqn8}) in Eq. (\ref{Eqn7}) and use the minor symmetry of the stiffness tensor, i.e., $\mathcal{C}_{ikjl} = \mathcal{C}_{iklj}$ to rewrite $\mathcal{C}_{ikjl}\llbracket\epsilon_{jl}\rrbracket = \mathcal{C}_{ikjl}a_{j}n_{l}$ from Eq. (\ref{Eqn6}). This yields
\begin{align}
\begin{split}
\left\{\mathcal{C}_{ikjl}^{\alpha}h(\phi) + \mathcal{C}_{ikjl}^{\beta}[1-h(\phi)]\right\}&a_{j}n_{l}n_{i} \\
&= - \left\{\mathcal{C}_{ikjl}^{\alpha}\left(\epsilon_{jl} - \epsilon_{jl}^{\star\alpha}\right)- \mathcal{C}_{ikjl}^{\beta}\left(\epsilon_{jl} - \epsilon_{jl}^{\star\beta}\right)\right\}n_{i}.
\end{split}
\label{EqnA1}
\end{align}
Now, we solve for the unknown $\boldsymbol{a}$ using Eq (\ref{EqnA1}). Rearranging the known quantities on the left-hand side of Eq. (\ref{EqnA5}) as a second-order tensor (denoted as $\boldsymbol{\mathcal{K}}$) yields
\begin{align}
\mathcal{K}_{kj}(\phi, \boldsymbol{n}) = n_{i}\left\{\mathcal{C}_{ikjl}^{\alpha}h(\phi) + \mathcal{C}_{ikjl}^{\beta}[1-h(\phi)]\right\}n_{l}.
\label{EqnA2}
\end{align}
Substituting Eq. (\ref{EqnA2}) in Eq. (\ref{EqnA1}) and using $\mathcal{C}_{ikjl} = \mathcal{C}_{kijl}$ yields 
\begin{align}
a_{j}(\phi, \boldsymbol{\epsilon}, \boldsymbol{n}) = - \mathcal{K}_{jk}^{-1}\left\{\llbracket \mathcal{C}_{kijl}\rrbracket\epsilon_{jl} - \left(\mathcal{C}_{kijl}^{\alpha}\epsilon_{jl}^{\star\alpha}-\mathcal{C}_{kijl}^{\beta}\epsilon_{jl}^{\star\beta}\right)\right\}n_{i},
\label{EqnA3}
\end{align}
where $\mathcal{K}_{jk}^{-1}$ denotes the inverse of tensor $\mathcal{K}_{kj}$ and $\llbracket \mathcal{C}_{kijl}\rrbracket = \left(\mathcal{C}_{kijl}^{\alpha} - \mathcal{C}_{kijl}^{\beta}\right)$. From Eq. (\ref{EqnA3}) we note that $\boldsymbol{a}$ depends on both the phase-field variable $\phi$ through the interpolation function $h(\phi)$ in Eq.(\ref{EqnA2}) and the total strain $\boldsymbol{\epsilon}$ due to the jump in stiffness  tensors. Consequently, it is useful to obtain these derivatives as well to calculate the Jacobian terms.

Thus, differentiating Eq. (\ref{EqnA3}) with respect to $\boldsymbol{\epsilon}$, using the identity $\partial \epsilon_{jl}/\partial \epsilon_{pq} = \left(\delta_{jp}\delta_{lq} + \delta_{lp}\delta_{jq}\right)/2$, the minor and major symmetries of stiffness tensors, i.e., $\llbracket\mathcal{C}_{ikpq}\rrbracket = \llbracket\mathcal{C}_{ikqp}\rrbracket$ and $\llbracket\mathcal{C}_{ikpq}\rrbracket= \llbracket\mathcal{C}_{pqik}\rrbracket = \llbracket\mathcal{C}_{pqki}\rrbracket$, yields
\begin{align}
\frac{\partial a_{j}}{\partial \epsilon_{pq}} &= -\mathcal{K}_{jk}^{-1}\left(\phi, \boldsymbol{n}\right)\llbracket \mathcal{C}_{pqki}\rrbracket n_{i}
\label{EqnA4}
\end{align}
Next, to calculate $\partial a_{j}/\partial \phi$, we take the implicit derivative of Eq. (\ref{EqnA1}) after using the expression for $\boldsymbol{\mathcal{K}}$. This gives
\begin{align}
\frac{\partial \mathcal{K}_{kj}}{\partial \phi}a_{j} +  \mathcal{K}_{km}\frac{\partial a_{m}}{\partial \phi} = 0 \implies \frac{\partial a_{j}}{\partial \phi} = - \mathcal{K}_{jk}^{-1}\left(\phi, \boldsymbol{n}\right)\left(\frac{\partial \mathcal{K}_{km}}{\partial \phi}a_{m}\right).
\label{EqnA5}
\end{align}
The inverse of $\boldsymbol{\mathcal{K}}$ is given in Eq. (\ref{EqnA2}) and its derivative with respect to $\phi$ is calculated by differentiating Eq. (\ref{EqnA2}) to obtain
\begin{align}
\frac{\partial \mathcal{K}_{km}}{\partial \phi} = n_{i}\left\{\mathcal{C}_{ikml}^{\alpha} - \mathcal{C}_{ikml}^{\beta} \right\}n_{l}\frac{\partial h}{\partial \phi}
\label{EqnA6}
\end{align}
Eqs. (\ref{EqnA3}), (\ref{EqnA4}) and (\ref{EqnA5}) are the main results of this appendix.

\renewcommand{\theequation}{B.\arabic{equation}}
\renewcommand{\thesubsection}{B.\arabic{subsection}}
\setcounter{subsection}{0}
\setcounter{equation}{0}
\section*{Appendix B}
\label{Appendix_B}
\section*{Derivation of stress and its derivatives}
In order to derive the overall stress $\boldsymbol{\sigma}$, we first note that the derivative of Eqs. (\ref{Eqn4}) and (\ref{Eqn5}) with respect to total strain $\boldsymbol{\epsilon}$ may be written as:
\begin{align}
\frac{\partial \epsilon_{ij}^{\alpha}}{\partial \epsilon_{pq}} &= \frac{\left(\delta_{ip}\delta_{jq} + \delta_{jp}\delta_{iq}\right)}{2} + h(\phi)\frac{\partial \llbracket\epsilon_{ij} \rrbracket}{\partial \epsilon_{pq}},
\label{EqnB1}\\
\frac{\partial \epsilon_{ij}^{\beta}}{\partial \epsilon_{pq}} &= \frac{\left(\delta_{ip}\delta_{jq} + \delta_{jp}\delta_{iq}\right)}{2} -[1- h(\phi)]\frac{\partial \llbracket\epsilon_{ij} \rrbracket}{\partial \epsilon_{pq}}.
\label{EqnB2}
\end{align}
Then, differentiating Eq. (\ref{Eqn13}) with respect to total strain yields
\begin{align}
\frac{\partial \omega_{bulk}}{\partial \epsilon_{pq}} = h(\phi)\frac{\partial \omega^{\beta}}{\partial \epsilon_{ij}^{\beta}}\frac{\partial \epsilon_{ij}^{\alpha}}{\partial \epsilon_{pq}} + [1-h(\phi)]\frac{\partial \omega^{\alpha}}{\partial \epsilon_{ij}^{\alpha}}\frac{\partial \epsilon_{ij}^{\alpha}}{\partial \epsilon_{pq}}
\label{EqnB3}
\end{align}
Using the definition $\partial \omega^{\theta}/\partial \epsilon_{ij}^{\theta} = \sigma_{ij}^{\theta}$, symmetry of stress tensor, i.e. $\sigma_{pq}^{\theta} = \sigma_{qp}^{\theta}$ and substituting Eqs. (\ref{EqnB1}) and (\ref{EqnB2}) in Eq. (\ref{EqnB3}) yields
\begin{align}
\frac{\partial \omega_{bulk}}{\partial \epsilon_{pq}} = \sigma_{pq}^{\alpha}[1-h(\phi)] + \sigma_{pq}^{\beta}h(\phi) + h(\phi)[1-h(\phi)]\left\{\sigma_{ij}^{\alpha} - \sigma_{ij}^{\beta}\right\}\frac{\partial \llbracket\epsilon_{ij}\rrbracket}{\epsilon_{pq}}
\label{EqnB4}
\end{align}
Moreover, from Eq. (\ref{Eqn6}) we see that
\begin{align}
\frac{\partial \llbracket \epsilon_{ij}\rrbracket}{\partial \epsilon_{pq}} = \frac{1}{2}\left(\frac{\partial a_{i}}{\partial \epsilon_{pq}}n_{j} +  n_{i}\frac{\partial a_{j}}{\partial \epsilon_{pq}}\right)
\label{EqnB5}
\end{align}
Again by using the symmetry of stress tensor it is easy to show that:
\begin{align}
\left\{\sigma_{ij}^{\alpha} - \sigma_{ij}^{\beta}\right\}\frac{\partial \llbracket\epsilon_{ij} \rrbracket}{\epsilon_{pq}}
 = \left\{\sigma_{ji}^{\alpha} - \sigma_{ji}^{\beta}\right\}n_{j}\frac{\partial a_{i}}{\partial \epsilon_{pq}}
\label{EqnB6}
\end{align}
By substituting Eq. (\ref{EqnB6}) in Eq. (\ref{EqnB4}) yields
\begin{align}
\frac{\partial \omega_{bulk}}{\partial \epsilon_{pq}} = \sigma_{pq}^{\alpha}[1-h(\phi)] + \sigma_{pq}^{\beta}h(\phi) + h(\phi)[1-h(\phi)]\left\{\sigma_{ji}^{\alpha} - \sigma_{ji}^{\beta}\right\}n_{j}\frac{\partial a_{i}}{\partial \epsilon_{pq}}
\label{EqnB7}
\end{align}
Because of Eq. (\ref{Eqn7}) we find that the last term in Eq. (\ref{EqnB7}) must be zero. Thus, we can simplify Eq. (\ref{EqnB7}) to 
\begin{align}
\frac{\partial \omega_{bulk}}{\partial \epsilon_{pq}} = \sigma_{pq} =  \sigma_{pq}^{\alpha}[1-h(\phi)] + \sigma_{pq}^{\beta}h(\phi) 
\label{EqnB8}
\end{align}
Eq.(\ref{EqnB8}) was also given by Kiefer \textit{et al.} \cite{Kiefer2017}, with the difference that $h(\phi)$ replaces $\phi$. Moreover, for numerical implementation, the derivatives of Eq. (\ref{EqnB7}) with respect to total strain $\boldsymbol{\epsilon}$ and the phase-field variable $\phi$ are needed. To this end, we differentiate Eq.(\ref{EqnB7}) and denote $\partial^{2}\omega_{bulk}/\partial \epsilon_{pq}\partial\epsilon_{mn}$ by $\partial\sigma_{pq}/\partial \epsilon_{mn}$
\begin{align}
\begin{split}
\frac{\partial \sigma_{pq}}{\partial \epsilon_{mn}} &= \frac{\partial \sigma_{pq}^{\alpha}}{\partial \epsilon_{rs}^{\alpha}}\frac{\partial \epsilon_{rs}^{\alpha}}{\partial \epsilon_{mn}}[1-h(\phi)] + \frac{\partial \sigma_{pq}^{\beta}}{\partial \epsilon_{rs}^{\beta}}\frac{\partial \epsilon_{rs}^{\beta}}{\partial \epsilon_{mn}} \\
&+h[1-h]\left\{\frac{\partial \sigma_{ji}^{\alpha}}{\partial \epsilon_{rs}^{\alpha}}\frac{\partial \epsilon_{rs}^{\alpha}}{\partial \epsilon_{mn}} - \frac{\partial \sigma_{ji}^{\beta}}{\partial \epsilon_{rs}^{\beta}}\frac{\partial \epsilon_{rs}^{\beta}}{\partial\epsilon_{mn}}\right\}n_{j}\frac{\partial a_{i}}{\partial \epsilon_{pq}}  
\end{split} 
\label{EqnB9}
\end{align}
Because of Eq. (\ref{EqnA4}) we must note that the second derivative of the magnitude of strain jump $\boldsymbol{a}$ is zero, i.e. $\partial^{2}a_{i}/\partial \epsilon_{pq}\partial \epsilon_{mn} = 0$.

Now, by using the definition $\partial \sigma_{pq}^{\theta}/\partial \epsilon_{rs}^{\theta} = \mathcal{C}_{pqrs}^{\theta}$, minor symmetry of stiffness tensors, i.e., $\mathcal{C}_{pqmn} = \mathcal{C}_{pqnm}$ and substituting Eqs. (\ref{EqnB1}) and (\ref{EqnB2}) in Eq. (\ref{EqnB9}) yields

\begin{equation}
\begin{split}
\frac{\partial \sigma_{pq}}{\partial \epsilon_{mn}} &= \mathcal{C}_{pqmn}^{\alpha}[1-h(\phi)] +  \mathcal{C}_{pqmn}^{\beta}h(\phi) + h(\phi)[1-h(\phi)]\left\{\mathcal{C}_{pqrs}^{\alpha} - \mathcal{C}_{pqrs}^{\beta}\right\}\frac{\partial \llbracket\epsilon_{rs} \rrbracket}{\partial \epsilon_{mn}}\\
&+h(\phi)[1-h(\phi)]n_{j}\frac{\partial a_{i}}{\partial \epsilon_{pq}}\left\{\mathcal{C}_{jimn}^{\alpha} - \mathcal{C}_{jimn}^{\beta}\right\} \\
&+ h(\phi)[1-h(\phi)]n_{j}\frac{\partial a_{i}}{\partial \epsilon_{pq}}\left\{h(\phi)\mathcal{C}_{jirs}^{\alpha} + [1-h(\phi)]\mathcal{C}_{jirs}^{\beta}\right\}\frac{\partial \llbracket \epsilon_{rs} \rrbracket}{\partial \epsilon_{mn}}
\end{split}
\label{EqnB10}
\end{equation}
All quantities on the right-hand side of Eq. (\ref{EqnB10}) are known. An expression for $\partial a_{i}/\partial \epsilon_{pq}$ has been obtained in Appendix A. To obtain $\partial \llbracket \epsilon_{rs} \rrbracket/\partial \epsilon_{mn}$, differentiate Eq. (\ref{Eqn6}) with respect to total strain $\boldsymbol{\epsilon}$. This yields 
\begin{align}
\frac{\partial \llbracket \epsilon_{rs} \rrbracket}{\partial \epsilon_{mn}} = \frac{1}{2}\left(\frac{\partial a_{r}}{\partial \epsilon_{mn}}n_{s} + n_{r}\frac{\partial a_{s}}{\partial \epsilon_{mn}}\right)
\label{EqnB10a}
\end{align}
Due to the minor symmetry of stiffness tensors and using Eq.(\ref{EqnB10a}), it can be shown that the second term in Eq.(\ref{EqnB10}) may be written as:
\begin{equation}
n_{j}\frac{\partial a_{i}}{\partial \epsilon_{pq}}\left\{\mathcal{C}_{jimn}^{\alpha} - \mathcal{C}_{jimn}^{\beta}\right\}  =  \frac{\partial \llbracket \epsilon_{ji}\rrbracket}{\partial \epsilon_{pq}}\left\{\mathcal{C}_{jimn}^{\alpha} - \mathcal{C}_{jimn}^{\beta}\right\}
\label{EqnB10b}
\end{equation}

Next, we want to calculate the derivative of Eq.(\ref{EqnB7}) with respect to the phase-field variable $\phi$. To this end, we first differentiate Eqs. (\ref{Eqn4}) and (\ref{Eqn5}) with respect to $\phi$. This gives
\begin{align}
\frac{\partial \epsilon_{ij}^{\alpha}}{\partial \phi} &= h^{\prime}(\phi)\llbracket \epsilon_{ij}\rrbracket + h(\phi)\frac{\partial \llbracket\epsilon_{ij}\rrbracket}{\partial \phi}\label{EqnB11}\\
\frac{\partial \epsilon_{ij}^{\beta}}{\partial \phi} &= h^{\prime}(\phi)\llbracket \epsilon_{ij}\rrbracket - \left[1-h(\phi)\right]\frac{\partial \llbracket\epsilon_{ij}\rrbracket}{\partial \phi}\label{EqnB12}
\end{align}

Differentiating Eq. (\ref{EqnB7}) with respect to $\phi$ and denoting $\partial^{2}\omega_{bulk}/\partial \epsilon_{pq}\partial \phi = \partial \sigma_{pq}/\partial \phi$ yields
\begin{align}
\begin{split}
\frac{\partial \sigma_{pq}}{\partial \phi} &= \left\{\sigma_{pq}^{\beta} - \sigma_{pq}^{\alpha}\right\}h^{\prime}(\phi) + [1-h(\phi)]\frac{\partial \sigma_{pq}^{\alpha}}{\partial \epsilon_{ij}^{\alpha}}\frac{\partial \epsilon_{ij}^{\alpha}}{\partial \phi} + h(\phi)\frac{\partial \sigma_{pq}^{\beta}}{\partial \epsilon_{ij}^{\beta}}\frac{\partial \epsilon_{ij}^{\beta}}{\partial \phi}\\
&+ h(\phi)[1-h(\phi)]\left\{\frac{\partial \sigma_{ji}^{\alpha}}{\partial \epsilon_{qr}^{\alpha}}\frac{\partial \epsilon_{qr}^{\alpha}}{\partial \phi} - \frac{\partial \sigma_{ji}^{\beta}}{\partial \epsilon_{qr}^{\beta}}\frac{\partial \epsilon_{qr}^{\beta}}{\partial \phi}\right\}n_{j}\frac{\partial a_{i}}{\partial \epsilon_{pq}}.
\label{EqnB13}
\end{split}
\end{align}
In the above equation, we note that we have omitted all terms that included the term $\left\{\sigma_{ji}^{\alpha} -\sigma_{ji}^{\beta}\right\}n_{j}$, since this quantity is zero, as mentioned before.
Finally, substituting Eqs (\ref{EqnB11}) and (\ref{EqnB12}) in Eq. (\ref{EqnB13}) we obtain
\begin{align}
\begin{split}
\frac{\partial \sigma_{pq}}{\partial \phi} &= \left\{\sigma_{pq}^{\beta} - \sigma_{pq}^{\alpha}\right\}h^{\prime}(\phi) + \left\{ [1-h(\phi)]\mathcal{C}_{pqij}^{\alpha} + h(\phi)\mathcal{C}_{pqij}^{\beta}\right\}h^{\prime}(\phi)\llbracket\epsilon_{ij}\rrbracket \\
&+ h(\phi)[1-h(\phi)]\left\{\mathcal{C}_{pqij}^{\alpha} - \mathcal{C}_{pqij}^{\beta}\right\}\frac{\partial \llbracket\epsilon_{ij}\rrbracket}{\partial \phi}\\
&+ h(\phi)[1-h(\phi)]\left\{\mathcal{C}_{jimn}^{\alpha} - \mathcal{C}_{jimn}^{\beta}\right\}\llbracket \epsilon_{mn}\rrbracket n_{j}\frac{\partial a_{i}}{\partial \epsilon_{pq}}\\
&+h(\phi)[1-h(\phi)]\left\{h(\phi)\mathcal{C}_{jimn}^{\alpha} + [1-h(\phi)]\mathcal{C}_{jimn}^{\beta} \right\}\frac{\partial \llbracket \epsilon_{mn}\rrbracket}{\partial \phi}n_{j}\frac{\partial a_{i}}{\partial \epsilon_{pq}}
\label{EqnB14}
\end{split} 
\end{align}
Eqs. (\ref{EqnB8}), (\ref{EqnB10}) and (\ref{EqnB14}) are the main results of this appendix.

\renewcommand{\thesubsection}{C.\arabic{subsection}}
\renewcommand{\theequation}{C.\arabic{equation}}
\setcounter{subsection}{0}
\setcounter{equation}{0}
\section*{Appendix C}
\label{Appendix_C}
\section*{Derivation of driving force and its derivatives}
In order to derive the driving force, we first differentiate Eq. (\ref{Eqn11}) with respect to the phase-field variable $\phi$. This gives
\begin{align}
\frac{\partial \omega_{bulk}}{\partial \phi} = h^{\prime}(\phi)\left\{ \omega^{\beta} - \omega^{\alpha}\right\} + h(\phi)\frac{\partial \omega^{\beta}}{\partial \epsilon_{ij}^{\beta}}\frac{\partial \epsilon_{ij}^{\beta}}{\partial \phi} + [1-h(\phi)]\frac{\partial \omega^{\alpha}}{\partial \epsilon_{ij}^{\alpha}}\frac{\partial \epsilon_{ij}^{\alpha}}{\partial \phi} 
\label{EqnC1}
\end{align}
Substituting Eqs. (\ref{EqnB11}) and (\ref{EqnB12}) in Eq. (\ref{EqnC1}) yields
\begin{align}
\begin{split}
\frac{\partial \omega_{bulk}}{\partial \phi} =& h^{\prime}(\phi)\left\{ \omega^{\beta} - \omega^{\alpha}\right\} + h^{\prime}(\phi)\left\{ h(\phi)\sigma_{ij}^{\beta} + [1-h]\sigma_{ij}^{\alpha}\right\}\llbracket\epsilon_{ij}\rrbracket\\
+& h(\phi)[1-h(\phi)]\left\{\sigma_{ij}^{\alpha} - \sigma_{ij}^{\beta}\right\}\frac{\partial \llbracket \epsilon_{ij} \rrbracket}{\partial \phi}
\label{EqnC2}
\end{split}
\end{align}
Moreover, by differentiating Eq. (\ref{Eqn6}) with respect to $\phi$ we obtain 
\begin{align}
\frac{\partial \llbracket\epsilon_{ij}\rrbracket}{\partial \phi} = \frac{1}{2}\left(\frac{\partial a_{i}}{\partial \phi}n_{j} + n_{i}\frac{\partial a_{j}}{\partial \phi}\right)
\label{EqnC3}
\end{align}
By substituting Eq.(\ref{EqnC3}) in Eq. (\ref{EqnC2}) and using the symmetry of stress tensor, i.e., $\sigma_{ij}^{\theta} = \sigma_{ji}^{\theta}$ yields
\begin{align}
\begin{split}
\frac{\partial \omega_{bulk}}{\partial \phi} =& h^{\prime}(\phi)\left\{ \omega^{\beta} - \omega^{\alpha}\right\} + h^{\prime}(\phi)\left\{ h(\phi)\sigma_{ij}^{\beta} + [1-h]\sigma_{ij}^{\alpha}\right\}\llbracket\epsilon_{ij}\rrbracket\\
+& h(\phi)[1-h(\phi)]\left\{\sigma_{ij}^{\alpha} - \sigma_{ij}^{\beta}\right\}n_{i}\frac{\partial a_{j}}{\partial \phi}
\label{EqnC4}
\end{split}
\end{align}
Because of Eq.(\ref{Eqn7}) we see that the last term in Eq.(\ref{EqnC4}) must be zero. Consequently, Eq.(\ref{EqnC4}) may be rewritten as
\begin{align}
\frac{\partial \omega_{b}}{\partial \phi} &=  h^{\prime}(\phi)\left[\left\{\omega_{b}^{\beta}\left(\boldsymbol{\tilde{\mu}},\boldsymbol{e}^{\beta}\right) - \omega_{b}^{\alpha}(\boldsymbol{\tilde{\mu}},\boldsymbol{e}^{\alpha})\right\} 
+ \left\langle \sigma_{ij}\right\rangle\llbracket\epsilon_{ij} \rrbracket\right]
\label{EqnC5}
\end{align}
where we have denoted $\left\langle \sigma_{ij}\right\rangle = h(\phi)\sigma_{ij}^{\beta} + [1-h]\sigma_{ij}^{\alpha}$. It must be pointed out that this equation is consistent with that given by Kiefer \textit{et al.} \cite{Kiefer2017} (see Eq. (33)), provided the chemical contribution is ignored and $\phi$ is replaced with $h(\phi)$. One can also show that, without the chemical contribution, Eq. (\ref{EqnC5}) is of the exact form as the ``driving traction" derived by Abeyaratne and Knowles \cite{Abeyaratne1990} in a sharp-interface setting. With the chemical contribution, Eq. (\ref{EqnC5}) was given by Larche and Cahn \cite{Larche1978} for a planar interface and by Johnson and Alexander \cite{Johnson1986} for a curved interface in a sharp-interface setting.

Furthermore, for numerical implementation we also require the derivative of Eq.(\ref{EqnC4}) with respect to total strain. Thus, differentiating Eq. (\ref{EqnC4}) 
\begin{align}
\begin{split}
\frac{\partial^{2}\omega_{b}}{\partial \epsilon_{ij}\partial \phi} &= h^{\prime}(\phi)\left\{\frac{\partial \omega^{\beta}}{\partial \epsilon_{mn}^{\beta}}\frac{\partial \epsilon_{mn}^{\beta}}{\partial \epsilon_{ij}} - \frac{\partial \omega^{\alpha}}{\partial \epsilon_{mn}^{\alpha}}\frac{\partial \epsilon_{mn}^{\alpha}}{\partial \epsilon_{ij}}\right\} \\
&+ h^{\prime}(\phi)\left\{ h(\phi)\frac{\partial \sigma_{pq}^{\beta}}{\partial \epsilon_{mn}^{\beta}}\frac{\partial \epsilon_{mn}^{\beta}}{\partial \epsilon_{ij}} + [1-h(\phi)]\frac{\partial \sigma_{pq}^{\alpha}}{\partial \epsilon_{mn}^{\alpha}}\frac{\partial \epsilon_{mn}^{\alpha}}{\partial \epsilon_{ij}}\right\}\llbracket \epsilon_{pq}\rrbracket\\
&+h^{\prime}(\phi)\left\{ h(\phi)\sigma_{pq}^{\beta} + [1-h(\phi)]\sigma_{pq}^{\alpha}\right\}\frac{\partial \llbracket\epsilon_{pq}\rrbracket}{\partial \epsilon_{ij}}\\
&+ h(\phi)[1-h(\phi)]\left\{\frac{\partial \sigma_{pq}^{\alpha}}{\partial \epsilon_{mn}^{\alpha}}\frac{\partial \epsilon_{mn}^{\alpha}}{\partial \epsilon_{ij}} - \frac{\partial \sigma_{pq}^{\beta}}{\partial \epsilon_{mn}^{\beta}}\frac{\partial \epsilon_{mn}^{\beta}}{\partial \epsilon_{ij}}\right\}n_{p}\frac{\partial a_{q}}{\partial \phi}
\end{split}
\label{EqnC6}
\end{align}
By substituting Eqs. (\ref{EqnB1}) and (\ref{EqnB2}) in Eq. (\ref{EqnC6}) yields
\begin{align}
\begin{split}
\frac{\partial^{2}\omega_{bulk}}{\partial \epsilon_{ij}\partial \phi} &= h^{\prime}(\phi)\left[\left\{\sigma_{ij}^{\beta} - \sigma_{ij}^{\alpha}\right\} + \left[1-2h(\phi)\right]\left\{\sigma_{mn}^{\alpha} -\sigma_{mn}^{\beta}\right\}\frac{\partial \llbracket \epsilon_{mn} \rrbracket}{\partial \epsilon_{ij}}\right]\\
&+h^{\prime}(\phi)\left\{h(\phi)\mathcal{C}_{ijpq}^{\beta} + [1-h(\phi)]\mathcal{C}_{ijpq}^{\alpha}\right\}\llbracket \epsilon_{pq}\rrbracket\\
&+ h^{\prime}(\phi)\left(\left\{\mathcal{C}_{pqmn}^{\alpha} - \mathcal{C}_{pqmn}^{\beta}\right\}\frac{\partial \llbracket \epsilon_{mn}\rrbracket}{\partial \epsilon_{ij}}\llbracket\epsilon_{pq}\rrbracket\right) h(\phi)[1-h(\phi)]\\
&+ h({\phi})[1-h(\phi)]\left\{\mathcal{C}_{pqij}^{\alpha} - \mathcal{C}_{pqij}^{\beta} \right\}n_{p}\frac{\partial a_{q}}{\partial \phi}\\
&+h(\phi)[1-h(\phi)] \left\{h(\phi)\mathcal{C}_{pqmn}^{\alpha} +  [1-h(\phi)]\mathcal{C}_{pqmn}^{\beta}\right\}\frac{\partial \llbracket \epsilon_{mn}\rrbracket}{\partial \epsilon_{ij}}n_{p}\frac{\partial a_{q}}{\partial \phi}
\end{split}
\label{EqnC7}
\end{align}
Moreover, by first multiplying Eq. (\ref{Eqn7}) with $\frac{\partial a_{q}}{\partial \phi}$, then differentiating it  with respect to total strain $\epsilon_{ij}$, and using Eqs. (\ref{EqnB1})-(\ref{EqnB2}) gives
\begin{align}
\left[\left\{\mathcal{C}_{pqij}^{\alpha} - \mathcal{C}_{pqij}^{\beta}\right\} + \left(h(\phi)\mathcal{C}_{pqmn}^{\alpha}  + [1-h(\phi)]\mathcal{C}_{pqmn}^{\beta}\right)\frac{\partial \llbracket \epsilon_{mn} \rrbracket}{\partial \epsilon_{ij}}\right]n_{p}\frac{\partial a_{q}}{\partial \phi} = 0
\label{EqnC8}
\end{align}
Because of Eq. (\ref{EqnC8}) we see that the sum of last two terms in Eq.(\ref{EqnC7}) must be zero. Consequently, Eq.(\ref{EqnC7}) may be simplified to
\begin{align}
\begin{split}
\frac{\partial^{2}\omega_{bulk}}{\partial \epsilon_{ij}\partial \phi} &= h^{\prime}(\phi)\left[\left\{\sigma_{ij}^{\beta} - \sigma_{ij}^{\alpha}\right\} + \left[1-2h(\phi)\right]\left\{\sigma_{mn}^{\alpha} -\sigma_{mn}^{\beta}\right\}\frac{\partial \llbracket \epsilon_{mn} \rrbracket}{\partial \epsilon_{ij}}\right]\\
&+h^{\prime}(\phi)\left\{h(\phi)\mathcal{C}_{ijpq}^{\beta} + [1-h(\phi)]\mathcal{C}_{ijpq}^{\alpha}\right\}\llbracket \epsilon_{pq}\rrbracket\\
&+ h^{\prime}(\phi)\left(\left\{\mathcal{C}_{pqmn}^{\alpha} - \mathcal{C}_{pqmn}^{\beta}\right\}\frac{\partial \llbracket \epsilon_{mn}\rrbracket}{\partial \epsilon_{ij}}\llbracket\epsilon_{pq}\rrbracket\right) h(\phi)[1-h(\phi)]
\end{split}
\label{EqnC9}
\end{align}
Eqs. (\ref{EqnC5}) and (\ref{EqnC9}) are the main results of this section.
\renewcommand{\thesubsection}{D.\arabic{subsection}}
\renewcommand{\theequation}{D.\arabic{equation}}
\setcounter{equation}{0}
\setcounter{subsection}{0}
\section*{Appendix D}
\section*{Analytical solutions}
In order to analytically solve the equations of mechanical equilibrium in any coordinate system, the position of the interface at any instant $t$, i.e., $x(t)$, must be given. In our validation results, we first perform a phase-field simulation for a given set of initial conditions and then use the numerically determined interface position at a given instant while comparing the analytical solution with the simulated solution. The interface position $x(t)$ is numerically obtained by tracking the phase-field variable $\phi = 0.5$. Therefore, we do not make any a priori assumption about the inclusion size.
\subsection{Solution for a flat interface}
\begin{figure}[ht]
\center
\includegraphics[scale=0.4]{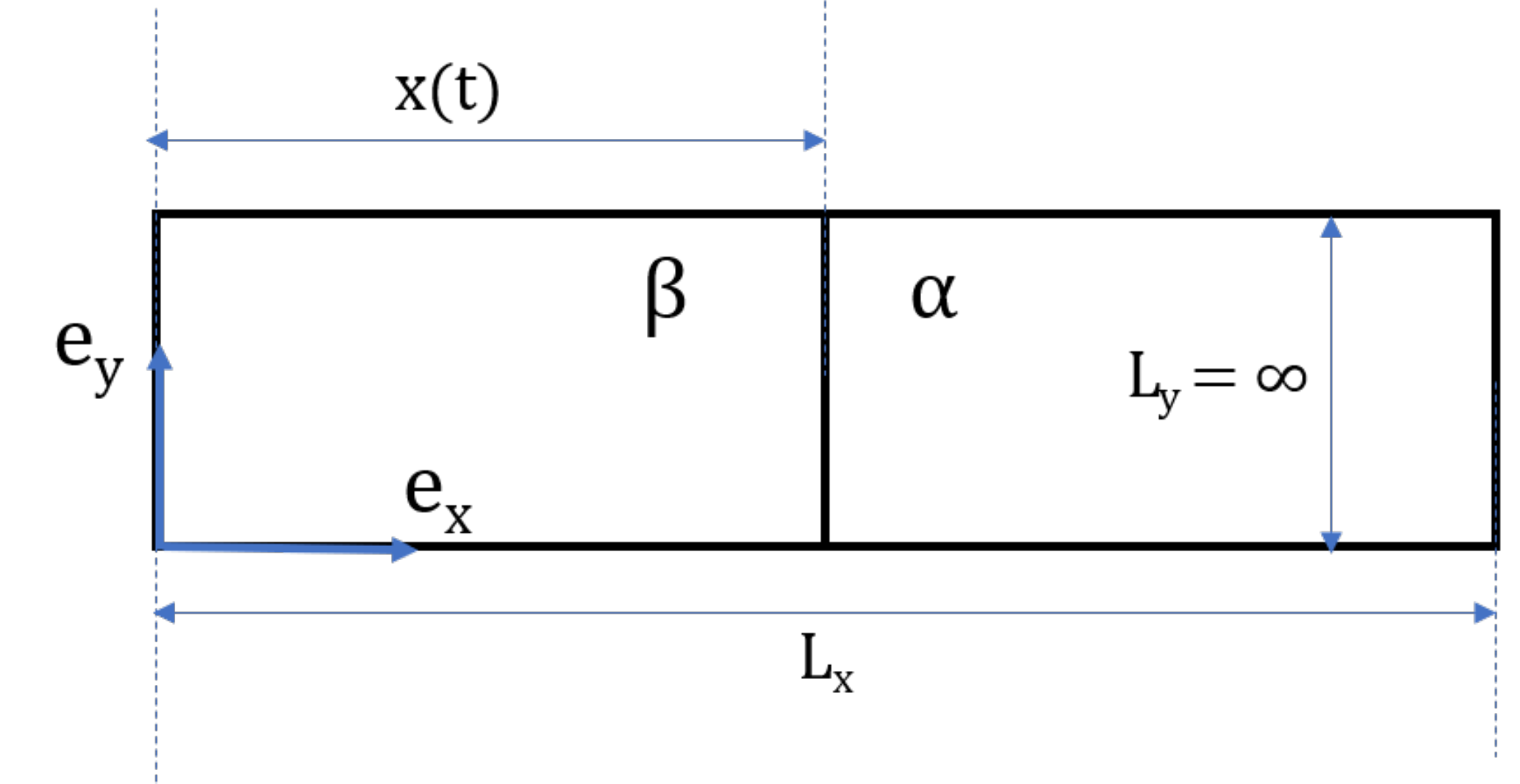}
\caption{Schematic of a two-phase solid with a flat interface}
\end{figure}
Consider a rectangular domain with a flat interface separating the $\beta$ and $\alpha$ phases. We further assume that the y-dimension is much greater than the x-dimension, which is finite and ranges from $\left[-L_{x}/2, L_{x}/2\right]$. Therefore, we enforce periodic boundary conditions along the y-direction and Dirichlet boundary conditions with respect to the displacement fields. For sake of simplicity, let only the x-component of displacement fields, i.e., $u^{l}_{x}$ and $u^{r}_{x}$, be non-zero at the left and right boundaries. Then, the x-component of the displacement field will be non-zero. That is
\begin{align}
\boldsymbol{u}(x,t) = \begin{cases}
u^{\beta}(x,t)\boldsymbol{e}_{x} \quad x < x(t)\\
u^{\alpha}(x,t)\boldsymbol{e}_{x} \quad x > x(t)
\end{cases}
\label{D01}
\end{align}
It immediately follows that the total \textit{phase} strain in phase $p$ is given by $\boldsymbol{\epsilon}^{p}(x,t) = du^{p}/dx\,\boldsymbol{e}_{x}\otimes \boldsymbol{e}_{x}$. Since the phase $\beta$ has a non-zero eigenstrain $\epsilon^{\star}$, the non-zero stress components are
\begin{align}
\sigma_{x}(x,t) &= 
\begin{cases}
\left(\lambda^{\beta} + 2\mu^{\beta}\right)\epsilon_{x} - 2(\lambda^{\beta} + \mu^{\beta})\epsilon^{\star},\quad x < x(t)\\
\left(\lambda^{\alpha} + 2\mu^{\alpha}\right)\epsilon_{x},\quad x > x(t)
\end{cases}
\label{D02}\\
\sigma_{y}(x,t) &= 
\begin{cases}
\lambda^{\beta} \epsilon_{x} - 2(\lambda^{\beta} + \mu^{\beta})\epsilon^{\star},\quad x < x(t)\\
\lambda^{\alpha}\epsilon_{x},\quad x > x(t)
\end{cases}
\label{D03}
\end{align}
Moreover, the equation of mechanical equilibrium simplifies to
\begin{align}
\frac{\partial \sigma_{x}^{p}}{\partial x} = 0
\label{D04}
\end{align}
where $p = \alpha,\beta$ indicates the phase. Substituting Eqs. (\ref{D02}) in Eq. (\ref{D04}) yields
\begin{align}
\left(\lambda^{p} + 2\mu^{p}\right)\frac{d^{2}u}{dx^{2}} = 0 \Leftrightarrow \frac{d^{2}u^{p}}{dx^{2}} = 0
\label{D05}
\end{align}
For this reason, the x-component of the displacement field must be linear within each phase
\begin{align}
u(x,t) = 
\begin{cases}
A^{\beta} x + B^{\beta}, \quad   x< x(t)\\
A^{\alpha} x + B^{\alpha}, \quad x> x(t)
\end{cases}
\label{D06}
\end{align}
where $A^{\beta}$, $B^{\beta}$, $A^{\alpha}$, $B^{\alpha}$ are the undetermined constants. From Eq.(\ref{D06}) and using the boundary conditions yields
\begin{align}
-A^{\beta}(L_{x}/2) + B^{\beta} &= u^{l}_{x} \implies B^{\beta} = u_{x}^{l} + A^{\beta}(L_{x}/2)\label{D07}\\
A^{\alpha}(L_{x}/2) + B^{\alpha}  &= u^{r}_{x}\implies B^{\alpha} = u_{x}^{r} - A^{\alpha}(L_{x}/2)\label{D08} 
\end{align}
Now, since the displacement field must be continuous at $x(t)$, i.e., $u_{x}^{\alpha} = u_{x}^{\beta}$. From Eqs. (\ref{D06}), (\ref{D07}) and (\ref{D08}) we have
\begin{align}
A^{\beta}\left[x(t) + L_{x}/2\right] - A^{\alpha}\left[x(t) - L_{x}/2\right]  = u_{x}^{r} - u_{x}^{l}
\label{D09}
\end{align}
Rearranging Eq. (\ref{D09}) yields
\begin{align}
 A^{\alpha} = \frac{A^{\beta}\left[x(t)+ L_{x}/2\right] - (u_{x}^{r} - u_{x}^{l})}{x(t) - L_{x}/2}
 \label{D10}
 \end{align}
Furthermore, the stress component normal to the interface plane must be continuous, i.e., $\sigma_{x}^{\alpha} = \sigma_{x}^{\beta}$. This yields
\begin{align}
\left(\lambda^{\beta} + 2\mu^{\beta}\right)A^{\beta} - \left(\lambda^{\alpha} + 2\mu^{\alpha}\right)A^{\alpha} = 2(\lambda^{\beta} + \mu^{\beta})\epsilon^{\star}
\label{D11}
\end{align} 
Substituting Eq. (\ref{D10}) in Eq. (\ref{D11}) and solving for $A^{\beta}$ yields
\begin{align}
A^{\beta} = \frac{2(\lambda^{\beta} + \mu^{\beta})\epsilon^{\star} - \left[\frac{(\lambda^{\alpha} + 2\mu^{\alpha})}{x(t) - L_{x}/2}\right]\left(u_{x}^{r} - u_{x}^{l}\right)}{\left( \lambda^{\beta} + 2\mu^{\beta}\right) - \left[\frac{x(t) + L_{x}/2}{x(t)- L_{x}/2}\right]\left( \lambda^{\alpha} + 2\mu^{\alpha}\right)}
\label{D12}
\end{align}
For the UO$_2$/void case, since we have enforced a displacement along the y-direction, the y-component of the displacement field and the shear strain are non-zero in this case. We also assume that the y-component of the displacement field is linear in both phases. Then,
\begin{align}
u_{y} = \begin{cases}
A_{y}^{\beta}x + B^{\beta}_{y},\quad x < x(t)\\
A_{y}^{\alpha}x + B^{\alpha}_{y},\quad x > x(t)
\end{cases}
\label{D13}
\end{align}
where $A_{y}^{\beta}$, $B_{y}^{\beta}$, $A_{y}^{\alpha}$, $B_{y}^{\alpha}$ are the undetermined constants. From Eq. (\ref{D13}) and using the boundary conditions yields
\begin{align}
-A^{\beta}_{y}(L_{x}/2) + B_{y}^{\beta} &= u^{l}_{y} \implies B_{y}^{\beta} = u_{y}^{l} + A^{\beta}_{y}(L_{x}/2)\label{D14}\\
A^{\alpha}_{y}(L_{x}/2) + B_{y}^{\alpha}  &= u^{r}_{y}\implies B_{y}^{\alpha} = u_{y}^{r} - A^{\alpha}_{y}(L_{x}/2)\label{D15} ,
\end{align}
where $u_{y}^{l}$ and $u_{y}^{r}$ are the y-components of the displacement field at the left and right boundaries. Similar to the above-mentioned derivation, by enforcing the equality of the y-component of the displacement field and solving for $A_{y}^{\alpha}$ yields
\begin{align}
 A_{y}^{\alpha} = \frac{A^{\beta}_{y}\left[x(t)+ L_{x}/2\right] - (u_{y}^{r} - u_{y}^{l})}{x(t) - L_{x}/2}
 \label{D16}
 \end{align}
Now, since the shear components of the stress tensor must also be equal, i.e. $\sigma_{xy}^{\alpha} = \sigma_{xy}^{\beta}$. Using Eq. (\ref{D16}) and solving for this equality yields
\begin{align}
A_{y}^{\beta} = \frac{\mu^{\alpha}\left(u_{y}^{r} - u_{y}^{l}\right)}{\mu^{\alpha}\left[x(t) + L_{x}/2\right] - \mu^{\beta}\left[x(t)- L_{x}/2\right]}
\end{align}

\subsection{Solution for a heterogeneous inclusion subject to zero radial stress at the outer boundary}
\begin{figure}[ht]
\center
\includegraphics[scale=0.4]{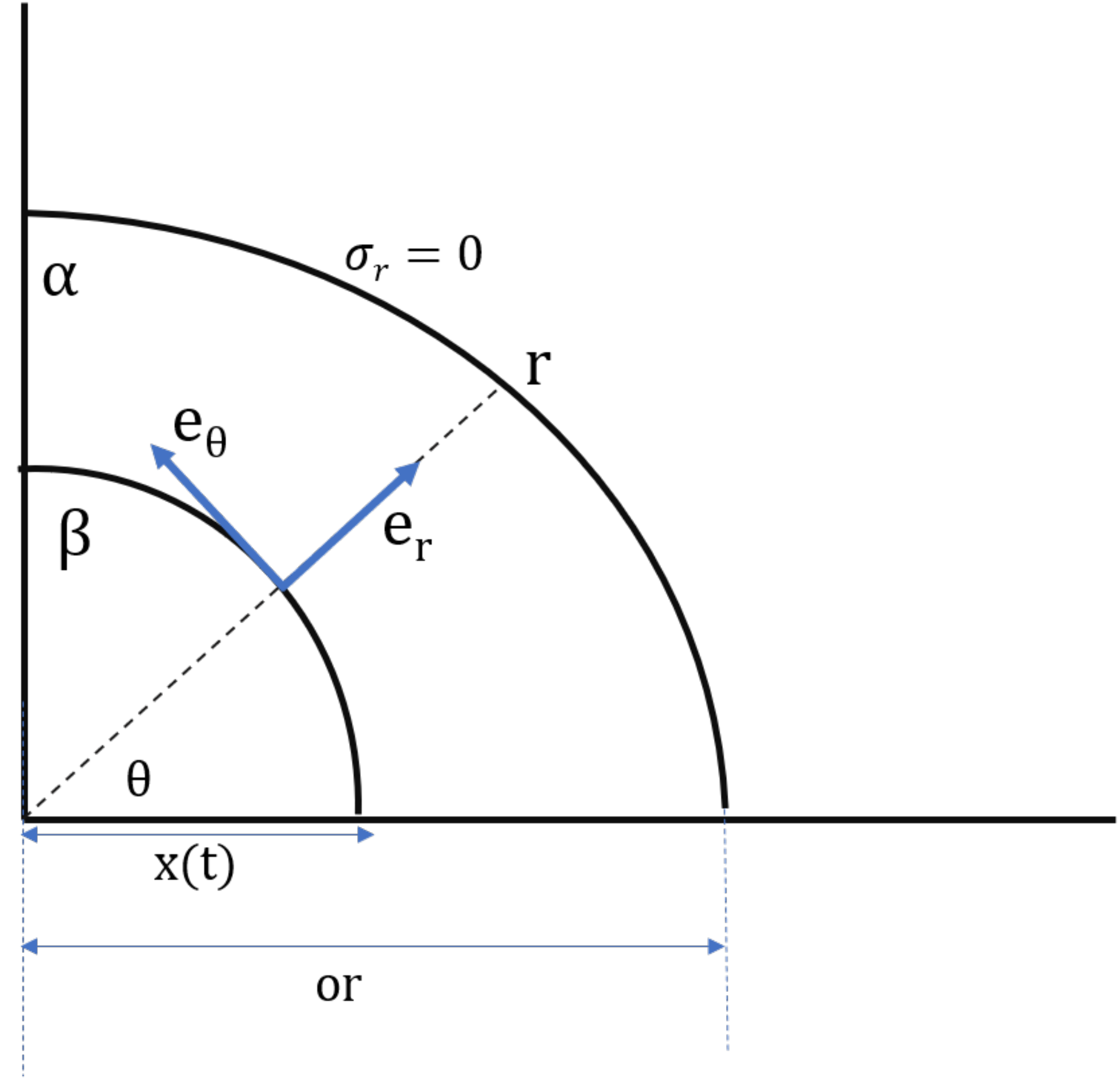}
\caption{Schematic of a two-phase solid with a curved interface}
\end{figure}

We assume that the displacement field is dependent on the radial direction only within each phase and is of the form $\boldsymbol{u}^{\alpha} = u_{r}^{\alpha}(r)\boldsymbol{e}_{r} + u_{\theta}^{\alpha}(r)\boldsymbol{e}_{\theta}$. Then the strain-displacement relations in polar coordinates for a given phase, say $\alpha$, can be written as
\begin{align}
\begin{split}
\epsilon_{r}^{\alpha}(r) &= du_{r}^{\alpha}(r)/dr, \\
\epsilon_{\theta}^{\alpha}(r) &= u_{r}^{\alpha}(r)/r,\\
\epsilon_{r\theta}^{\alpha}(r) &= 0 
\label{EqnD1}
\end{split}
\end{align}
For the simplicity's sake, we first assume an elastically homogeneous and isotropic two-phase system. Then, the stress-strain relations  take the following simple form
\begin{align}
\begin{split}
\sigma_{r}^{\alpha}(r) &= \lambda (\epsilon^{\alpha}_{r} + \epsilon^{\alpha}_{\theta}) + 2\mu\epsilon^{\alpha}_{r}(r)\\
\sigma_{\theta}^{\alpha}(r) &= \lambda (\epsilon^{\alpha}_{r} + \epsilon^{\alpha}_{\theta}) + 2\mu\epsilon^{\alpha}_{\theta}(r)\\
\sigma_{r\theta}^{\alpha}(r) &= 2\mu \epsilon_{r\theta}^{\alpha} = 0
\label{EqnD2}
\end{split}
\end{align}
Note that for the inhomogeneous case, the Lame's parameters, $\lambda$ and $\mu$, will be denoted with a superscript indicating the stress state within that phase. Since the elastic stresses depend on the radial position only, the equation of mechanical equilibrium then reduces to
\begin{align}
\frac{d\sigma_{r}^{\alpha}}{d r} + \left(\frac{\sigma_{r}^{\alpha} - \sigma_{\theta}^{\alpha}}{r}\right) = 0
\label{EqnD3}
\end{align}
Substituting Eqs. (\ref{EqnD1}) \& (\ref{EqnD2}) in Eq. (\ref{EqnD3}) yields
\begin{align}
\left(\lambda + 2\mu\right)\left[\frac{d^{2}u_{r}^{\alpha}}{dr^{2}} +  \frac{1}{r}\frac{du_{r}^{\alpha}}{dr} - \frac{u_{r}^{\alpha}}{r^{2}}\right] = 0 \Leftrightarrow \frac{d}{dr}\left[\frac{1}{r}\frac{d}{dr}\left( u_{r}^{\alpha} r\right)\right] = 0
\label{EqnD4}
\end{align}
From Eq.(\ref{EqnD4}) we see that $u_{r}^{\alpha}$ takes the following general form
\begin{align}
u_{r}(r) =
\begin{cases}
A^{\beta} r + B^{\beta}/r \quad r\leq x(t)\\
A^{\alpha} r + B^{\alpha}/r \quad r> x(t)
\end{cases}
\label{EqnD5}
\end{align}
where $A^{\alpha}$, $A^{\beta}$, $B^{\alpha}$ and $B^{\beta}$ are the undetermined constants. Since the radial displacement must be bounded, therefore $B^{\beta}$ must be zero when $r=0$. To model finite domains, we assume that the radial stress at the outer radius $r = or$ is zero.
\begin{align}
\sigma_{r}(r = or) = 0
\label{EqnD6}
\end{align}
Using Eqs. (\ref{EqnD5}) and (\ref{EqnD2}) in Eq. (\ref{EqnD6}) we see that
\begin{align}
A^{\alpha} = \frac{B^{\alpha}\mu}{or^{2}(\lambda + \mu)}
\label{EqnD7}
\end{align}
For an inhomogeneous inclusion, it can be shown that $A^{\alpha} = \frac{B^{\alpha}\mu^{\alpha}}{or^{2} \left(\lambda^{\alpha} + \mu^{\alpha}\right)}$. Moreover, since the radial displacement field must be continuous at $r= x(t)$. Using Eqs. (\ref{EqnD5})  and (\ref{EqnD7}) yields
\begin{align}
A^{\beta} = B^{\alpha}\left[\frac{\mu}{or^2(\lambda + \mu)} + \frac{1}{x^{2}} \right]
\label{EqnD8}
\end{align}
For an inhomogeneous inclusion, it can be shown that $A^{\beta} = B^{\alpha}\left[\frac{\mu^{\alpha}}{or^2(\lambda^{\alpha} + \mu^{\alpha})} + \frac{1}{x^{2}} \right]$. Since the radial stress must be continuous, using Eq. (\ref{EqnD2}), we see that
\begin{align}
(2\mu + \lambda)\left(\epsilon_{r}^{\alpha} - \epsilon_{r}^{\beta}\right) + \lambda \left(\epsilon^{\alpha}_{\theta} - \epsilon^{\beta}_{\theta}\right) = -2(\lambda + \mu)\epsilon^{\star}
\label{EqnD9}
\end{align}
Using Eqs. (\ref{EqnD1}), (\ref{EqnD5}), (\ref{EqnD7}) and (\ref{EqnD8}) in Eq. (\ref{EqnD9})  and solving for $B^{\alpha}$ yields
\begin{align}
B^{\alpha} = \frac{(\lambda + \mu)\epsilon^{\star} x^{2}}{2\mu + \lambda}
\end{align}
Using expressions for $A^{\beta}$ and $A^{\alpha}$ for the inhomogeneous case, it can be shown that for an inhomogeneous inclusion, $B^{\alpha}$ is equal to
\begin{align}
B^{\alpha} = \frac{-(\lambda^{\beta} + \mu^{\beta})\epsilon^{\star}}{\frac{\mu^{\alpha}}{or^{2}}\left[1-\left(\frac{\lambda^{\beta} + \mu^{\beta}}{\lambda^{\alpha} + \mu^{\alpha}}\right) \right] - \frac{1}{x^{2}}(\mu^{\alpha} + \lambda^{\beta} + \mu^{\beta}) } 
\label{EqnD10}
\end{align}
In summary, the elastic quantities for the homogeneous case within each phase can be written as:
\begin{align}
u_{r}(r) =
\begin{cases}
A^{\beta}, r  \quad r\leq x(t)\\
\frac{B^{\alpha}\mu}{or^{2}(\lambda + \mu)} r + B^{\alpha}/r, \quad r> x(t)
\end{cases}
\label{EqnD11}
\end{align}

\begin{align}
\epsilon_{r}(r) &=
\begin{cases}
A^{\beta},  \quad r\leq x(t)\\
\frac{B^{\alpha}\mu}{or^{2}(\lambda + \mu)} - B^{\alpha}/r^{2}, \quad r> x(t)
\end{cases}
\\ 
\epsilon_{\theta}(r) &=
\begin{cases}
A^{\beta},  \quad r\leq x(t)\\
\frac{B^{\alpha}\mu}{or^{2}(\lambda + \mu)} + B^{\alpha}/r^{2}, \quad r> x(t)
\end{cases}
\label{EqnD12}
\end{align}

\begin{align}
\sigma_{r}(r) &=
\begin{cases}
\lambda(\epsilon_{r} + \epsilon_{\theta}) + 2\mu\epsilon_{r} - 2(\lambda + \mu)\epsilon^{\star} \quad r\leq x(t)\\
\lambda(\epsilon_{r} + \epsilon_{\theta}) + 2\mu\epsilon_{r} \quad r> x(t)
\end{cases}
\\
\sigma_{\theta}(r) &=
\begin{cases}
\lambda(\epsilon_{r} + \epsilon_{\theta}) + 2\mu\epsilon_{\theta} - 2(\lambda + \mu)\epsilon^{\star}  \quad r\leq x(t)\\
\lambda(\epsilon_{r} + \epsilon_{\theta}) + 2\mu\epsilon_{\theta} \quad r> x(t)
\end{cases}
\label{EqnD13}
\end{align}
For the inhomogeneous case, we have
\begin{align}
u_{r}(r) =
\begin{cases}
A^{\beta}, r  \quad r\leq x(t)\\
\frac{B^{\alpha}\mu^{\alpha}}{or^{2}(\lambda^{\alpha} + \mu^{\alpha})} r + B^{\alpha}/r, \quad r> x(t)
\end{cases}
\label{EqnD14}
\end{align}

\begin{align}
\epsilon_{r}(r) &=
\begin{cases}
A^{\beta},  \quad r\leq x(t)\\
\frac{B^{\alpha}\mu^{\alpha}}{or^{2}(\lambda^{\alpha} + \mu^{\alpha})} - B^{\alpha}/r^{2}, \quad r> x(t)
\end{cases}
\\ 
\epsilon_{\theta}(r) &=
\begin{cases}
A^{\beta},  \quad r\leq x(t)\\
\frac{B^{\alpha}\mu^{\alpha}}{or^{2}(\lambda^{\alpha} + \mu^{\alpha})} + B^{\alpha}/r^{2}, \quad r> x(t)
\end{cases}
\label{EqnD15}
\end{align}

\begin{align}
\sigma_{r}(r) &=
\begin{cases}
\lambda^{\beta}(\epsilon_{r} + \epsilon_{\theta}) + 2\mu^{\beta}\epsilon_{r} - 2(\lambda^{\beta} + \mu^{\beta})\epsilon^{\star} \quad r\leq x(t)\\
\lambda^{\alpha}(\epsilon_{r} + \epsilon_{\theta}) + 2\mu^{\alpha}\epsilon_{r} \quad r> x(t)
\end{cases}
\\
\sigma_{\theta}(r) &=
\begin{cases}
\lambda^{\beta}(\epsilon_{r} + \epsilon_{\theta}) + 2\mu^{\beta}\epsilon_{\theta} - 2(\lambda^{\beta} + \mu^{\beta})\epsilon^{\star}  \quad r\leq x(t)\\
\lambda^{\alpha}(\epsilon_{r} + \epsilon_{\theta}) + 2\mu^{\alpha}\epsilon_{\theta} \quad r> x(t)
\end{cases}
\label{EqnD16}
\end{align}

\subsection{Solution for an elastically heterogeneous system subject to radial strain at the outer boundary}
The radial displacement field within both phases, say $\alpha$ and $\beta$, must satisfy Eq.(\ref{EqnD5}). Following similar arguments, as discussed in the previous case, the value of  $B^{\beta}$ must be zero; for the radial displacement field to be finite. Consequently, for this problem, the radial displacement field takes the following form:
\begin{align}
u_{r}(r) =
\begin{cases}
A^{\beta} r  \quad r\leq x(t)\\
A^{\alpha} r + B^{\alpha}/r \quad r> x(t)
\end{cases}
\label{EqnD17}
\end{align}
Since the radial strain at the outer boundary is specified, say $\epsilon_{r}^{g}$, then using Eq.(\ref{EqnD1}), it follows that 
\begin{align}
A^{\alpha} = \epsilon_{r}^{g} + B^{\alpha}/or^{2}
\label{EqnD18}
\end{align}
Substituting Eq. (\ref{EqnD18}) in Eq. (\ref{EqnD17}) and enforcing the continuity of radial displacement at the interface yields
\begin{align}
A^{\beta} = \epsilon_{r}^{g} + B^{\alpha}\left(\frac{1}{or^{2}}  + \frac{1}{x^{2}}\right)
\label{EqnD19}
\end{align}
Now, by using the inhomogeneous form of  Eq. (\ref{EqnD2}) and solving for the equality of radial stresses at the interface yields
\begin{align}
\left(\lambda^{\alpha} + \mu^{\alpha}\right)A^{\alpha} - \frac{\mu^{\alpha}B^{\alpha}}{x^{2}} = \left(\lambda^{\beta} + \mu^{\beta}\right)A^{\beta}
\label{EqnD20}
\end{align}
Substituting Eqs. (\ref{EqnD18}) and (\ref{EqnD19}) in Eq.(\ref{EqnD20}) and solving for $B^{\alpha}$ yields
\begin{align}
B^{\alpha} = \frac{\left[\left(\lambda^{\beta} + \mu^{\beta}\right) - \left(\lambda^{\alpha} + \mu^{\alpha}\right)\right]\epsilon_{r}^{g}}{\frac{1}{or^{2}}\left[(\lambda^{\alpha} + \mu^{\alpha}) - (\lambda^{\beta} + \mu^{\beta})\right] - \frac{1}{x^{2}}\left( \mu^{\alpha} + \mu^{\beta} + \lambda^{\beta}\right)}
\end{align}
Thus by substituting $B^{\alpha}$, $B^{\beta}$, and $A^{\alpha}$ in Eq. (\ref{EqnD17}) yields the radial displacement for this case.

\subsection{Solute mole fractions in an elastically stressed solid}
To calculate the equilibrium mole fractions,  $X_{B}^{\alpha}$ and $X_{B}^{\beta}$, in a stressed two-phase binary alloy, a set of two equations is required. The first equation represents the local thermodynamic driving force of the transformation, and the second equation is due to the assumption of \textit{local} thermodynamic equilibrium at the interface. Specifically,
\begin{align}
\omega^{\beta} (\tilde{\mu}^{\beta}, \epsilon^{\beta}) - \omega^{\alpha}(\tilde{\mu}^{\alpha}, \epsilon^{\alpha})  + \sigma_{ij}^{\beta}\left(\epsilon_{ij}^{\alpha} - \epsilon_{ij}^{\beta}\right) &= 0 \label{EqnD36}\\
\tilde{\mu}^{\beta} - \tilde{\mu}^{\alpha} &= 0 \label{EqnD37}
\end{align}
By solving coupled Eqs. (\ref{EqnD36}) and (\ref{EqnD37}), $X^{\alpha}_{B}$ and $X^{\beta}_{B}$ can be determined. It must be noted that this solution requires that the stress and strain in the bulk phases are known. From Eq. (\ref{EqnD37}), we see that $\tilde{\mu}^{\alpha} = \tilde{\mu}^{\beta} = \tilde{\mu}$. Consequently, by Taylor expansion about the unstressed equilibrium diffusion potential, i.e. $\tilde{\mu} = \tilde{\mu}^{eq}$, the grand-potentials of a phase $\theta$ can be written as:
\begin{align}
\omega^{\theta} = \omega^{\theta}(\tilde{\mu}^{eq}) + \frac{\partial \omega^{\theta}} {\partial \mu}(\tilde{\mu} - \tilde{\mu}^{eq}) + \underbrace{\frac{1}{2}\sigma_{ij}^{\theta}\left(\epsilon_{ij}^{\theta} -\epsilon_{ij}^{\star\theta}\right)}_{f_{el}^{\theta}}
 \label{EqnD38}
\end{align}
Substituting Eq. (\ref{EqnD38}) in Eq. (\ref{EqnD36}), and using the relations $\omega^{\alpha}(\tilde{\mu}^{eq}) = \omega^{\beta}(\tilde{\mu}^{eq})$ and $\partial \omega^{\theta}/\partial \tilde{\mu} = - X_{B}^{\theta}/V_{m}^{\theta}$, Eq. (\ref{EqnD36}) can be rewritten as:
\begin{align}
(\tilde{\mu} - \tilde{\mu}^{eq}) = \frac{-V_{m}^{\theta}\left[ \left(f_{el}^{\beta} - f_{el}^{\alpha}\right) + \sigma_{ij}^{\beta}\left(\epsilon_{ij}^{\alpha} - \epsilon_{ij}^{\beta}\right) \right]}{X_{B}^{\alpha} - X_{B}^{\beta}}
\end{align}
This equation represents the deviation of the unstressed equilibrium diffusion potential from the stressed case. Following this, the equilibrium mole fraction in the stressed solid can be easily obtained by Taylor approximating the mole fractions about the unstressed equilibrium mole fractions:
\begin{align}
X_{B}^{\alpha} = X_{B}^{\alpha, eq} + \left.\frac{\partial X_{B}^{\alpha,eq}}{\partial \tilde{\mu}_{B}}\right|_{\tilde{\mu}_{B}^{eq}}\left(\tilde{\mu}_{B} - \tilde{\mu}_{B}^{eq}\right)\\
X_{B}^{\beta} = X_{B}^{\beta, eq} + \left.\frac{\partial X_{B}^{\beta,eq}}{\partial \tilde{\mu}_{B}}\right|_{\tilde{\mu}_{B}^{eq}}\left(\tilde{\mu}_{B} - \tilde{\mu}_{B}^{eq}\right)
 \end{align}
 where $\frac{\partial X_{B}^{\theta, eq}}{\partial \tilde{\mu}_{B}} $ represents the inverse of a thermodynamic factor of component B in phase $\theta$. 
\renewcommand{\thesubsection}{E.\arabic{subsection}}
\renewcommand{\theequation}{E.\arabic{equation}}
\setcounter{equation}{0}
\setcounter{subsection}{0}
\section*{Appendix E}
\section*{Method to calculate diffusion potential}
\color{black}
Diffusion potential is calculated numerically by solving Eq. (\ref{Eqn24}) in the MOOSE (Multiphysics Object-Oriented Simulation Environment) framework \cite{Gaston2009}. Specifically, this equation is implemented as a MOOSE \texttt{Kernel} \cite{Gaston2009}. In this section, we discuss how this \texttt{Kernel} is implemented, and also provide a link to the source code.

To write a \texttt{Kernel}, two terms are needed: a residual vector $\mathcal{R}_{i}$ and a Jacobian matrix $\mathcal{J}_{ij}$ \cite{Gaston2009}. Further, the residual vector is obtained by first deriving the weak form \cite{Gaston2009}. Specifically, for an independent diffusing component $k$, the weak form of Eq. (\ref{Eqn24}) is
\begin{align}
\int_{V}\delta\mu_{k}\left\{h(\phi)X_{k}^{\beta}(\boldsymbol{\tilde{\mu}}) + [1-h(\phi)]X_{k}^{\alpha}(\boldsymbol{\tilde{\mu}}) - X_{k} \right\}dv = 0,
\label{EqnRV3}
\end{align}
where $V$ is the volume of the domain; $\delta \mu_{k}$ is a suitable test function; $h(\phi)$ is the interpolation function; $X_{k}^{\theta=\alpha,\beta}$ are the \textit{phase} mole fractions of $\alpha$ and $\beta$ phases; and $X_{k}$ is the overall mole fraction. It should be emphasized that $X_{k}^{\beta}(\boldsymbol{\mu})$ and $X_{k}^{\alpha}(\boldsymbol{\mu})$ are prerequisite input properties that are needed as functions of diffusion potentials. Moreover, these properties can be calculated either analytically \cite{Plapp2011} or numerically \cite{Chatterjee2021} depending on the solution model. Hence, these properties are implemented as \texttt{Material} objects in MOOSE \cite{Gaston2009}. On the other hand, $X_{k}(\boldsymbol{x},t)$ is a field that has to be determined by solving  the composition evolution Eq. (\ref{Eqn26}). 

Subsequently, we discretize Eq. (\ref{EqnRV3}) using the Galerkin finite element method \cite{Gaston2009}. It can be shown that the $i^{th}$ component of the residual vector associated with this \texttt{Kernel} is
\begin{align}
\mathcal{R}_{i}\left(\boldsymbol{\mu}^{h}, \phi^{h}, X_{k}^{h}\right) = \int_{V} N_{i}\left\{h(\phi^{h})X_{k}^{\beta}(\boldsymbol{\tilde{\mu}}^{h}) + [1-h(\phi)]X_{k}^{\alpha}(\boldsymbol{\tilde{\mu}}^{h}) - X_{k}^{h}\right\}dv,
\label{EqnRV4}
\end{align}
where $N_{i}$ is the shape function at the $i^{th}$ node and $\boldsymbol{\tilde{\mu}}^{h}$, $X_{k}^{h}$ \& $\phi^{h}$ are the discretized field variables such that

\begin{align}
\tilde{\boldsymbol{\mu}} \approx \tilde{\boldsymbol{\mu}}^{h} = \sum_{j=1}^{\mathcal{N}}\tilde{\boldsymbol{\mu}}_{j}N_{j},\quad X_{k} \approx X_{k}^{h} = \sum_{j=1}^{\mathcal{N}}X_{kj}N_{j},\quad \phi \approx \phi^{h} = \sum_{j=1}^{\mathcal{N}}\phi_{j}N_{j}.
\label{EqnRV5}
\end{align}
Here, $\mathcal{N}$ is the total number of nodes in an element, $\tilde{\boldsymbol{\mu}}_{j}$ and $\phi_{j}$ are the nodal values of diffusion potential and phase-field variables at node $j$, and $X_{kj}$ is the nodal value of mole fraction variable of $k^{th}$ component at node $j$. Now,  in order to iteratively solve Eq.(\ref{EqnRV4}), the Jacobian matrix has to be derived.

Notice that the residual vector is a function of $\boldsymbol{\tilde{\mu}}^{h}$, $\phi^{h}$ \& $X_{k}^{h}$, and  consequently three derivatives are needed to construct the Jacobian matrix. First, the components of the Jacobian matrix associated with the variable $\boldsymbol{\tilde{\mu}}^{h}$ are
\begin{align}
\mathcal{J}_{ij}\left(\boldsymbol{\mu}^{h}, \phi^{h}\right) = \frac{\partial R_{i}}{\partial \tilde{\mu}_{rj}} = \int_{V}N_{i}\left\{h(\phi^{h})\chi_{kr}^{\beta}\left(\boldsymbol{\tilde{\mu}^{h}}\right) + [1-h(\phi^{h})]\chi_{kr}^{\alpha}\left(\boldsymbol{\tilde{\mu}^{h}}\right) \right\}N_{j}dv,
\label{EqnRV6}
\end{align}
where $\tilde{\mu}_{rj}$ is the nodal value of the diffusion potential variable of the $r^{th}$ chemical component at node $j$ and $\chi_{kr}^{\theta=\beta,\alpha} = {\partial X_{k}^{\alpha,\beta}}/{\partial \tilde{\mu}_{r}}$ are the components of the susceptibility matrix \cite{Plapp2011}. This matrix can be calculated by inverting the thermodynamic factor 
matrix \cite{Chatterjee2021}, and is implemented as a MOOSE \texttt{Material} object. Second, the components of the Jacobian matrix associated with the variable $\phi^{h}$ are
\begin{align}
\mathcal{J}_{ij}\left(\boldsymbol{\mu}^{h}, \phi^{h}\right) = \frac{\partial \mathcal{R}_{i}}{\partial \phi_{j}} = \int_{V} N_{i}\left\{X_{k}^{\beta}(\boldsymbol{\tilde{\mu}}^{h}) - X_{k}^{\alpha}\left(\boldsymbol{\tilde{\mu}}^{h}\right)\right\}h^{\prime}(\phi^{h})N_{j}dv,
\label{EqnRV7}
\end{align}
where $h^{\prime}(\phi^{h})$ is the first derivative of the interpolation function $h(\phi^{h})$. 

Lastly, we take the derivative of Eq. (\ref{EqnRV4}) with respect to the nodal values $X_{kj}$. This yields
\begin{align}
\mathcal{J}_{ij} = -\int_{V}N_{i}N_{j}dv.
\label{EqnRV7b}
\end{align}

By implementing Eqs. (\ref{EqnRV4}), (\ref{EqnRV6}), (\ref{EqnRV7}) and (\ref{EqnRV7b}) we can obtain the diffusion potential of a component $k$. For a binary alloy, it is implemented as a C++ object: \texttt{BinaryPhaseConstraintMu.C}. This code is publicly available at https://github.com/souravmat-git/gibbs.

It should be noted that so far we have tacitly assumed that the eigenstrains and elastic modulus are constants. Consequently, the supplied \textit{phase} mole fractions are not explicitly dependent on the elastic fields \textit{within} the bulk phases (see Eq.\ref{Eqn16}). This however does not imply that the diffusion potential, and consequently the phase mole fractions, are independent of elastic fields. The phase compositions \textit{at the interface} are still dependent on the elastic fields, and consequently on the interface position. More concretely, in case of a sharp interface model, this dependence was analytically shown by Johnson \cite{Johnson1987} (see Eq. 17). As a result of this dependence, the growth velocity depends on the elastic fields \cite{Johnson1987}.

However, for cases when the eigenstrains and elastic modulus are dependent on composition, there is an explicit dependence of phase mole fractions on elastic fields \textit{within} the bulk phases. As an example, for a binary A-B alloy to model only composition-dependent eigenstrains, using Eq. (18) we obtain
\begin{align}
X_{B}^{\theta}(\mu_{B}, \boldsymbol{\epsilon}^{\theta})  =  X_{B}^{\theta}(\tilde{\mu}_{B}) + V_{m}\left(\frac{\partial \boldsymbol{\epsilon}^{\star\theta}}{\partial \tilde{\mu}_{B}}:\boldsymbol{\sigma}^{\theta}(\boldsymbol{\epsilon}^{\theta})\right).
\label{EqnRV8a}
\end{align}
Note that as a result of the second term there is a now a dependence of phase compositions on elastic fields even within the bulk phases. Since it is difficult to determine the dependence of eigenstrains on diffusion potentials, we can use chain rule to write
\begin{align}
\frac{\partial \boldsymbol{\epsilon}^{\star\theta}}{\partial \tilde{\mu_{B}}} = \frac{\partial \boldsymbol{\epsilon}^{\star\theta}}{\partial X_{B}^{\theta}}\frac{\partial X_{B}^{\theta}}{\partial \tilde{\mu_{B}}} = \frac{\partial \boldsymbol{\epsilon}^{\star\theta}}{\partial X_{B}^{\theta}}\chi_{BB}^{\theta},
\label{EqnRV8b}
\end{align}
where $\chi_{BB}^{\theta}$ is the solute susceptibility of phase $\theta$. Assuming a first order Taylor dependence of eigenstrains on composition \cite{Wu2001}, we obtain
\begin{align}
\frac{\partial \boldsymbol{\epsilon}^{\star\theta}}{\partial X_{B}} = \boldsymbol{\epsilon^{\star\theta}}\left[X_{B}^{\theta}(\boldsymbol{\tilde{\mu}^{h}}) - X_{B}^{0}\right],
\label{EqnRV8}
\end{align}
where $X_{B}^{0}$ is the overall (or average) alloy mole fraction. Eq. (\ref{EqnRV8}) indicates that the eigenstrains are dependent on the the local compositions within the bulk phases. Moreover, Eqs. (\ref{EqnRV8a}), (\ref{EqnRV8b}) and (\ref{EqnRV8}) are implemented as \texttt{Material} objects  namely:  \texttt{StrainDependentTaylorApproximation.C}, \texttt{EigenStrainPhaseMaterial.C} and \texttt{EigenStrainPrefactor.C}, respectively, and are freely available at https://github.com/souravmat-git/gibbs.

\renewcommand{\thesubsection}{F.\arabic{subsection}}
\renewcommand{\theequation}{F.\arabic{equation}}
\renewcommand{\thefigure}{F.\arabic{figure}}
\setcounter{equation}{0}
\setcounter{subsection}{0}
\setcounter{figure}{0}
\section*{Appendix F}
\section*{Discussion on the non-variational term}
\color{black}
For a discussion on the role of non-variational term in case of the partial rank-one homogenization scheme, we have divided this appendix into two sections. First, we show the influence of the variational term on simulation accuracy for three of the considered simulations. We also show that in one of the simulations the inclusion of this term leads to a non-converging solution. Following this, we discuss possible reasons for this non-convergence.

\subsection{Problem}
To understand the effect of the variational term on simulation accuracy, we repeated the non-planar single particle $\gamma/\gamma^{\prime}$ simulation with an interface width of $0.10$ $\mu$m with the variational term. Fig.\ref{Fig.R1} shows that the presence of this term has no effect on the variation of interface position as a function of square root of time. We further note that this behaviour was also observed in case of the planar Ni-Al $\gamma/\gamma^{\prime}$ simulation. However, for the case of non-planar single particle UO$_2$/void simulation having an interface width of $0.10$ $\mu$m, we found that due to the presence of the variational term the simulations were not converging. Fig.\ref{Fig.R2}  shows that, although the simulation accuracy remains unaffected, the run time in the presence of variational term is significantly lower compared to the simulation without this term. Moreover, the computation time is approximately $6$ timer higher in the presence of the variational term compared to the simulation without this term (Fig.\ref{Fig.R4}). Contrastingly, for the planar UO$_2$/void case with the same ratio of Young's moduli, we found that this term had no influence on simulation accuracy for the same interface width value $0.10$ $\mu$m  (Fig.\ref{Fig.R4a}). 

The reason for the increase in computation time, in case of non-planar UO$_2$/void simulation with the variational term, is the concurrent effect of increase in the number of non-linear iterations (Fig.\ref{Fig.R5}) and the decrease in time step size (Fig.\ref{Fig.R6}), as compared to the simulation without this term. Nevertheless, it is noteworthy that there was no speed-up in terms of computation time  (Fig.\ref{Fig.R3}) and time step size  (Fig.\ref{Fig.R4b}) due to the exclusion of this term in case of non-planar Ni-Al and planar UO$_2$/Va cases, respectively.  Next, we attempt to explain this poor convergence for the case of non-planar UO$_2$/void simulation with the variational term.
\subsection{Reason}
For sake of discussion, we note that the variational term is the divergence of a vector quantity, $\partial \omega_{bulk}/\partial \nabla \phi$, which has dimensions of J/m$^2$. So in order to determine the influence of the variational term, we look at the magnitude of this vector quantity. Since this quantity is temporally and spatially dependent, we calculate its local and average values. For our simulations, we define the average value of a quantity, $\boldsymbol{\Phi}$, as 
\begin{align}
\Phi_{avg}(t) = \frac{\int_{V}\boldsymbol{\Phi}(\boldsymbol{x},t)dv}{\int_{V}dv}
\label{EqnAvg}
\end{align}

Additionally, since we employ non-dimensional quantities, the vector term can be written in non-dimensional form as
\begin{align}
\frac{\partial \overline{\omega_{bulk}}}{\partial \overline{\nabla} \phi} &=  \frac{\mu_{el}}{m}\left[h(\phi)[1-h(\phi)]\left\{\overline{\sigma_{jk}^{\alpha}} - \overline{\sigma_{jk}^{\beta}}\right\}\frac{\partial n_{k}}{\partial \phi_{,i}}a_{j}\right],
\label{EqnRV14}
\end{align}
where $\mu_{el}$ [J/m$^{3}$] is a characteristic modulus used to non-dimensionalize the stresses and $m$ [J/m$^{3}$] is the barrier height. Note that $\overline{\boldsymbol{\sigma}}$ indicates non-dimensional values. As discussed in the text, we have taken $\mu_{el}$ to be the shear modulus of the $\gamma^{\prime}$ precipitate and UO$_2$ matrix in the Ni-Al and UO$_2$/vacancy simulations, respectively. Further, it can be shown that
\begin{align}
\frac{\partial n_{k}}{\partial \phi_{,i}} = -\frac{\delta_{ki}}{\norm{\nabla \phi}} + \frac{\phi_{,k}\phi_{,i}}{\norm{\nabla \phi}^{3}},
\label{EqnRV15}
\end{align}
where $\norm{\nabla \phi}$ is the norm of $\nabla \phi$. By substituting Eq. (\ref{EqnRV15}) in  Eq.(\ref{EqnRV14}) we can rewrite the total vector term as a vector sum of two vectors, $\mathbf{V1}$ and  $\mathbf{V2}$, such that
\begin{align}
\frac{\partial \overline{\omega_{bulk}}}{\partial \overline{\nabla} \phi} = \mathbf{V1} + \mathbf{V2},
\label{EqnRV16}
\end{align}
where
\begin{align}
\mathbf{V1} &= -\frac{\mu_{el}}{m}\left[h(\phi)[1-h(\phi)]\left\{\overline{\sigma_{ji}^{\alpha}} - \overline{\sigma_{ji}^{\beta}}\right\}\frac{a_{j}}{\norm{\nabla \phi}}\right],\label{EqnRV16a}\\
\mathbf{V2} &= \frac{\mu_{el}}{m}\left[h(\phi)[1-h(\phi)]\left\{\overline{\sigma_{jk}^{\alpha}} - \overline{\sigma_{jk}^{\beta}}\right\}\frac{\phi_{,k}\phi_{,i}}{\norm{\nabla \phi}^{3}}a_{j}\right].\label{EqnRV16b}
\end{align}
It is important to note that due to the factor $h(\phi)[1-h(\phi)$, the vector terms in Eqs. (\ref{EqnRV16}) (\ref{EqnRV16a}) \& (\ref{EqnRV16b}) are non-zero only in the interfacial regions. Moreover, since we have defined the unit normal to the interface as $\boldsymbol{n} = -\phi_{,k}/\norm{\nabla \phi}$, and by employing PRH scheme static compatibility at the interface is ensured, we have 
\begin{align}
\left\{\sigma_{jk}^{\alpha} - \sigma_{jk}^{\beta}\right\}\frac{\phi_{,k}}{\norm{\nabla \phi}} = 0
\label{EqnRV17}.
\end{align}
As a consequence of Eq. (\ref{EqnRV17}), we note that the vector term $\mathbf{V2}$ in Eq.(\ref{EqnRV16b}) must be vanishingly small. To test this, we determined the average value of the magnitude of these vectors, i.e., Eqs. (\ref{EqnRV16}) (\ref{EqnRV16a}) \& (\ref{EqnRV16b}),  using Eq.(\ref{EqnAvg}). 

We find that the average value of the magnitude of $\norm{\mathbf{V2}}_{avg}$ is vanishingly small for both $\gamma/\gamma^{\prime}$ and UO$_{2}$/void simulations (Figs. \ref{Fig.R9} and \ref{Fig.R11}). This implies that $\norm{\partial \overline{\omega}_{bulk}/\partial \overline{\nabla} \phi}$ is nearly equal to $\norm{\mathbf{V1}}$. This is also evident from Figs. (\ref{Fig.R8}) and (\ref{Fig.R10}) in the case of $\gamma/\gamma^{\prime}$ and UO$_{2}$/void simulations. Moreover, it is important to observe that the magnitude of  $\norm{\mathbf{V1}}_{avg}$ is at least six orders of magnitude higher in case of UO$_2$/void compared to $\gamma/\gamma^{\prime}$ simulation. Since $\norm{\mathbf{V1}}_{avg} \gg \norm{\mathbf{V2}}_{avg}$ in both cases, we can neglect $\mathbf{V2}$ in our analysis. 

\begin{figure}[!ht]
\begin{subfigure}{0.5\textwidth}
\includegraphics[keepaspectratio,width=\textwidth]{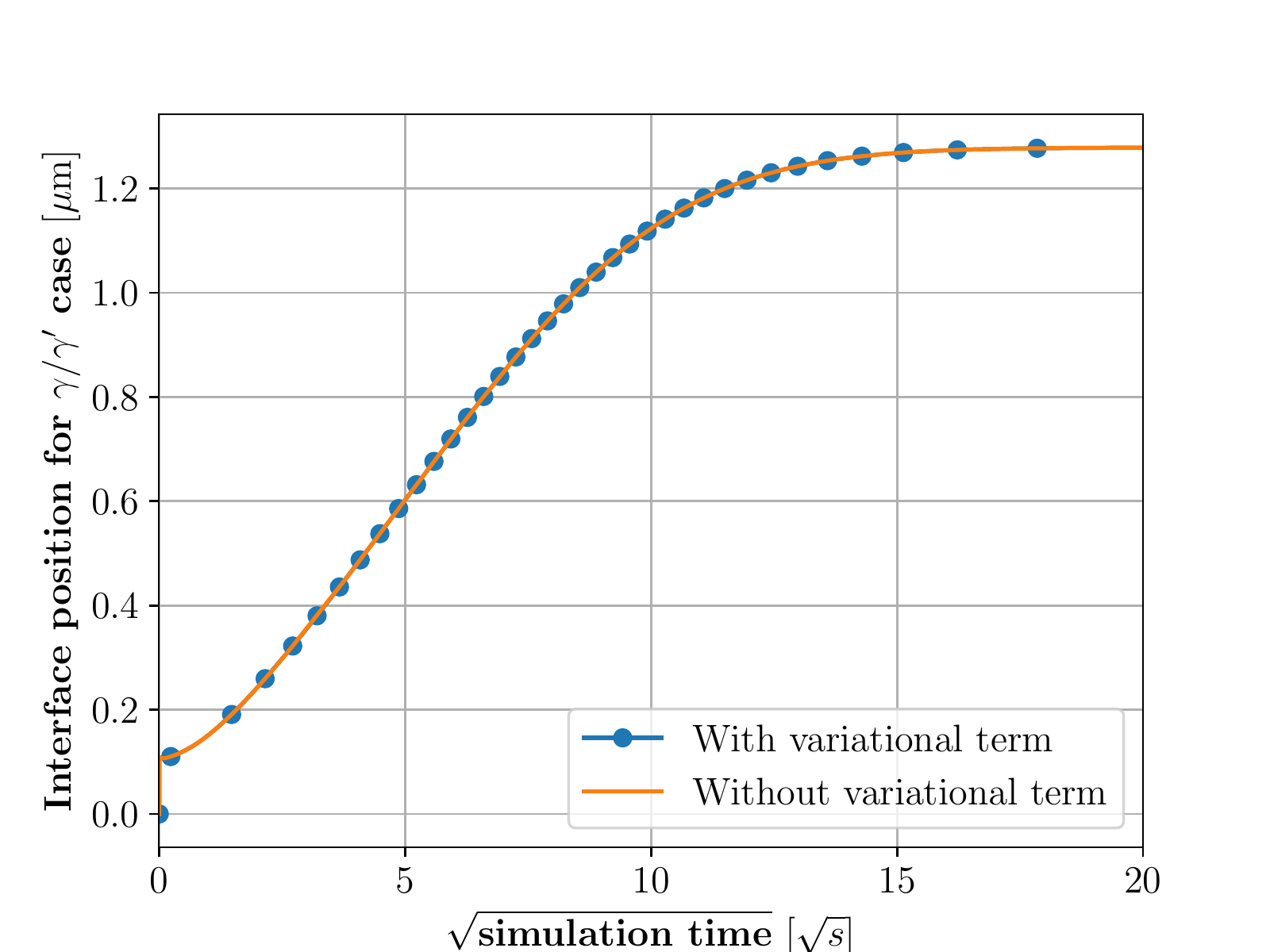}
\caption{}
\label{Fig.R1}
\end{subfigure}
\begin{subfigure}{0.5\textwidth}
\includegraphics[keepaspectratio,width=\textwidth]{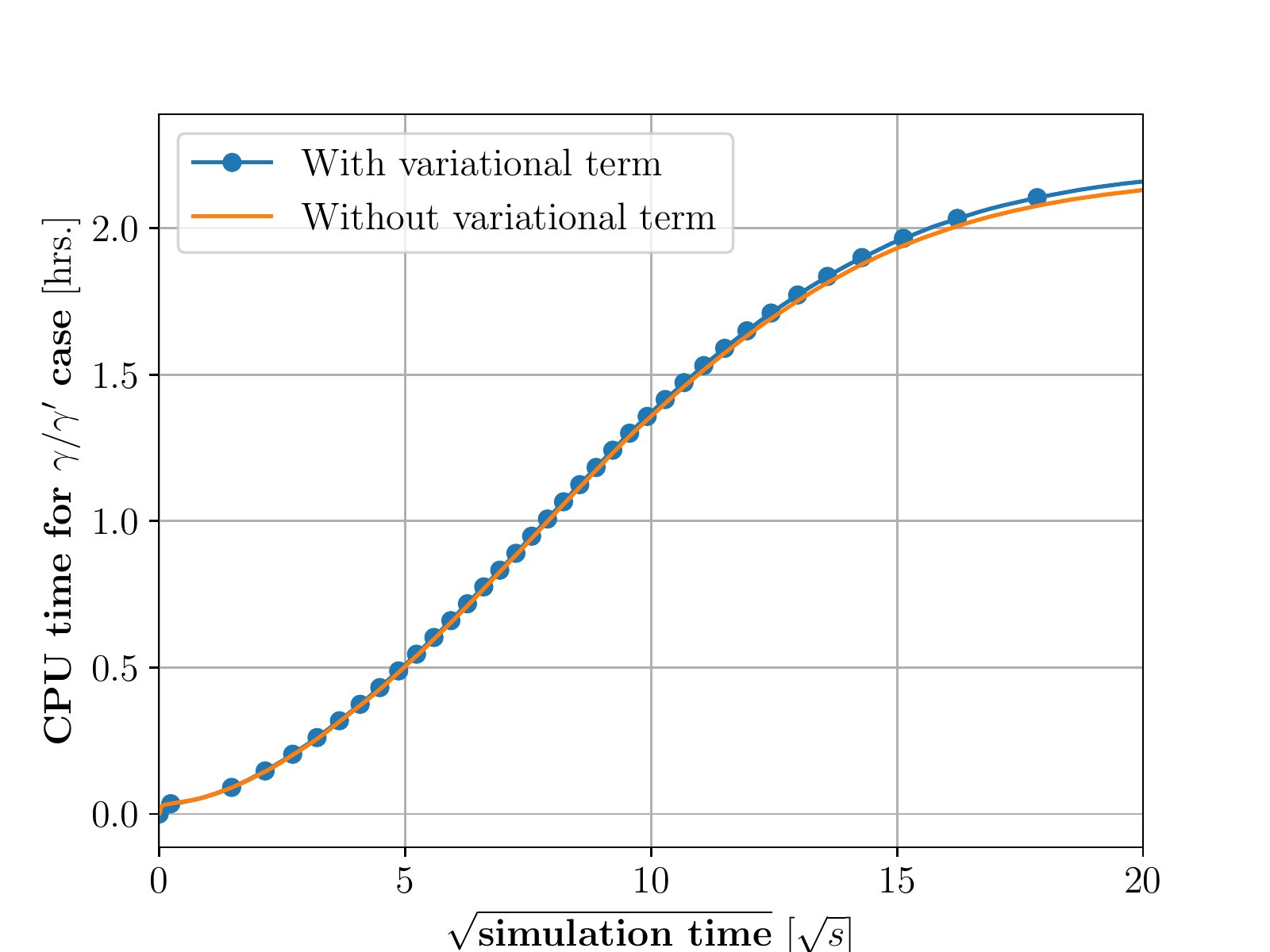}
\caption{}
\label{Fig.R3}
\end{subfigure}
\begin{subfigure}{0.5\textwidth}
\includegraphics[keepaspectratio,width=\textwidth]{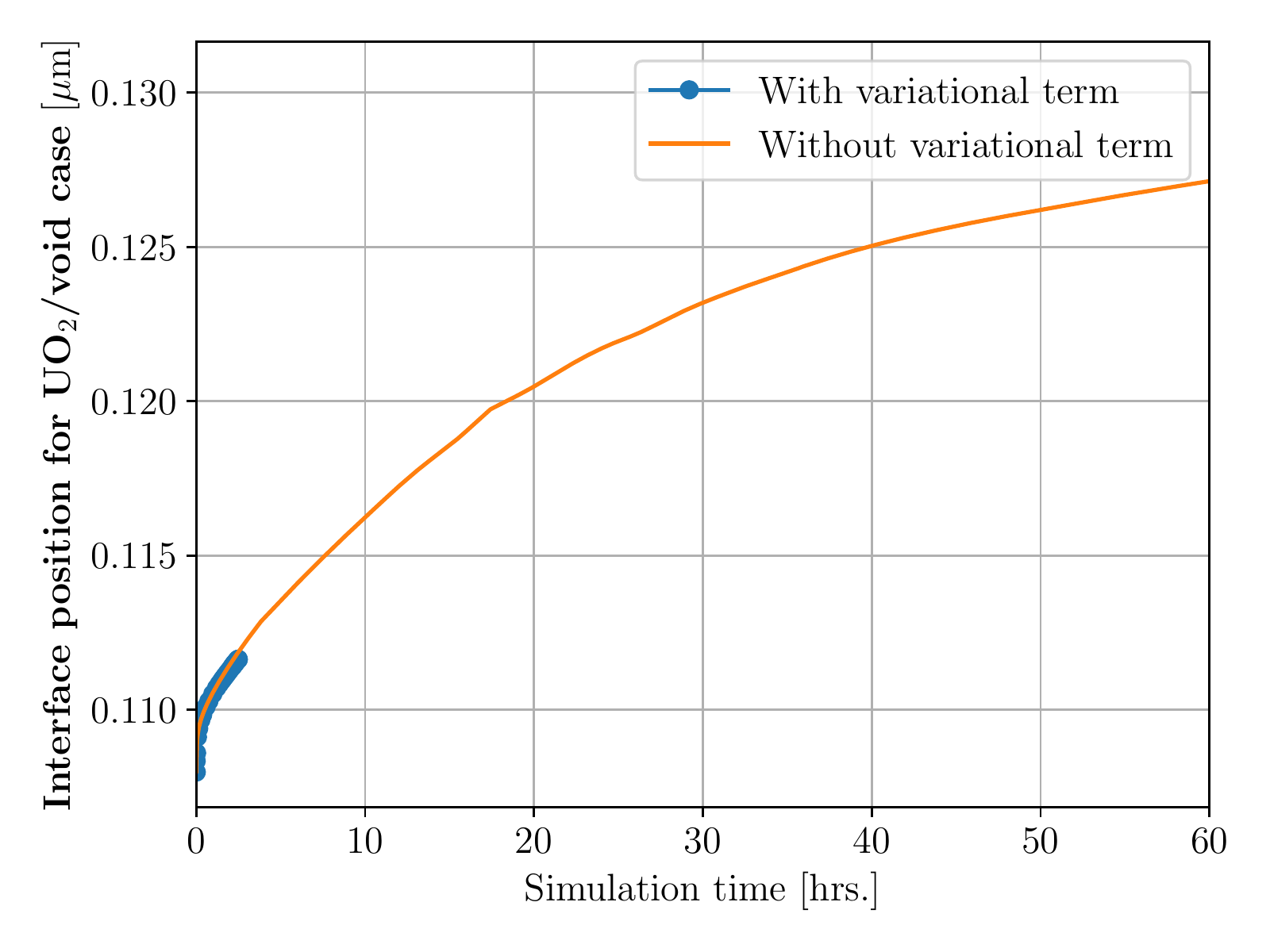}
\caption{}
\label{Fig.R2}
\end{subfigure}
\begin{subfigure}{0.5\textwidth}
\includegraphics[keepaspectratio,width=\textwidth]{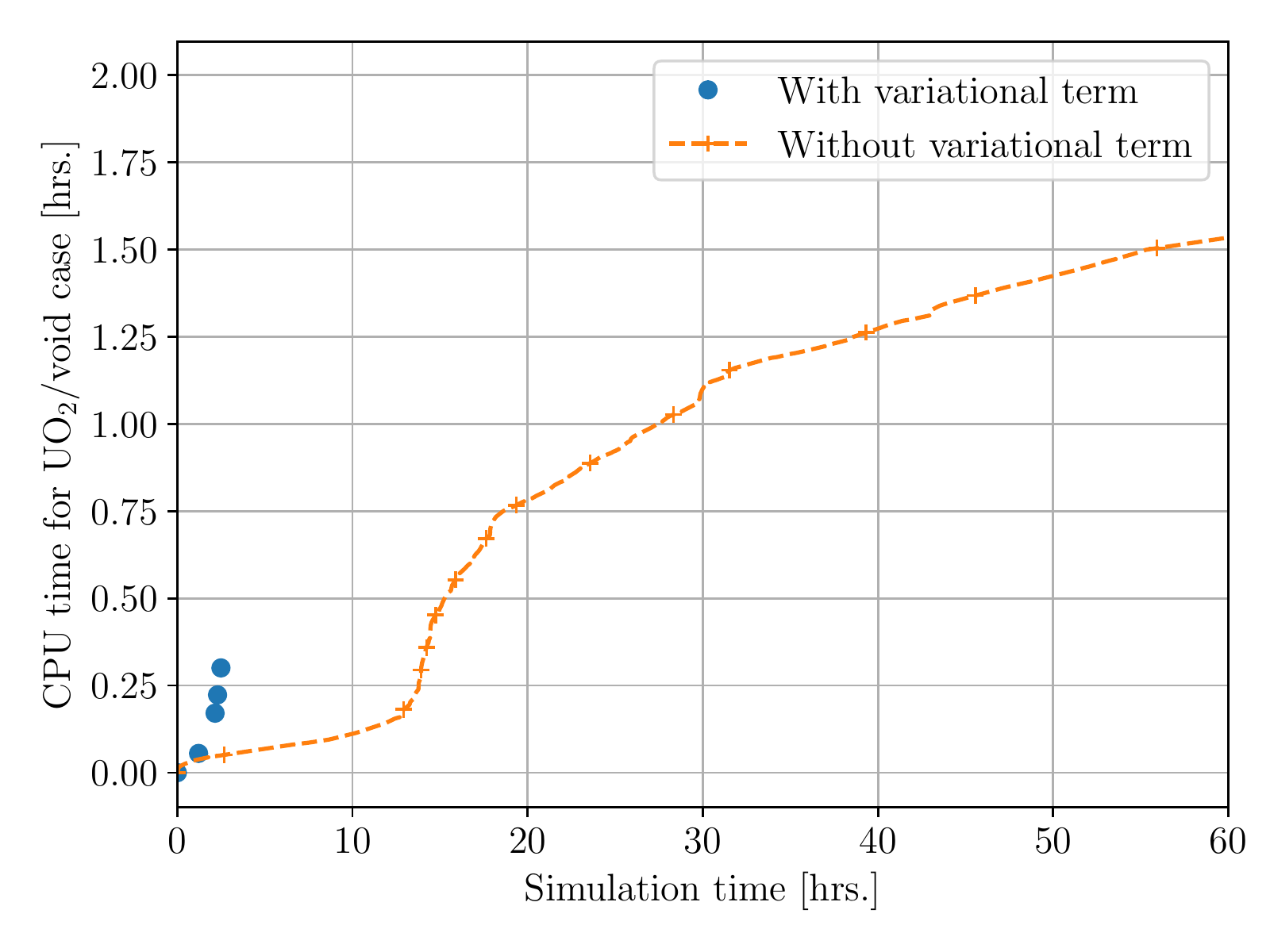}
\caption{}
\label{Fig.R4}
\end{subfigure}
\begin{subfigure}{0.5\textwidth}
\includegraphics[keepaspectratio,width=\textwidth]{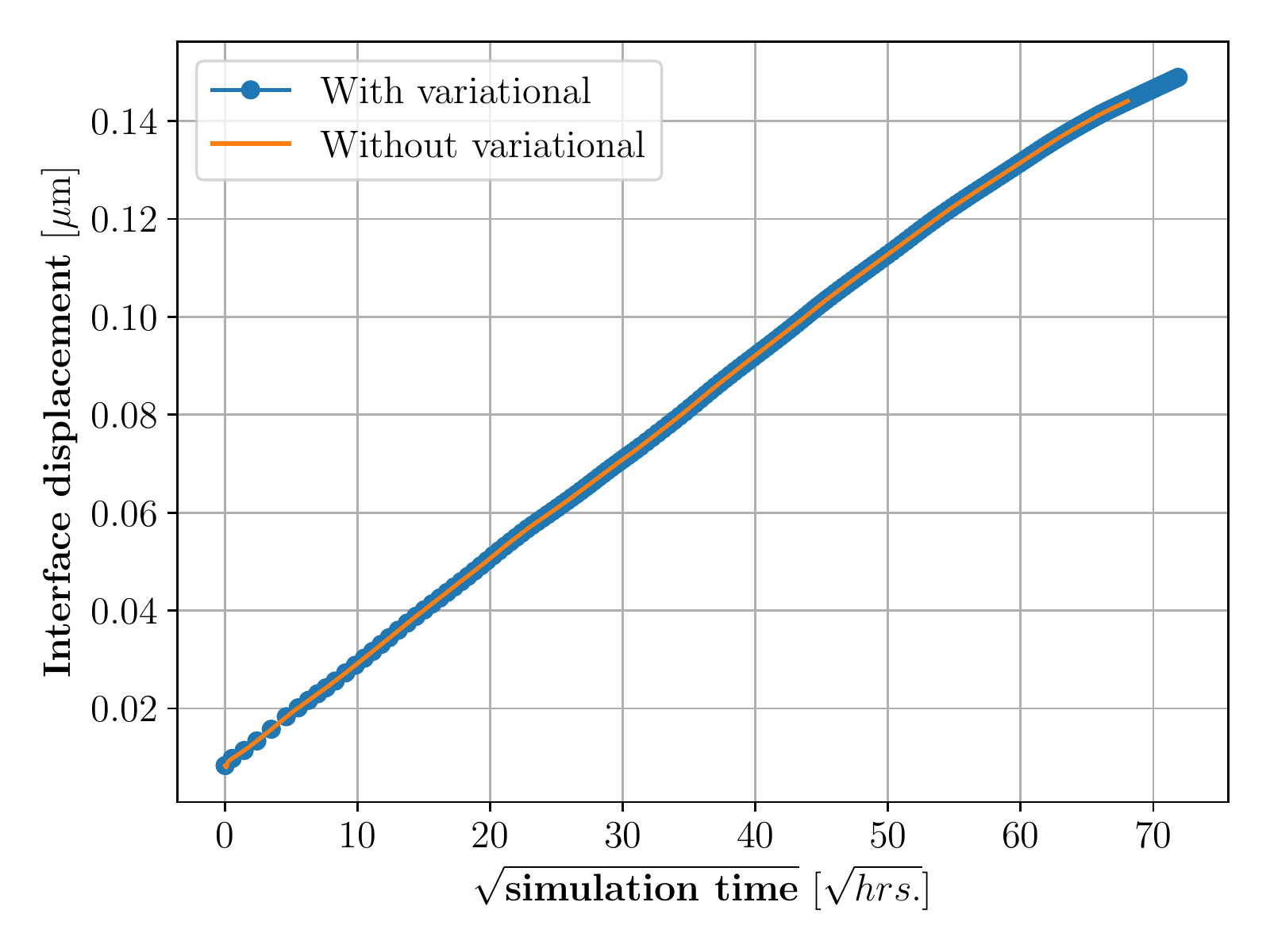}
\caption{}
\label{Fig.R4a}
\end{subfigure}
\begin{subfigure}{0.5\textwidth}
\includegraphics[keepaspectratio,width=\textwidth]{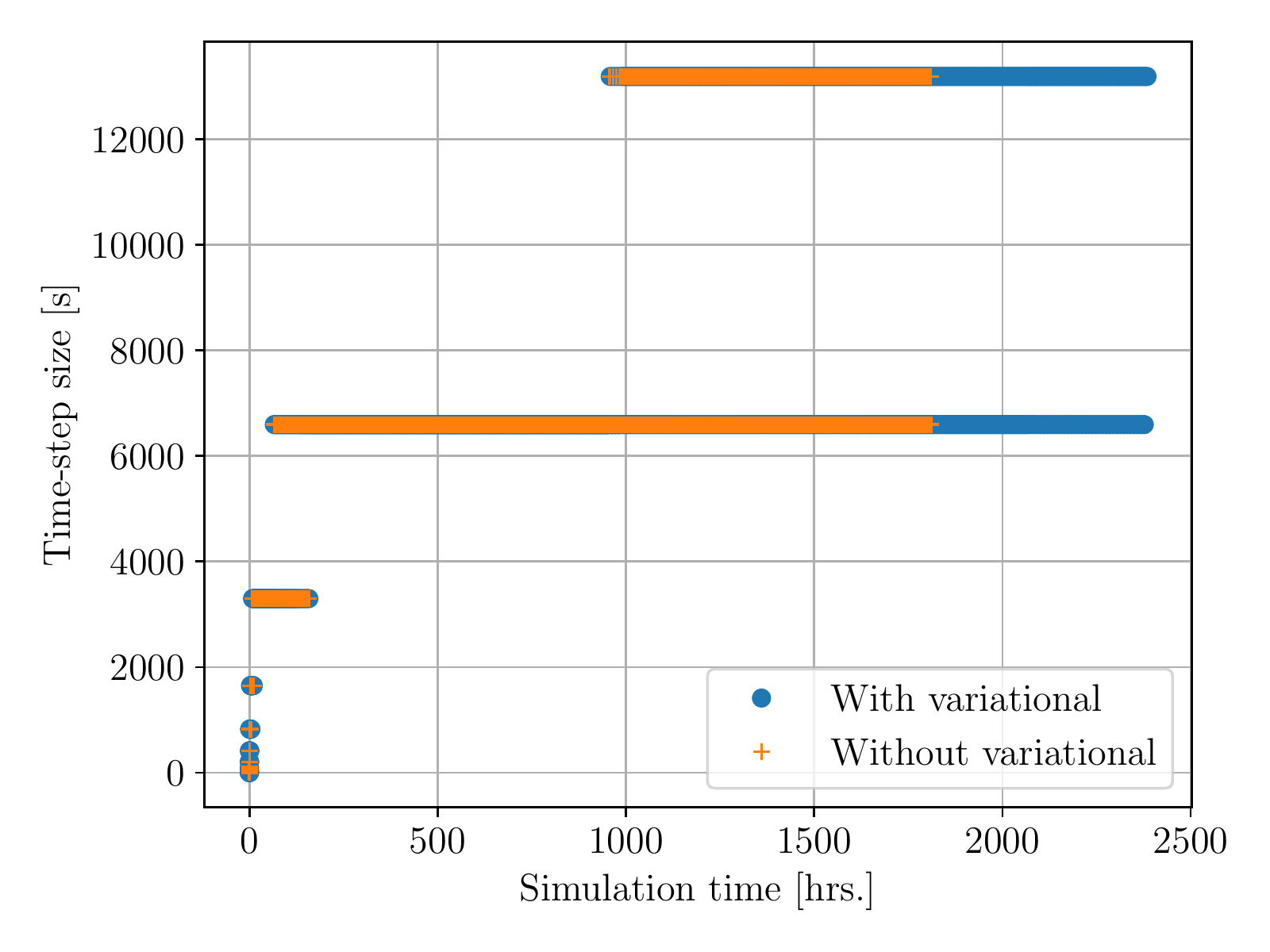}
\caption{}
\label{Fig.R4b}
\end{subfigure}
\caption{\textcolor{black}{For the case of non-planar $\gamma/\gamma^{\prime}$ simulation, a) variation in interface position as a function of square root of simulation time, and b) computation time as a function of simulation time with and without the variational term. Similarly, for the case of non-planar UO$_2$/void, c) variation in interface position as a function of simulation time; and d) computation time as a function of simulation time with and without the variational term. Lastly, for the case of planar UO$2$/void system,  e) variation in interface displacement as a function of square root of simulation time and b) time step size as a function of simulation time with and without the variational term.}}
\end{figure}

\begin{figure}[!ht]
\begin{subfigure}{0.5\textwidth}
\includegraphics[keepaspectratio,width=\textwidth]{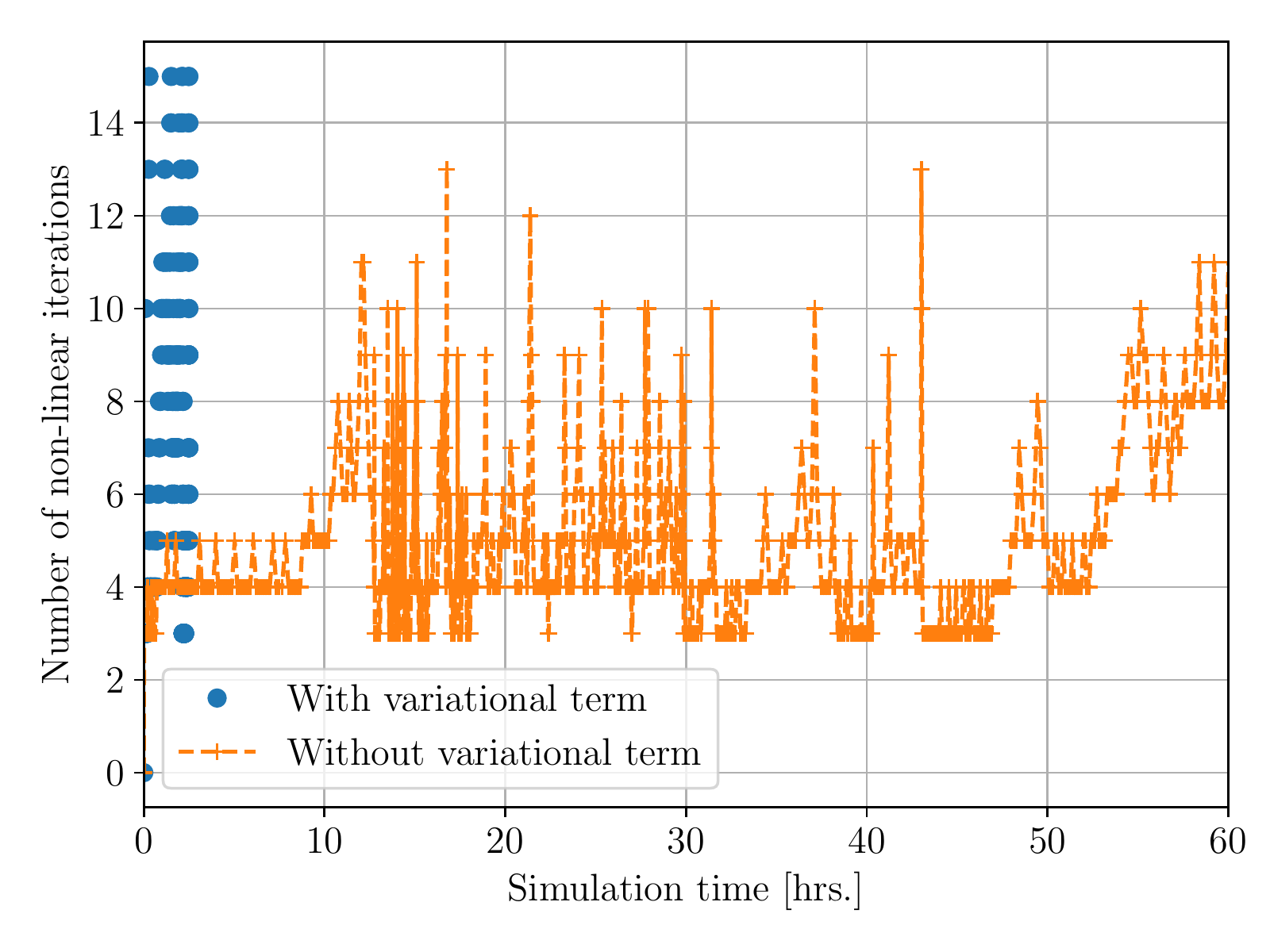}
\caption{}
\label{Fig.R5}
\end{subfigure}
\begin{subfigure}{0.5\textwidth}
\includegraphics[keepaspectratio,width=\textwidth]{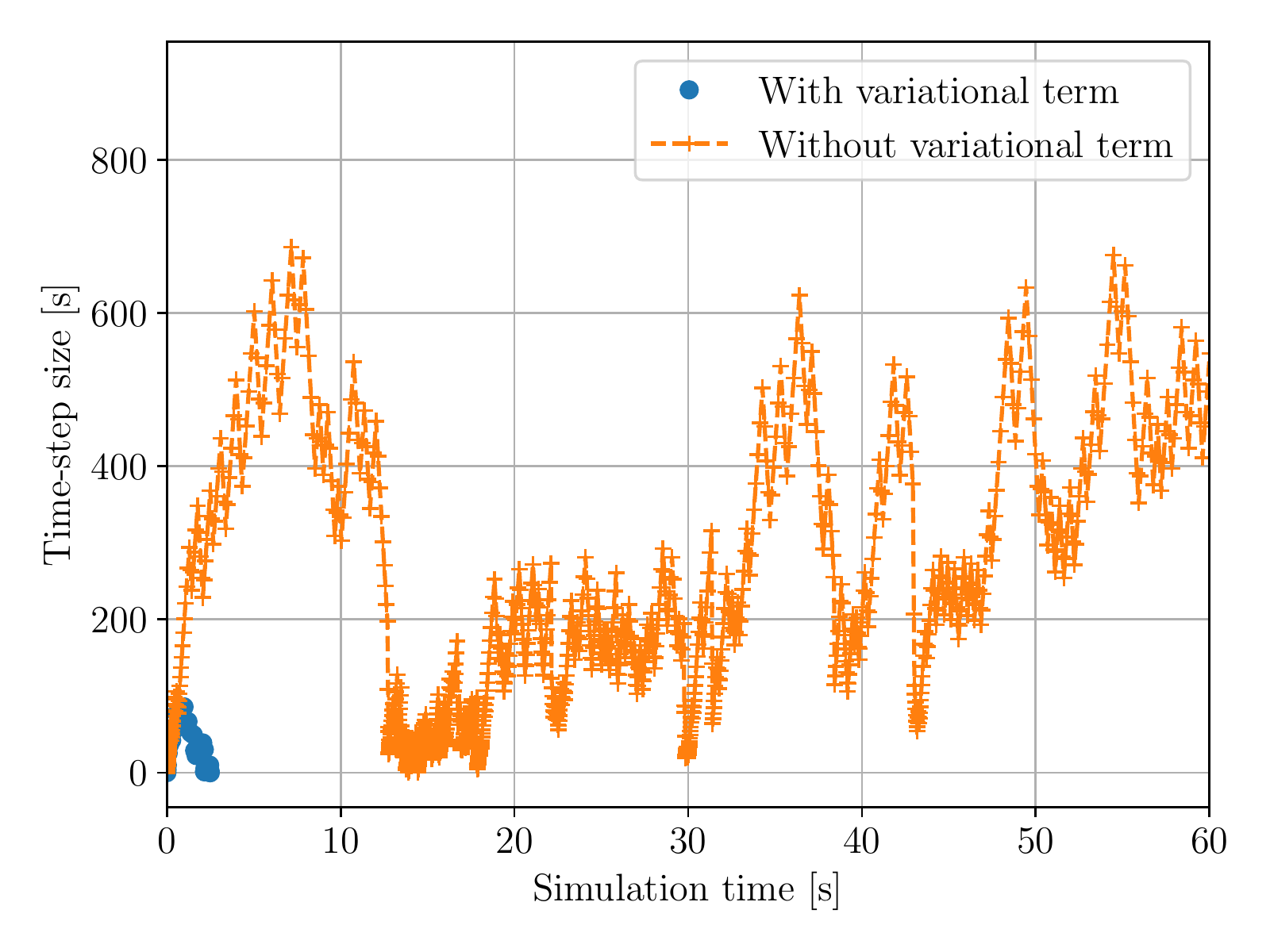}
\caption{}
\label{Fig.R6}
\end{subfigure}
\caption{\textcolor{black}{For the case of non-planar UO$_2$/ void system, a) number of non-linear iterations as a function of simulation time, and b) time step size with and without the variational term.}}
\end{figure}

\begin{figure}[!ht]
\begin{subfigure}{0.5\textwidth}
\includegraphics[keepaspectratio,width=\textwidth]{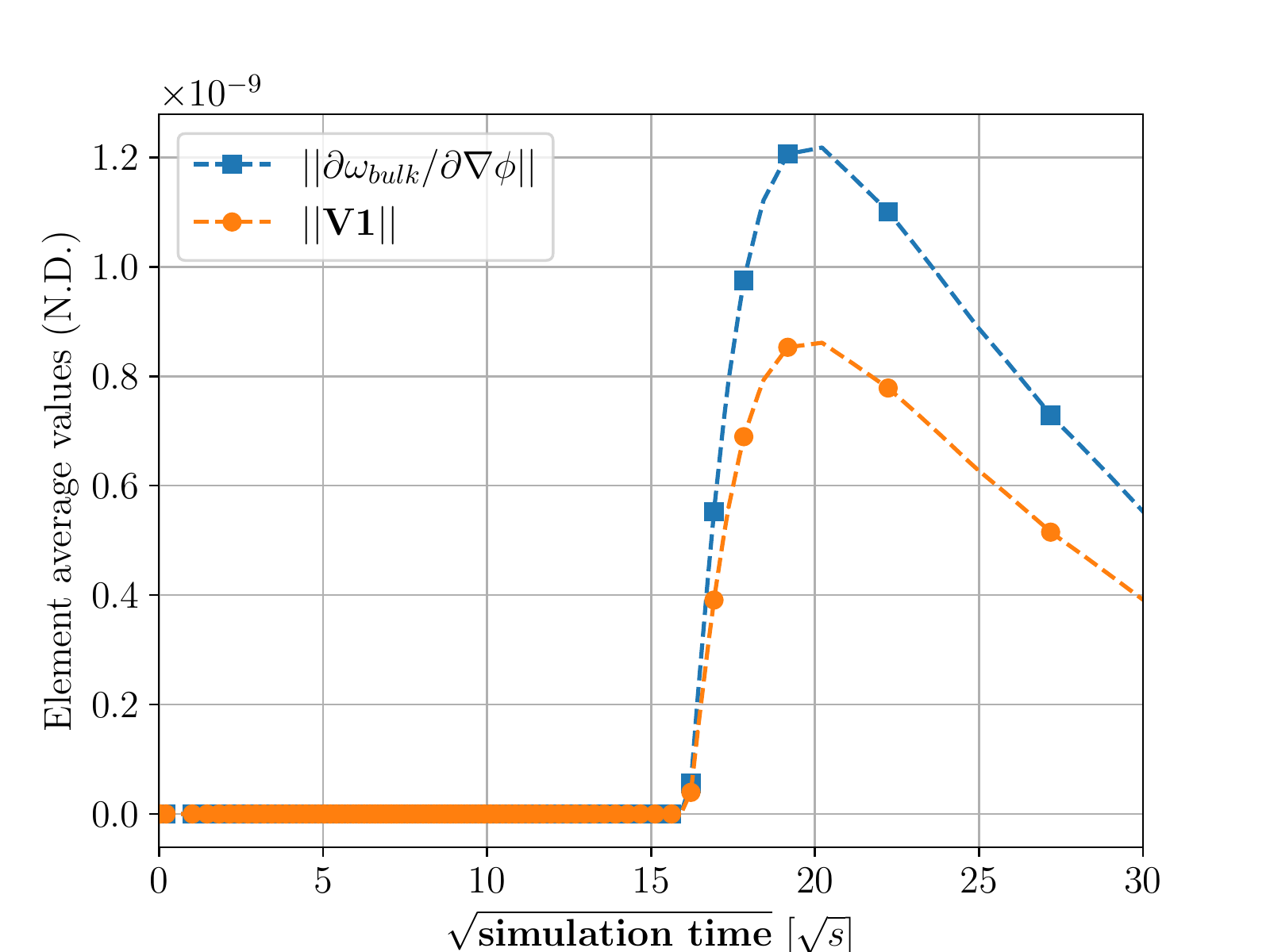}
\caption{}
\label{Fig.R8}
\end{subfigure}
\begin{subfigure}{0.5\textwidth}
\includegraphics[keepaspectratio,width=\textwidth]{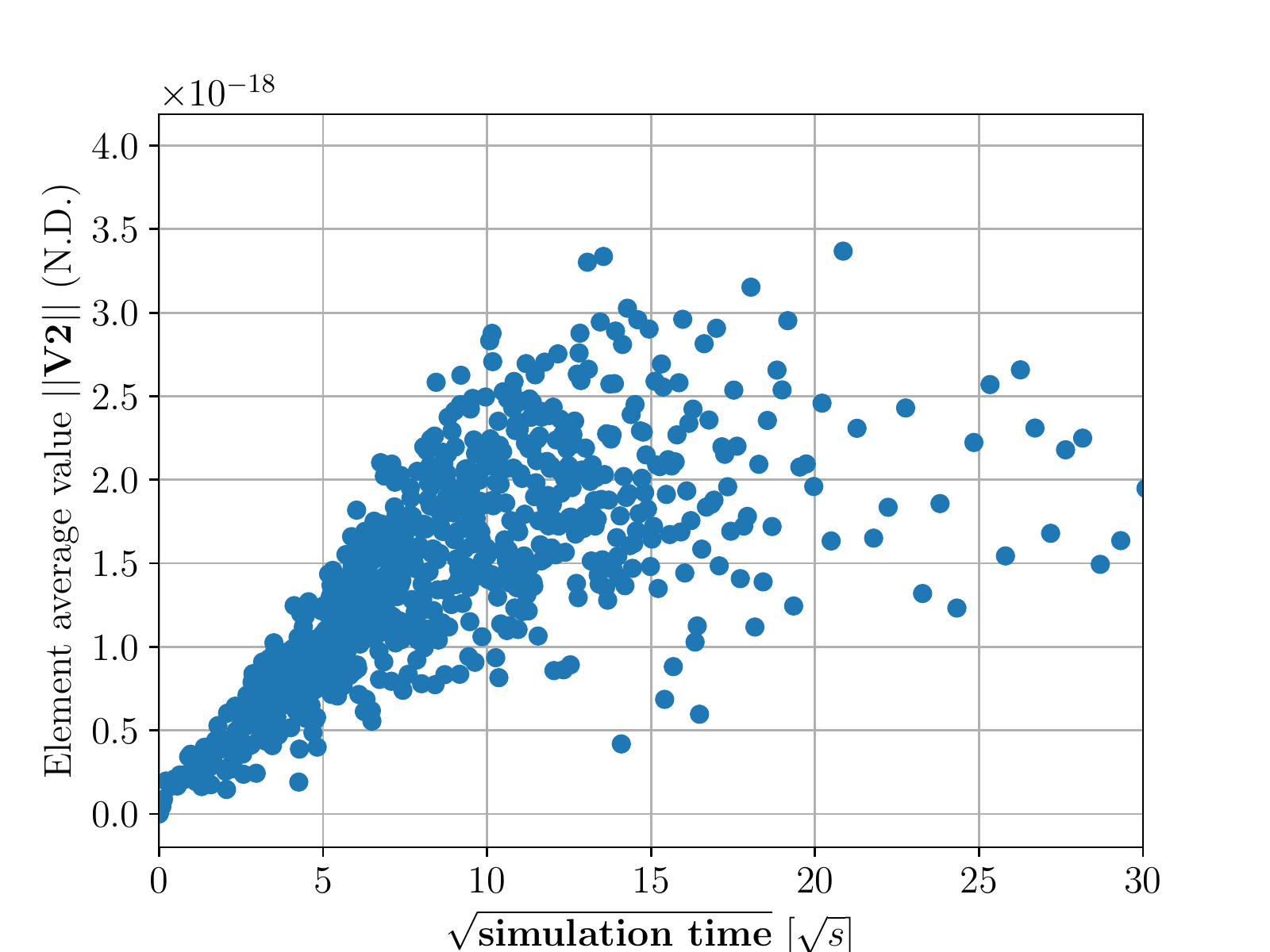}
\caption{}
\label{Fig.R9}
\end{subfigure}
\begin{subfigure}{0.5\textwidth}
\includegraphics[keepaspectratio,width=\textwidth]{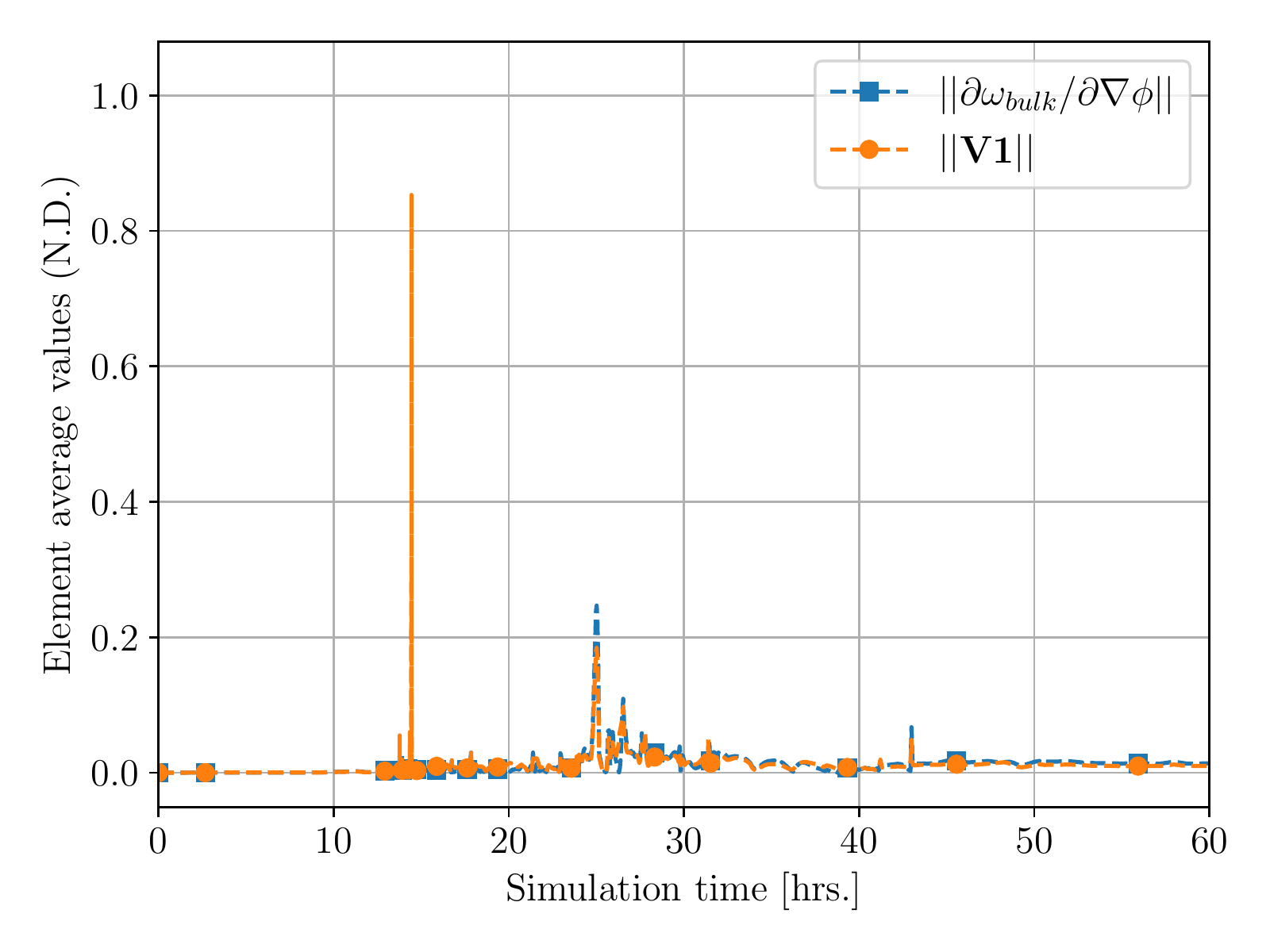}
\caption{}
\label{Fig.R10}
\end{subfigure}
\begin{subfigure}{0.5\textwidth}
\includegraphics[keepaspectratio,width=\textwidth]{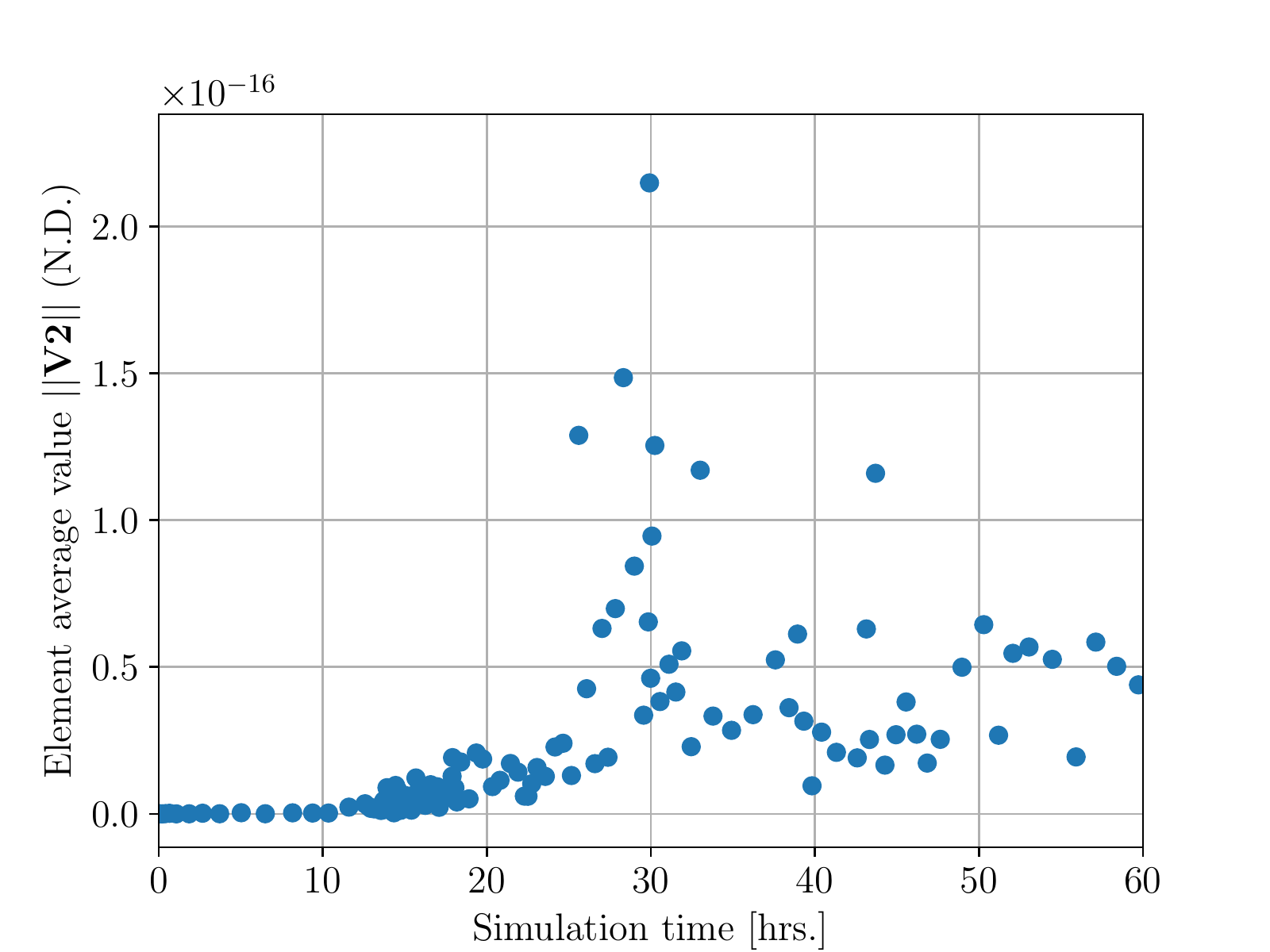}
\caption{}
\label{Fig.R11}
\end{subfigure}
\caption{\textcolor{black}{Temporal variation in the average values of the magnitude of the vectors contributing to the variational term see Eqs. (\ref{EqnRV16}) -(\ref{EqnRV17}). Here, N.D. denotes non-dimensional values. Note that the average value of the magnitude of vector $\norm{\mathbf{V2}}$ is negligible compared to  $\norm{\mathbf{V1}}$.}}
\end{figure}

\begin{figure}
\begin{subfigure}{0.5\textwidth}
\includegraphics[keepaspectratio,width=\textwidth]{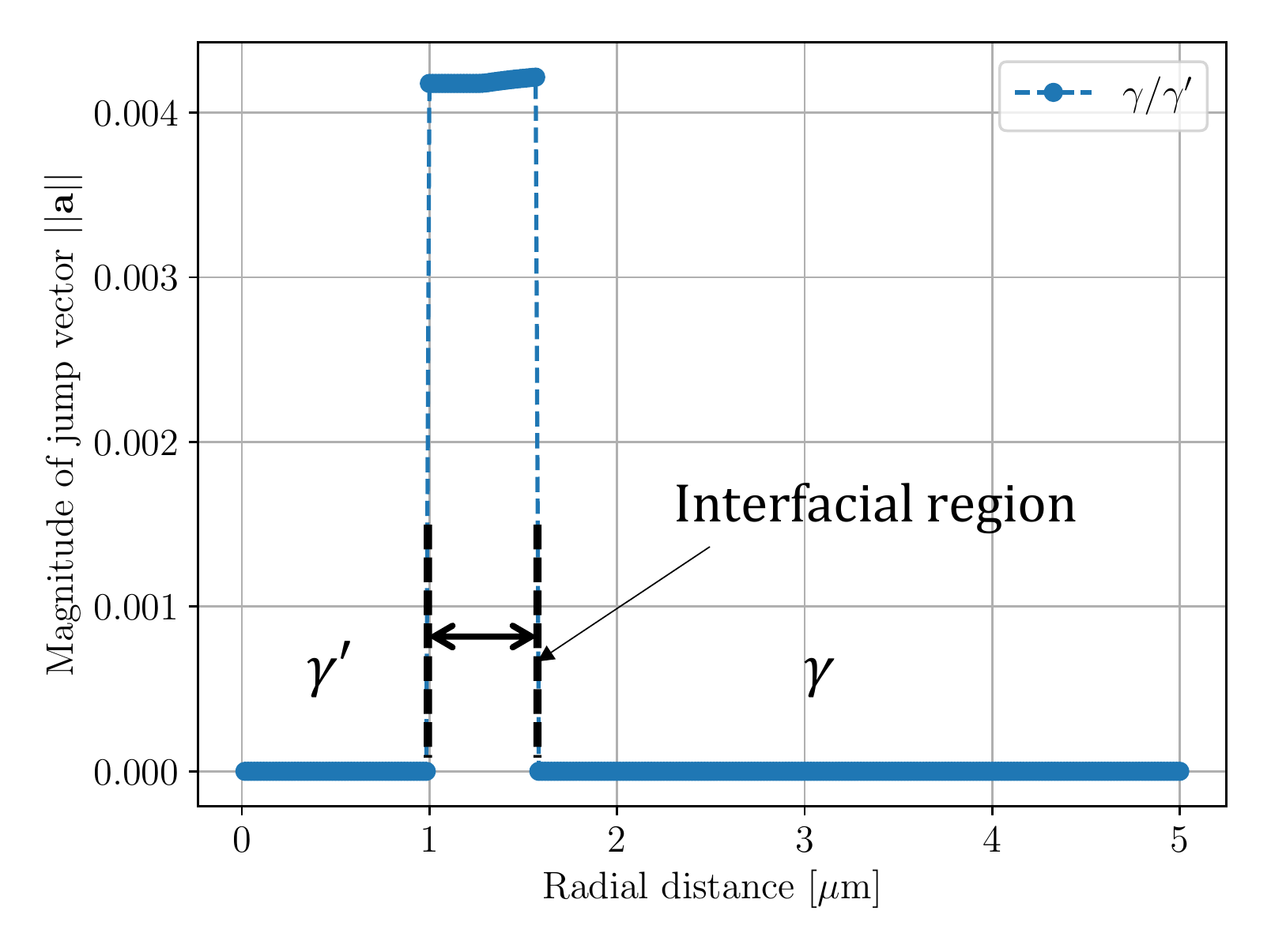}
\caption{}
\label{Fig.R10}
\end{subfigure}
\begin{subfigure}{0.5\textwidth}
\includegraphics[keepaspectratio,width=\textwidth]{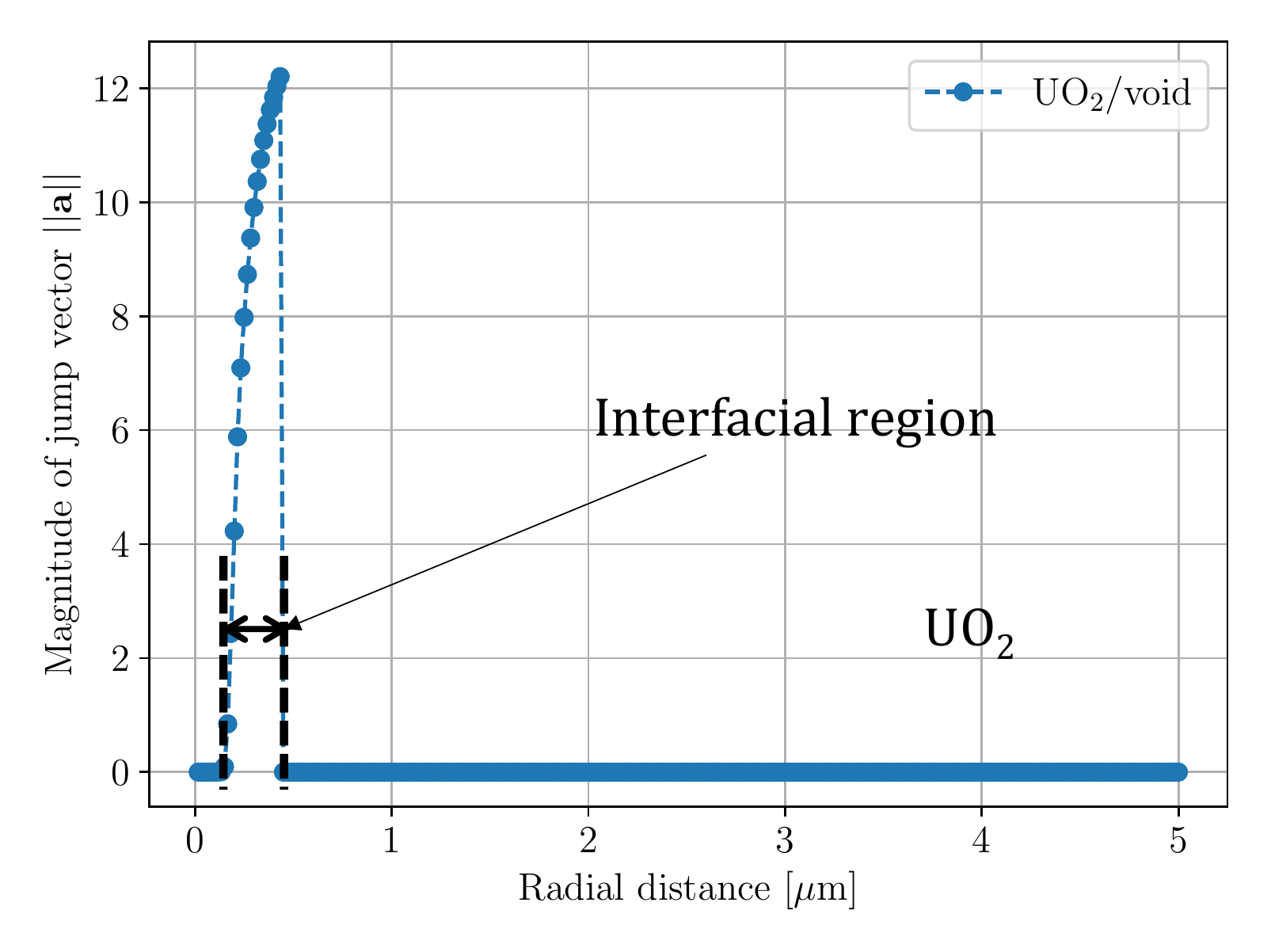}
\caption{}
\label{Fig.R11}
\end{subfigure}
\begin{center}
\begin{subfigure}{0.5\textwidth}
\includegraphics[keepaspectratio,width=\textwidth]{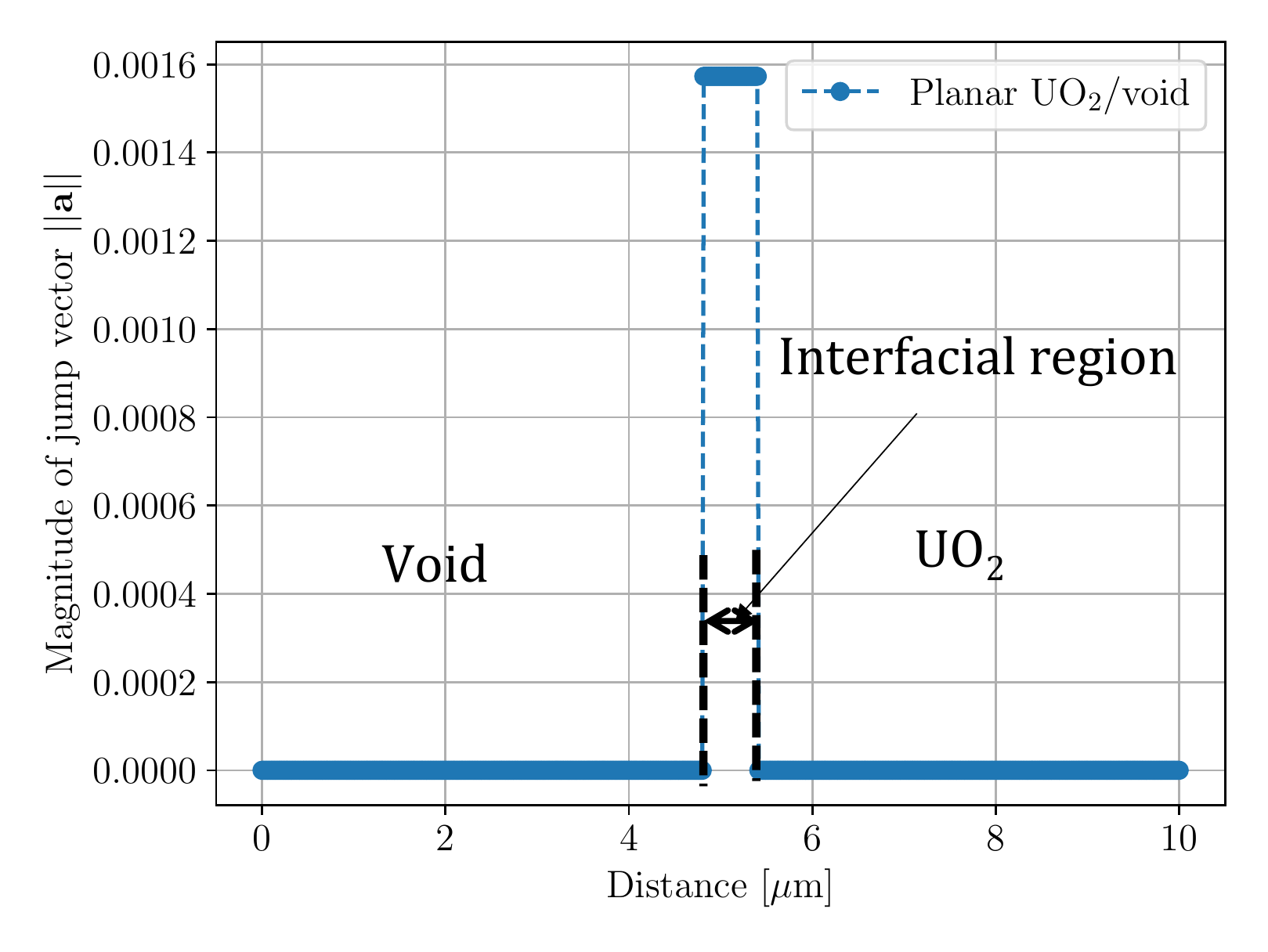}
\caption{}
\label{Fig.R13a}
\end{subfigure}
\end{center}
\caption{\textcolor{black}{Spatial variation in the magnitude of jump vector $\norm{\boldsymbol{a}}$  in case of a) non-planar $\gamma/\gamma^{\prime}$, b) non-planar UO$_2$/void and c) planar UO$_2$/void system. For the non-planar cases, the jump was calculated along the radial direction, while for the planar case, it was normal to the interface. The dotted lines mark the interfacial regions where the strain jump vector $\boldsymbol{a}$ is non-zero. This interfacial region is based on the cut-off value set to distinguish the bulk from the interfacial regions. For our simulations, it was set to $\norm{\nabla \phi}^{2} < 1\mathrm{e}{-18}$.}}
\end{figure}

Now, since $\mathbf{V1}$ is directly proportional to the strain jump vector $\boldsymbol{a}$, we determined its local and average value in both cases. Figs. \ref{Fig.R10}  and \ref{Fig.R11} show the local variation in the jump magnitude,$\norm{\boldsymbol{a}}$, for the case of $\gamma/\gamma^{\prime}$ and UO$_2$/void simulations, respectively. We find that the jump magnitude in the UO$_2$/void case (see Fig. \ref{Fig.R11}) is nearly 300 times higher compared to the $\gamma/\gamma^{\prime}$ case (see Fig. \ref{Fig.R10}). This is due to the high phase contrast in the former case as compared to the latter. Moreover, it is due to this high magnitude of $\norm{\boldsymbol{a}}$ that the presence of the variational term leads to a non-convergent solution. For this particular case, it is therefore difficult to exactly ascertain its effect on simulation accuracy for longer computation time. Nevertheless, despite having the same ratio of Young's moduli, in case of the planar UO$_2$/void simulation this term has no effect on solution accuracy (Fig.\ref{Fig.R4a}). This can again explained by calculating the strain jump magnitude $\norm{\boldsymbol{a}}$. We find that the strain jump magnitude in the planar case is at least four orders of magnitude lower compared to the non-planar UO$_2$/void case (compare Figs. \ref{Fig.R13a} \& \ref{Fig.R11}). This value is lower in the planar case possibly because the strain jump vector $\boldsymbol{a}$ depends on both the difference in Young's moduli between the phases and the total strain $\boldsymbol{\epsilon}$ at the interface (Eq. 9).

Moreover, it should be emphasized that the driving force due to bulk contributions, $\delta \omega_{bulk}/\delta \phi$, consists of two terms in case of the partial rank-one scheme: the non-variational term $\partial \omega_{bulk}/\partial \phi$ and the variational contribution (Eq. 22 in the revised manuscript). Although not shown here, we found that the vector term was significantly lower in case of Ni-Al simulations compared to the non-variational term. Therefore, the variational term had no effect on the accuracy of the solution, and the interface was driven solely by the non-variational term. Contrastingly, this term was not insignificant compared to the non-variational term for the case of UO$_2$/void. We think because of the high magnitude of this term relative to the non-variation term the simulations were not converging. It is worth noting  that ``thin-interface'' phase-field models where a relation between model parameters and interfacial quantities can be obtained are particularly suited when the driving forces due to bulk contributions are small \cite{NProvatas2011}. It also partially explains why in case of both planar and non-planar UO$_2$/void cases, the variation in interface position showed a dependence on interface width (Figs. 4c and 9c) but not in case of Ni-Al simulations.

\clearpage
\bibliography{paper1}

\end{document}